\documentclass[12pt]{iopart}

\usepackage[utf8]{inputenc}
\usepackage{graphicx,times}             
\usepackage{natbib}
\usepackage{url}
\usepackage{amsmath}
\usepackage{amssymb}
\usepackage{iopams}  
\usepackage{footnote}
\usepackage{lipsum}
\usepackage{hyperref}
 \usepackage{xcolor} 

\providecommand{\newblock}{\hskip .11em\@plus.33em\@minus.07em}
\makeatletter
\@ifundefined{thebibliography}{%
}{}
\makeatother

\begin{document}


\title[The First Infrared Portrait of A Solar-Like Star]{The First Infrared Portrait of A Solar-Like Host Star with Debris Disk: Pioneering High-Resolution H- and K- Band Spectroscopy of HD\,115617 with Comparative Optical Spectrum Analysis}

\author{Sena Aleyna \c{S}entürk}
\address{Department of Space Sciences and Technologies, 
Faculty of Science, Blok-B, Akdeniz University, Antalya 07058, Türkiye}

\author{Timur \c{S}ahin}
\address{Department of Space Sciences and Technologies, 
Faculty of Science, Blok-B, Akdeniz University, Antalya 07058, Türkiye}
\ead{timursahin@akdeniz.edu.tr}

\author{Cenk Kayhan}
\address{Scientific Research Projects Coordination Unit, Kayseri University, Kayseri 38280, Türkiye}

\vspace{10pt}
\begin{indented}
\item[]November 2025
\end{indented}

\begin{abstract}
We present the first high-resolution near-infrared spectroscopic analysis of the solar analog HD\,115617 (61 Virginis), complemented by optical spectroscopy, asteroseismology, and spectral energy distribution modeling. Using ESPRESSO and IGRINS spectra with a newly calibrated NIR line list, we derived atmospheric parameters that revealed notable differences between spectral regions: the optical analysis yielded \(T_{\text{eff}} = 5500 \pm 140\) K, \(\log g = 4.40 \pm 0.16\), and solar metallicity, whereas the NIR yielded \(T_{\text{eff}} = 5750 \pm 140\) K. We tested this 250 K discrepancy using the independent line depth ratio (LDR) method for both spectra. When applied to the optical lines, LDR confirmed the cooler scale (5553$\pm$73 K); for the NIR lines, it provided an intermediate temperature (5636$\pm$15 K). Asteroseismic scaling with TESS data yielded a radius of \(0.98 \pm 0.09\,R_\odot\), consistent with SED fitting and confirming the star's main-sequence solar-like status. However, the age estimates diverged between methods, with optical and NIR analyses yielding ages of 10.97 and 8.04 Gyr, respectively. Critically, a condensation temperature analysis revealed no significant trend, confirming the star's bulk solar-like composition and showing no chemical signature of planetary formation processes. Kinematic diagnostics place HD\,115617 in the thin Galactic disk, with a birth radius of \(\sim 5.7–8.0\) kpc. Although the spectral differences may be linked to the star's multi-planet system or debris disk, our analysis highlights the critical challenge of distinguishing such effects from methodological systematics in multi-wavelength studies. Consequently, we propose a systematic, homogeneous optical-NIR survey of solar-type stars to resolve this ambiguity, which could ultimately inform novel indirect methods for characterizing stellar environments.
\end{abstract}

\section{Introduction}
Understanding the chemical composition and evolutionary pathways of stars is essential for reconstructing the sequence of events that shaped the Milky Way. The elemental patterns imprinted in stars act as fossil records of the Galaxy’s formation history \citep{Gibson2003}, and modern models of Galactic evolution rely on these chemical signatures as key constraints. High-resolution spectroscopy plays a central role in this endeavor, enabling precise measurements of atmospheric parameters and detailed abundance patterns. However, the accuracy of such analyses strongly depends on both the quality of the observed spectra and the reliability of the underlying atomic data.

Although optical spectroscopy has traditionally dominated abundance studies, it is restricted by atmospheric transmission gaps and performs poorly in regions affected by significant interstellar extinction. In contrast, near-infrared observations probe longer wavelengths that penetrate dust more effectively, allowing access to chemically informative lines in heavily obscured environments such as the Galactic center. Therefore, NIR spectrographs provide unique leverage for characterizing stellar populations that are inaccessible at optical wavelengths.

Despite these advantages, robust abundance work in the NIR has been hindered by the lack of a unified and well-validated atomic line list for NIR wavelengths. Many existing compilations suffer from issues such as telluric contamination, incorrect line identifications, and inconsistent oscillator strengths, particularly for Fe\,{\sc ii} lines. To address these limitations, \cite{Senturk2024} constructed an expanded atomic line list spanning the Y to K bands (1.0–2.5 µm). Their log gf values for Fe\,{\sc i}, Fe\,{\sc ii}, and selected $\alpha$-element transitions were recalibrated using a high-quality telluric-corrected solar atlas and the IGRINS spectrum of the well-characterized solar analog HD\,76151. This improvement resolves discrepancies found in earlier lists, including APOGEE DR17 \citep{Abdurrouf2022}, and permits a consistent ionization balance in the NIR that is comparable to that achieved in the optical spectrum. Additionally, the agreement between the kinematic/orbital analyses and abundance trends in that study further demonstrated the internal coherence of the revised line list.

Driven by major advances in instrumentation and data-handling capabilities, large spectroscopic surveys have expanded rapidly in recent years, driven by major advances in instrumentation and data-handling capabilities. Programs such as RAVE \citep{Steinmetz2006}, APOGEE \citep{Majewski2017}, SEGUE \citep{Yanny2009, Eisenstein2011}, the Gaia-ESO Survey \citep{Gilmore2012}, WEAVE \citep{Dalton2012}, GALAH \citep{DeSilva2015}, LAMOST \citep{Luo2015}, the Maunakea Spectroscopic Explorer \citep{McConnachie2016}, and \textit{Gaia} RVS \citep{Cropper2018} collectively provide an unprecedented view of the Galaxy’s chemical and dynamical structure, enabling increasingly refined models of Galactic evolution.

A substantial fraction of these efforts has focused on FGK-type stars, whose broad range of temperatures, masses, and evolutionary stages make them powerful tracers of the Milky Way’s formation history (\citealt{Casagrande2011}; \citealt{Bensby2014}). Their long main-sequence lifetimes, particularly for G and K dwarfs, preserve chemical signatures over billions of years, allowing a detailed reconstruction of abundance trends and population gradients across the disk and halo \citep{Santos2004, Adibekyan2012}. Consequently, FGK stars form the backbone of modern studies on chemical evolution.

As a special subset of the FGK class, solar-type stars possess model atmospheric parameters ($T_{\rm eff}$, $\log g$, [Fe/H]), mass, evolutionary status, and age ranges similar to those of the Sun \citep{RamirezMelendez}. In this context, three categories are prominent in the literature: "solar twins," whose parameters are nearly identical to the Sun's; "solar analogs," defined with broader tolerances for similar atmospheric properties; and "solar-like" stars, which are close to the Sun only in terms of basic parameters. These stars serve as natural laboratories for a better understanding of the Sun's evolutionary process and for making inferences about the formation conditions of the Solar System \citep{Nissen2015}. They also attract intense interest because of their potential for astrobiology, habitable zone analyses, and exoplanet hosting (\citealp{Kasting1993}; \citealp{Howard2012}).

The classification of solar-type stars also intersects with questions regarding the birth environment of the Sun. For instance, N-body simulations by \cite{Zwart2009} suggest that the Sun likely originated in a massive open cluster of order $10^{4}$ stars, which has since dispersed through dynamical processes. If this scenario is correct, the Sun’s siblings are now distributed throughout the Galactic disk and may be identifiable through chemical tagging. Supporting this idea, \cite{Liu2014} demonstrated that the detailed chemical patterns of solar twins encode information about their natal environments, while \cite{Melendez2020} showed that $\alpha$-element and s-process abundances are particularly informative for distinguishing candidate co-natal populations.

Motivated by the potential to probe both stellar astrophysics and Galactic history through chemical signatures, we conducted a detailed case study. Here, we present the first high-resolution near-infrared spectroscopic investigation of the solar-type multi-planet host star HD 115617 (61 Virginis). Motivated by the advantages offered by the newly calibrated NIR line list of \cite{Senturk2024}, we analyzed the atmospheric and chemical properties of the star across the full 400–2500 nm wavelength range. 

Our dataset combines optical spectra from ESPRESSO (R $\approx$ 140\,000; \citealt{Pepe2014}) with IGRINS near-infrared spectra (R $\approx$ 45\,000; \citealt{Park2018}), enabling a systematic comparison of the elemental abundances derived from both regions of the spectra. This comparative evaluation allowed us to discuss the spectral line behavior and abundance consistency. Our analysis reveals significant tension between the atmospheric parameters (e.g., $\Delta$$T_{\rm eff}$ $\approx$ \textit{250 K}) and age estimates derived from the optical and NIR spectra. These tensions suggest possible wavelength-dependent systematics or astrophysical processes, such as stellar activity or interactions with known planetary systems, which merit closer examination. Collectively, these findings reinforce the need to revisit the classification methodologies for solar-type stars and critically evaluate the assumptions made in atmospheric models.

The remainder of this paper is organized as follows. In Section \ref{sec:obs}, we detail the acquisition and reduction of high-resolution spectroscopic data from ESPRESSO (optical) and IGRINS (near-infrared). Section \ref{sec:spec} outlines the chemical abundance analysis, including the methodology used to determine the stellar atmospheric parameters and the resulting elemental composition. In Section \ref{sec:sed}, we derive the fundamental photometric parameters through spectral energy distribution (SED) modeling using the ARIADNE code. Section \ref{sec:astero} discusses the structural properties of the star using asteroseismic results derived from TESS observations. Section \ref{age} addresses the determination of stellar age using two independent methods: MCMC-based isochrone fitting (Section \ref{MCMC}) and evolutionary modeling via the MESA code (Section \ref{MESA}). The kinematic properties and Galactic orbital dynamics of HD\,115617 are presented in Section \ref{sec:kinematic}. In Section \ref{evo}, we determine the evolutionary status of the star by combining precise \textit{Gaia} DR3 parallaxes with multi-band photometry, including Johnson ($B, V$), \textit{Gaia} ($G, G_{\text{BP}}, G_{\text{RP}}$), and TESS magnitudes. Section \ref{sec:dis} provides a comprehensive discussion of the results, addressing the implications of the observed optical-NIR discrepancy and summarizing our conclusions. Finally, Section \ref{sec:concluding} presents our concluding remarks, summarizing the key findings and their broader astrophysical implications.

\begin{table*}
\small
\caption{Model atmosphere parameters of the HD\,115617 (this study), HD\,76151, and Sun \citep{Senturk2024}}
\centering
\resizebox{0.9\textwidth}{!}{
\begin{tabular}{l|c|c|c|c|c|c}
\hline
\hline
Star & Spectrograph & Wavelength Range & $T_{\rm eff}$ & $\log g$ & [Fe/H] & $\xi$ \\
 ~   &      ~       &       (\AA)      & (K)  & (cgs) & (dex)  & (kms$^{-1}$)\\
\hline
\hline
HD\,115617 & ESPRESSO & 3800 - 7880   & 5500$\pm$140 & 4.40$\pm$0.16 &  0.02$\pm$0.10 & 0.52$\pm$0.50 \\
HD\,115617 & IGRINS   & 14500 - 24500 & 5750$\pm$140 & 4.30$\pm$0.34 &  0.13$\pm$0.10 & 2.30$\pm$0.50 \\
HD\,115617 & E + I    & 3800 - 24500  & 5675$\pm$132 & 4.65$\pm$0.16 &  0.12$\pm$0.12 & 0.43$\pm$0.50 \\
\hline
Sun & KPNO       & 4000 - 9000  & 5770$\pm$130 & 4.40$\pm$0.19 & 0.00$\pm$0.09 & 0.66$\pm$0.50 \\
Sun & ESO        & 9780 - 50050 & 5780$\pm$55  & 4.40$\pm$0.22 & 0.00$\pm$0.03 & 1.08$\pm$0.50 \\
Sun & KPNO + ESO & 4000 - 50050 & 5790$\pm$118 & 4.44$\pm$0.20 & 0.00$\pm$0.08 & 0.90$\pm$0.50 \\
\hline
HD\,76151 & HARPS  & 3780 - 6900   & 5780$\pm$88  & 4.35$\pm$0.16 & 0.14$\pm$0.08 & 0.69$\pm$0.50 \\
HD\,76151 & IGRINS & 14500 - 24500 & 5780$\pm$178 & 4.31$\pm$0.25 & 0.19$\pm$0.17 & 2.10$\pm$0.50 \\
HD\,76151 & H + I  & 3780 - 24500  & 5790$\pm$170 & 4.35$\pm$0.18 & 0.24$\pm$0.09 & 0.30$\pm$0.50 \\
\hline
\hline
\end{tabular}
\label{tab1}
}
\end{table*}

\section{Observation}
\label{sec:obs}

For this study, we assembled a set of high-resolution, high signal-to-noise (S/N) spectra in both the optical and near-infrared (NIR) regimes to characterize the spectrum of HD \, 115617. The use of such a broad wavelength range serves two primary goals: (1) to examine whether the atmospheric parameters and elemental abundances inferred from the optical and NIR domains are consistent with each other, thereby revealing any wavelength-dependent systematic errors. (2) It allows us to obtain a more robust chemical analysis by combining information from different spectral regions, which mitigates the uncertainties that arise when relying on a single band wavelength.

HD\,115617 (61 Vir) is an ideal candidate for this approach. As a nearby solar analog with a well-established multi-planet system and circumstellar debris, it offers a rich laboratory environment for comprehensive spectroscopic studies. Radial-velocity studies have revealed the presence of at least three exoplanets in the system. The two innermost planets, with minimum masses of 5.2 and 18.7 M\(_\oplus\), orbit at distances of 0.05 and 0.22 AU, respectively, as reported by \citet{Vogt2010}. Furthermore, this system possesses a rich circumstellar environment. A debris disk, detected via an infrared excess, has been spatially resolved using Herschel PACS observations \citep{Wyatt2012, Greaves2014}. The disk is well-aligned with the stellar equator, with an inclination of \(77^\circ \pm 4^\circ\) \citep{Greaves2014}, indicating a system that has largely evolved in a coplanar fashion \citep{2025MNRAS.tmp.2118M}.

The optical spectrum was obtained from the ESPRESSO instrument via the European Southern Observatory (ESO) public archive\footnote{PI: F. Pepe; Project ID 1102.C-0744; MJD 58550.32327; 700 s integration.}. ESPRESSO provides coverage from 3800 to 7880 \AA. The NIR data were taken from the IGRINS spectral library \citep{Park2018} (MJD 57380; 40 s exposure; $S/N=$235 at 2.2 $\mu$), which delivers high-quality H- and K-band spectra (1490–1800 nm and 1960–2460 nm) at a resolving power of R $\approx$ 45\,000, already corrected for telluric absorption and shifted to the stellar rest frame.

Both datasets required preprocessing prior to abundance analysis. Using the iSpec framework \citep{Blanco-Cuaresma2014}, we normalized each spectrum to its local continuum and applied the radial velocity corrections. For the ESPRESSO data, the radial velocity was measured via cross-correlation with a high-resolution NARVAL solar reference spectrum, yielding -7.81$\pm$0.06 km s$^{-1}$. This agrees well with the \textit{Gaia} DR3 value of –7.86$\pm$0.13 km s$^{-1}$ \citep{DR3}. For the IGRINS data, we adopted the radial velocity provided in the library entry (–8.13 km s$^{-1}$; \citealt{Park2018}), which is consistent with the optical determination within the expected uncertainties.

\begin{figure*}
    \centering
    \includegraphics[width=0.49\linewidth]{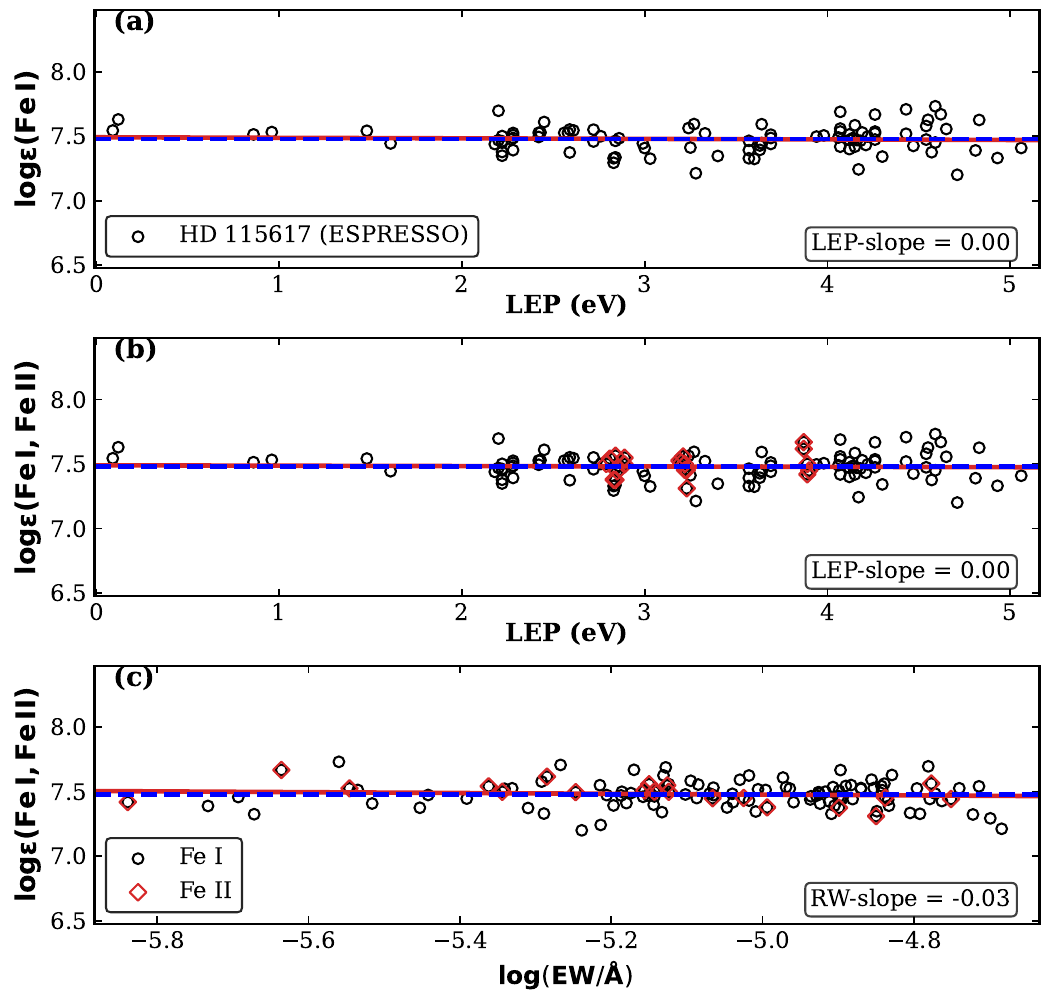}
    \includegraphics[width=0.49\linewidth]{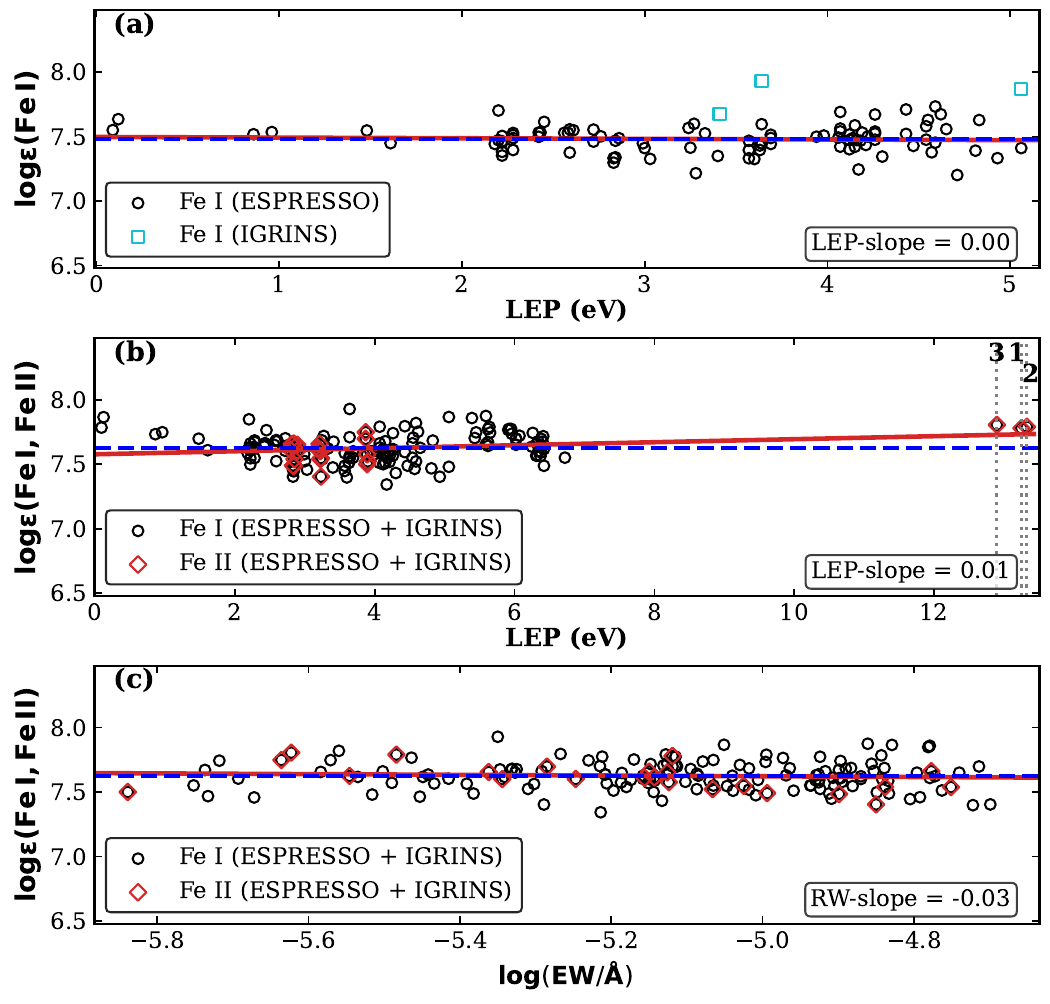}
    \caption{Determination of atmospheric parameters for HD\,115617 from ESPRESSO and IGRINS spectra. The lines indicated with numbers present Fe\,{\sc ii} lines at 15\,531 (1) \AA, 15\,928 (2) \AA, and 15\,952 (3) \AA.}
    \label{fig1}
\end{figure*}

\begin{figure*}
    \centering
    \includegraphics[width=0.79\linewidth]{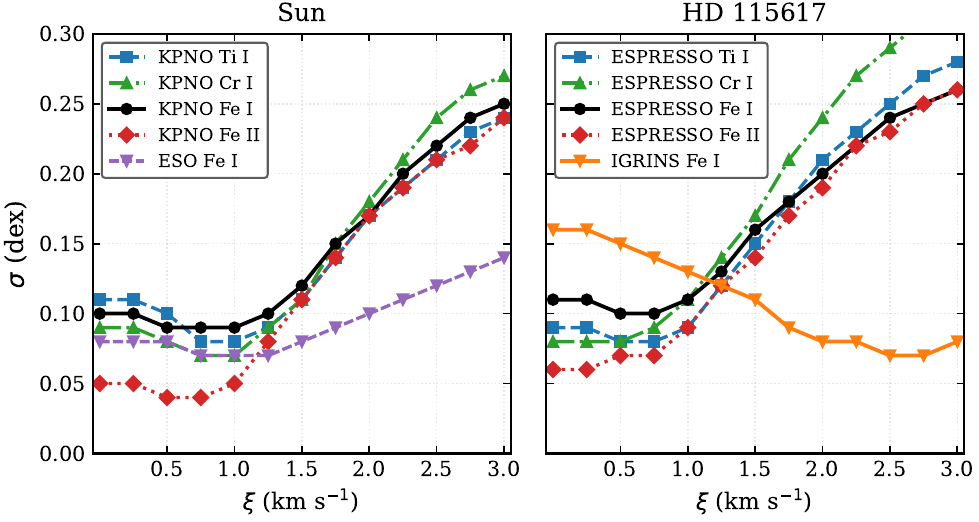}
    \caption{The standard deviation of the Ti, Cr, and Fe abundances from the suite of Ti\,{\sc i}, Cr\,{\sc i}, Fe\,{\sc i}, and Fe\,{\sc ii} lines as a function of $\xi$.}
    \label{fig2}
\end{figure*}

Following the completion of all preprocessing steps, both the ESPRESSO and IGRINS spectra were ready for equivalent width measurements and for carrying out the differential chemical abundance analysis.

Accurate atomic line lists are crucial for reliable atmospheric modeling and abundance determination of chemical species. This accuracy is especially important in the infrared region, where line identifications and atomic parameters have not been vetted as extensively as in the optical region. Furthermore, the NIR region is complicated by strong telluric absorption and molecular transitions produced in the Earth’s atmosphere. These transitions impose additional demands on the reliability of the adopted atomic data sets. Owing to these challenges, selecting and validating the line list plays a central role in ensuring the reliability of the chemical abundances derived from the NIR spectra. As with most analyses of F-, G-, and K-type stars, the Sun provides an indispensable reference owing to its well-established atmospheric parameters and thoroughly studied abundance patterns. These attributes make the solar spectrum an ideal testbed for verifying theoretical line predictions, particularly in the sparsely explored NIR, and for assessing the sensitivity of atomic data used in the calculations.

In this study, we used the solar spectrum to calibrate and validate the line list used in our analysis. Following the approach outlined by \citet{Sahin2024}, we compiled a composite solar reference that covered both the optical and NIR ranges. For wavelengths between approximately 4000 and 5000 \AA, we utilized the high-resolution KPNO solar atlas \citep{Kurucz1984} with R~$\approx$~700\,000. For the interval from 5000–10000 \AA, we employed the telluric-corrected Göttingen FTS disk-integrated flux atlas from \citet{Baker2020}, which provides an even higher spectral resolution (R~$\approx$~1\,000\,000). These atlases were selected specifically for their exceptional resolving power and reliability, which are essential for identifying unblended lines and obtaining precise equivalent width measurements of the lines. We also verified the agreement of the line profiles and flux levels in their overlapping wavelength region (5000–6800 \AA), ensuring that the composite solar reference remained homogeneous across the entire optical range. This rigorous solar-based calibration enables a robust evaluation of the applied atomic data and minimizes systematic offsets when extending the analysis into the comparatively less characterized infrared region for HD\,115617.

\begin{figure*}
    \centering
    \includegraphics[width=0.90\linewidth]{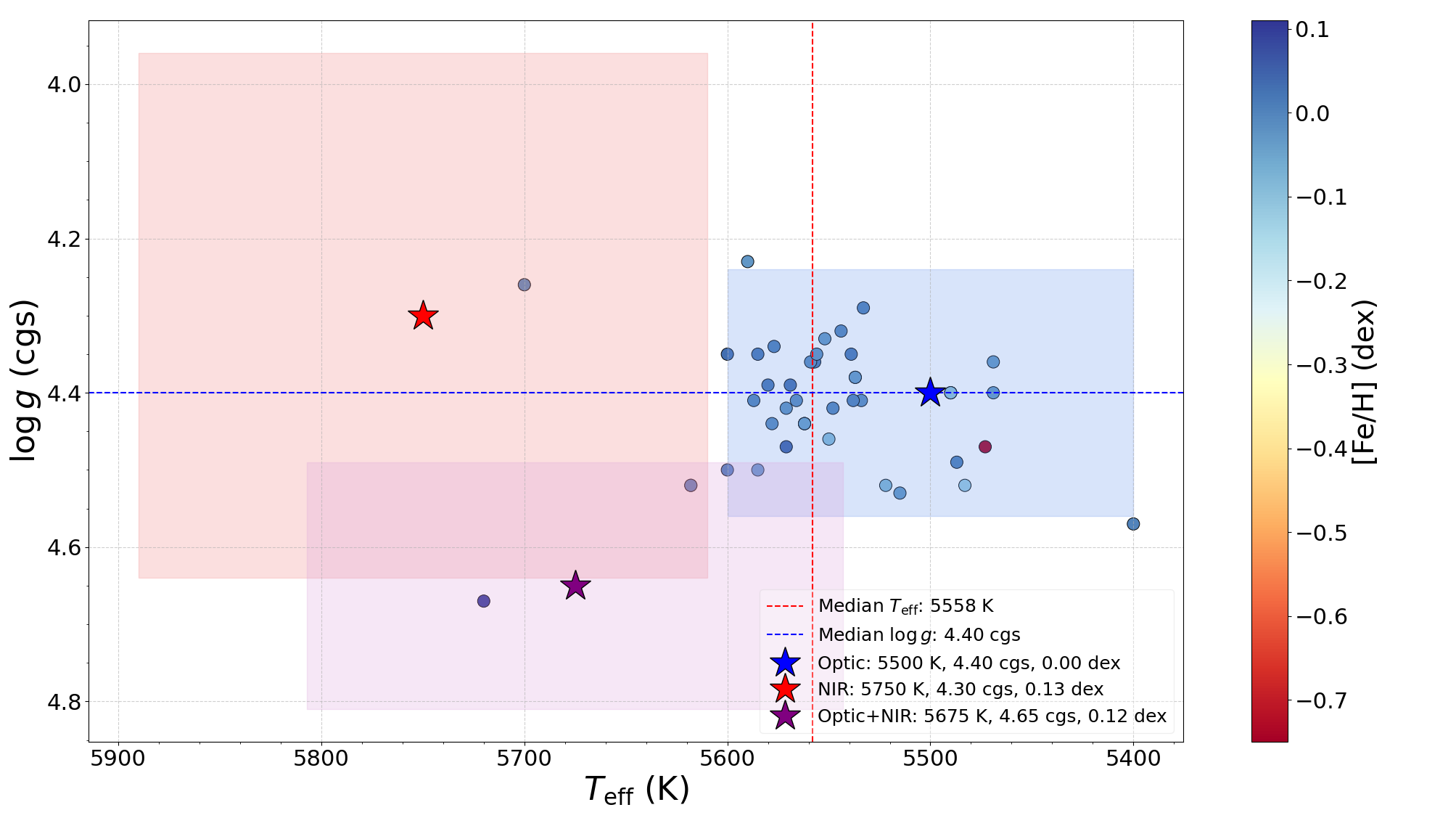}
    \caption{The faint blue area in the image represents the errors in the model parameters obtained from ESPRESSO spectrum of HD\,115617 in the optical region (filled blue star symbol), while the faint red area represents the errors in the model parameters obtained from the IGRINS spectrum of HD\,115617 in the IR (filled red star symbol). The model parameters from both the optical and IR data are represented by filled purple star symbols. The circular symbols denote the parameter values for HD\,115617 as reported in the literature.}
    \label{fig3}
\end{figure*}

To apply the validated atomic line lists in practice, we measured the equivalent widths (EWs) of the selected transitions in both HD\,115617 and the solar reference spectrum using the LIME code \citep{Sahin2017}. Using continuum-normalized spectra, LIME determines the equivalent widths (EWs) by fitting Gaussian profiles to individual absorption features. When selecting spectral lines for measurement, we prioritized transitions that were cleanly resolved, symmetric, and free from blending with neighboring stellar or telluric features to minimize systematic uncertainties in the abundance analysis.

For each line, LIME requires essential atomic parameters: wavelength, species, oscillator strength ($\log gf$), and lower-level excitation potential, which were obtained from the comprehensive line list of \citet{Senturk2024}. Their compilation extended the APOGEE DR17 list \citep{Majewski2017}, broadening its coverage from the original 1500.0–1700.7 nm interval to the full Y, J, H, and K bands (1000–2500 nm). All candidate transitions were tested against a high-resolution (R $\approx$ 700\,000), telluric-corrected KPNO solar spectrum to ensure that the features were well-reproduced and suitable for EW-based abundance measurements. Lines that met these criteria were cross-validated with entries in established atomic databases, including NIST, the Kurucz database, and VALD, with special attention paid to lines in the infrared region, where identifications are historically less secure.

Through this multistep verification process, which involved testing against solar spectra, confirming atomic data, and rejecting blended or unreliable features, we constructed a final set of transitions optimized for precise abundance analysis. The atomic dataset adopted for HD\,115617 in this study represents a rigorously vetted selection of lines, ensuring the highest possible level of reliability for subsequent spectroscopic analysis. In effect, our study incorporates and re-evaluates all line lists available since 1999, consolidating them into a carefully vetted selection suitable for the accurate infrared abundance determination of HD\,115617.

\section{Spectroscopic Analyses}
\label{sec:spec}

With a rigorously vetted line list in hand, we proceeded with the spectroscopic abundance analysis. This was performed using the ATLAS9 model atmospheres \citep{Castelli2004}, computed under the assumption of local thermodynamic equilibrium (LTE), using the ODFNEW method. The model atmospheric parameters ($T_{\rm eff}$, $\log g$, [Fe/H], and $\xi$) were determined using the MOOG code \citep{moog}, which is widely employed for spectral-line analysis. For comprehensive methodological details on stellar abundance analyses, readers are referred to extensive studies by our group. Initial studies focused on fundamental abundance patterns \citep{SahinLambert,Sahin2011}, while \cite{SahinBilir} examined these stars within the context of Galactic dynamics. More recently, \cite{Sahin2023, Sahin2024} and \cite{Marismak2024} have pushed for higher precision using state-of-the-art spectrographs, providing the methodological framework for the current NIR analysis presented in \cite{Senturk2024}.

Representative diagrams illustrating the methodology used to determine the model atmospheric parameters for both HD\,115617 and the Sun are shown in Figure \ref{fig1}. The upper panels show the relationship between iron abundance ($log\, \epsilon$(Fe)) and the lower-level excitation potential (LEP). Accordingly, the effective temperature ($T_{\rm eff}$) was determined under the condition that the abundances derived from the Fe\,{\sc i} lines were independent of their LEP values (i.e., spectroscopic excitation technique\footnote{It is sensitive to neutral spectral lines with a broad range of excitation potentials.}). The microturbulence velocity ($\xi$) was determined by ensuring that the abundances obtained from neutral and first-ionized atoms (Fe\,{\sc i} and Fe\,{\sc ii}) were independent of their reduced equivalent widths (log (EW/$\lambda$; i.e., the LTE
principle).

The microturbulence parameter was further tested independently using Ti\,{\sc i}, Cr\,{\sc i}, Fe\,{\sc i}, and Fe\,{\sc ii} lines. Different values in the range $\xi$ = 0–3 km s $^{-1}$ were tested for HD\,115617 and the Sun, and the dispersion ($\sigma$) of Ti, Cr, and Fe abundances was calculated for each value. The results are shown in Figure \ref{fig2}. The dispersion test applied to the ESPRESSO spectrum of HD\,115617 yielded a value of $\xi \approx$0.5 km s $^{-1}$ from the Fe\,{\sc i} lines. The same method provided a consistent result of $\approx$0.5 km s $^{-1}$ for the Fe\,{\sc ii}, Ti\,{\sc i}, and Cr\,{\sc i} lines. Evaluating both methods together, the uncertainty in the microturbulence velocity for the ESPRESSO spectrum was estimated to be 0.52$\pm$0.5 km s $^{-1}$. In contrast, for the IGRINS spectrum, the microturbulence velocity determined from the neutral iron lines, obtained by minimizing the slope based on the LTE principle, was found to be $\xi =$ 2.30$\pm$0.5 km s $^{-1}$.

\begin{table*}
\caption{The Abundances of the Observed Species for HD\,115617.  Simultaneously, the solar abundances obtained from the KPNO-OPTICAL solar spectrum in \cite{Senturk2024}  and those obtained by \cite{Asplund2021} are provided. Bold values indicate elemental abundances measured in the KPNO-NIR solar and IGRINS spectra of HD\,115617. The abundances in bold type face was determined using the spectrum synthesis method.}
\centering
\resizebox{0.85\textwidth}{!}{

\begin{tabular}{l|c|c|c|c|c|c|c}
\hline
\hline
~       &                  \multicolumn{3}{c|}{HD\,115617}                   & \multicolumn{2}{c|}{\cite{Sahin2024}} & \multicolumn{2}{c}{\cite{Asplund2021}} \\
\hline
Species & log$\epsilon$(X) $\pm \sigma$ & [X/Fe]$\pm \sigma_{[X/Fe]}$ & n &  log$\epsilon_{\rm \odot}$(X) $\pm \sigma$      &         n   & log$\epsilon_{\rm \odot}$(X) $\pm \sigma$ & $\Delta$log$\epsilon_{\rm \odot}$(X)\\
~       & (dex) & (dex) & ~ & (dex) & ~ & (dex) & (dex) \\ 
\hline
O I* & 8.80$\pm$0.02 & -0.03$\pm$0.11 & 3 & 8.81$\pm$0.03 & 3 & 8.69$\pm$0.04 & 0.12 \\
Na I & 6.19$\pm$0.00 & -0.05$\pm$0.16 & 1 & 6.22$\pm$0.12 & 4 & 6.22$\pm$0.00 & 0.00 \\
Mg I* & 7.70$\pm$0.04 & 0.06$\pm$0.11 & 2 & 7.62$\pm$0.03 & 5 & 7.55$\pm$0.03 & 0.07 \\ 
Al I* & 6.44$\pm$0.01 & -0.01$\pm$0.10 & 5 & 6.43$\pm$0.03 & 8 & 6.43$\pm$0.03 & 0.00 \\  
Si I & 7.52$\pm$0.04 & 0.00$\pm$0.14 & 7 & 7.50$\pm$0.09 & 21 & 7.51$\pm$0.03 & -0.01 \\  
\textbf{Si I} & \textbf{7.65$\pm$0.00} & \textbf{-0.03$\pm$0.15} & \textbf{1} & \textbf{7.50$\pm$0.11} & \textbf{28} & 7.51$\pm$0.03 & -0.01 \\  
Ca I & 6.32$\pm$0.07 & 0.01$\pm$0.15 & 10 & 6.29$\pm$0.09 & 21 & 6.30$\pm$0.04 & -0.01 \\  
Sc II & 3.15$\pm$0.13 & -0.01$\pm$0.17 & 3 & 3.14$\pm$0.02 & 12 & 3.14$\pm$0.04 & 0.00 \\  
Ti I & 4.91$\pm$0.08 & -0.04$\pm$0.14 & 39 & 4.93$\pm$0.05 & 63 & 4.97$\pm$0.05 & -0.04 \\  
Ti II & 5.02$\pm$0.09 & -0.01$\pm$0.14 & 8 & 5.01$\pm$0.05 & 11 & 4.97$\pm$0.05 & 0.04 \\  
\textbf{Ti II} & \textbf{5.18$\pm$0.00} & \textbf{0.03$\pm$0.16} & \textbf{1} & \textbf{4.82$\pm$0.00} & \textbf{1} & 4.97$\pm$0.05 & -0.15 \\  
V I* & 3.88$\pm$0.04 & -0.04$\pm$0.11 & 5 & 3.90$\pm$0.03 & 5 & 3.90$\pm$0.08 & 0.00 \\  
Cr I & 5.65$\pm$0.07 & -0.05$\pm$0.15 & 17 & 5.68$\pm$0.09 & 29 & 5.62$\pm$0.04 & 0.06 \\  
Cr II & 5.65$\pm$0.11 & -0.01$\pm$0.18 & 4 & 5.64$\pm$0.11 & 4 & 5.62$\pm$0.04 & 0.02 \\  
Mn I* & 5.45$\pm$0.04 & -0.02$\pm$0.13 & 5 & 5.45$\pm$0.08 & 14 & 5.42$\pm$0.06 & 0.03 \\  
Fe I & 7.52$\pm$0.10 & 0.00$\pm$0.18 & 98 & 7.50$\pm$0.11 & 252 & 7.46$\pm$0.04 & 0.04 \\  
\textbf{Fe I} & \textbf{7.63$\pm$0.07} & \textbf{0.04$\pm$0.14} & \textbf{40} & \textbf{7.46$\pm$0.07} & \textbf{204} & 7.46$\pm$0.04 & 0.00 \\  
Fe II & 7.52$\pm$0.08 & 0.00$\pm$0.16 & 20 & 7.50$\pm$0.09 & 32 & 7.46$\pm$0.04 & 0.04 \\  
\textbf{Fe II*} & \textbf{7.63$\pm$0.00} & \textbf{0.00$\pm$0.12} & \textbf{3} & \textbf{7.50$\pm$0.06} & \textbf{10} & 7.46$\pm$0.04 & 0.04 \\  
Co I* & 4.94$\pm$0.07 & -0.03$\pm$0.14 & 6 & 4.95$\pm$0.06 & 8 & 4.94$\pm$0.05 & 0.01 \\  
Ni I & 6.23$\pm$0.07 & -0.04$\pm$0.16 & 45 & 6.25$\pm$0.10 & 66 & 6.20$\pm$0.04 & 0.05 \\  
Cu I* & 4.25$\pm$0.06 & 0.03$\pm$0.13 & 2 & 4.20$\pm$0.06 & 4 & 4.18$\pm$0.05 & 0.02 \\  
Zn I* & 4.66$\pm$0.02 & 0.01$\pm$0.10 & 2 & 4.63$\pm$0.02 & 2 & 4.56$\pm$0.05 & 0.08 \\  
Sr I* & 2.87$\pm$0.00 & 0.01$\pm$0.10 & 1 & 2.84$\pm$0.00 & 1 & 2.83$\pm$0.06 & 0.01 \\  
Y II* & 2.30$\pm$0.08 & 0.00$\pm$0.13 & 2 & 2.28$\pm$0.02 & 2 & 2.21$\pm$0.05 & 0.07 \\  
Zr I* & 2.54$\pm$0.00 & -0.01$\pm$0.10 & 1 & 2.53$\pm$0.00 & 1 & 2.59$\pm$0.04 & -0.06 \\  
Zr II* & 2.58$\pm$0.00 & -0.05$\pm$0.10 & 1 & 2.61$\pm$0.02 & 2 & 2.59$\pm$0.04 & 0.02 \\  
Ba II* & 2.26$\pm$0.03 & -0.08$\pm$0.11 & 2 & 2.32$\pm$0.02 & 2 & 2.27$\pm$0.05 & 0.05 \\  
La II* & 1.12$\pm$0.02 & -0.04$\pm$0.11 & 2 & 1.14$\pm$0.05 & 3 & 1.11$\pm$0.04 & 0.03 \\  
Ce II* & 1.63$\pm$0.03 & 0.01$\pm$0.11 & 2 & 1.60$\pm$0.04 & 3 & 1.58$\pm$0.04 & 0.02 \\  
Nd II* & 1.41$\pm$0.01 & 0.03$\pm$0.10 & 2 & 1.36$\pm$0.03 & 3 & 1.42$\pm$0.04 & 0.06 \\  
\hline
\hline
\end{tabular}
\label{tab2}
}
\end{table*}

\begin{table*}
\small
\caption{Sensitivity of the derived abundances to uncertainties in the model atmospheric arameters from the KPNO-OPTICAL Solar spectrum and the ESPRESSO spectrum of HD\,115617 (top panel). The bottom panel presents the spectra from the KPNO-NIR solar spectrum and IGRINS spectrum of HD\,115617. The uncertainties for species in bold type face were obtained via spectrum synthesis.}
\centering
\resizebox{0.88\textwidth}{!}{
\begin{tabular}{cccccc|ccccc}
\hline
\hline
\multicolumn{11}{c}{$\Delta log \epsilon (\textbf{OPTICAL})$} \\ 
\hline
\multicolumn{6}{c}{Sun} & \multicolumn{5}{c}{HD\,115617}\\
\hline
 & 5790 & 4.40 & 0.00 & 0.66 &  &  &5500 & 4.40 & 0.02 & 0.52 \\
Species & $\Delta\,T_{eff}$ & $\Delta\,log\,g$ & $\Delta[Fe/H]$ & $\Delta \xi$ &  & &$\Delta\,T_{eff}$ & $\Delta\,log\,g$ & $\Delta[Fe/H]$ & $\Delta \xi$ \\
  & (+45 K) & (+0.09 cgs) & (+0.04 dex) & (+0.50 dex) &  &  &(+140 K) & (+0.16 cgs) & (+0.10 dex) & (+0.50 dex) \\
\hline
O\,{\sc i}*   & +0.04 &   -0.07 & +0.02  & -0.03  & ~ & ~ & -0.07 & -0.03 & +0.01 & -0.02 \\
Na\,{\sc i}  & +0.07 &   -0.06 & +0.00  & -0.04  & ~ & ~ & +0.05 & -0.01 & +0.00 & -0.02 \\
Mg\,{\sc i}*  & +0.05 &   -0.05 & +0.00  & -0.06  & ~ & ~ & +0.03 & +0.00 & +0.01 & -0.04 \\
Al\,{\sc i}*  & +0.04 &   -0.02 & +0.00  & -0.02  & ~ & ~ & +0.03 & +0.01 & +0.01 & -0.03 \\
Si\,{\sc i}  & +0.03 &   +0.01 & +0.01  & -0.03  & ~ & ~ & +0.01 & +0.01 & +0.01 & -0.03 \\
Ca\,{\sc i}  & +0.09 &   -0.05 & +0.01  & -0.09  & ~ & ~ & +0.08 & -0.03 & +0.01 & -0.10 \\
Sc\,{\sc ii} & +0.00 &   +0.07 & +0.03  & -0.08  & ~ & ~ & +0.00 & +0.06 & +0.03 & -0.06 \\
Ti\,{\sc i}  & +0.13 &   -0.02 & -0.01  & -0.10  & ~ & ~ & +0.11 & -0.01 & +0.01 & -0.12 \\
Ti\,{\sc ii} & +0.01 &   +0.06 & +0.03  & -0.10  & ~ & ~ & +0.00 & +0.05 & +0.03 & -0.10 \\
 V\,{\sc i}*  & +0.14 &   +0.00 & +0.00  & -0.05  & ~ & ~ & +0.08 & +0.03 & +0.02 & -0.03 \\
 Cr\,{\sc i} & +0.11 &   -0.03 & +0.00  & -0.12  & ~ & ~ & +0.10 & -0.03 & +0.01 & -0.15 \\
 Cr\,{\sc ii}& -0.03 &   +0.07 & +0.03  & -0.12  & ~ & ~ & -0.03 & +0.04 & +0.02 & -0.11 \\
 Mn\,{\sc i}* & +0.10 &   -0.05 & +0.00  & -0.15  & ~ & ~ & +0.07 & -0.02 & +0.02 & -0.10 \\
 Fe\,{\sc i} & +0.09 &   -0.03 & +0.00  & -0.11  & ~ & ~ & +0.07 & -0.01 & +0.01 & -0.12 \\ 
 Fe\,{\sc ii}& -0.03 &   +0.07 & +0.03  & -0.11  & ~ & ~ & -0.03 & +0.05 & +0.03 & -0.11 \\
 Co\,{\sc i}* & +0.09 &   +0.00 & +0.00  & -0.07  & ~ & ~ & +0.08 & -0.03 & +0.01 & -0.09 \\
 Ni\,{\sc i} & +0.08 &   -0.02 & +0.01  & -0.10  & ~ & ~ & +0.05 & -0.01 & +0.02 & -0.09 \\
Cu\,{\sc i}*  & +0.07 &   -0.03  & +0.02 & -0.13  & ~ & ~ & +0.03 & +0.01 & +0.01 & -0.06 \\
 Zn\,{\sc i}* & +0.08  & -0.01  & +0.01  & -0.19  & ~ & ~ & +0.02  & +0.00 & +0.03 & -0.21 \\
 Sr\,{\sc i}* & +0.09  & -0.02  & +0.01  & -0.23  & ~ & ~ &  +0.09  & -0.01  & +0.00 & -0.25 \\
 Y\,{\sc ii}* & +0.06  & +0.04  & +0.01  & -0.20  & ~ & ~ & +0.01  & +0.05  & +0.03 & -0.26 \\
 Zr\,{\sc i }*& +0.07  & -0.02  & +0.02  & -0.15  & ~ & ~ & +0.02 & +0.04 & +0.03 & -0.17 \\ 
 Zr\,{\sc ii}* & +0.05 & +0.03  & +0.00  & -0.17  & ~ & ~ & +0.02  & +0.06  & +0.03 & -0.24 \\
 Ba\,{\sc ii}* & +0.10 & +0.04  & +0.03  & -0.15  & ~ & ~ & +0.04  & +0.02  & +0.05 & -0.21 \\
 Ce\,{\sc ii}* & +0.08 & +0.03  & +0.02  & -0.09  & ~ & ~ & +0.02  & +0.07  & +0.03 & -0.09 \\
 Nd\,{\sc ii}* & +0.11 & +0.05  & +0.03  & -0.04  & ~ & ~ & +0.03  & +0.08  & +0.03 &  -0.03\\
\hline \hline 
\multicolumn{11}{c}{$\Delta log \epsilon (\textbf{NIR})$} \\ 
\hline
\multicolumn{6}{c}{Sun} & \multicolumn{5}{c}{HD\,115617}\\
\hline
 & 5780 & 4.40 & 0.00 & 1.08 &  &  &5750 & 4.30 & 0.13 & 2.30 \\
Species & $\Delta\,T_{eff}$ & $\Delta\,log\,g$ & $\Delta[Fe/H]$ & $\Delta \xi$ &  & &$\Delta\,T_{eff}$ & $\Delta\,log\,g$ & $\Delta[Fe/H]$ & $\Delta \xi$ \\
  & (+55 K) & (+0.22 cgs) & (+0.08 dex) & (+0.50 dex) &  &  &(+140 K) & (+0.34 cgs) & (+0.10 dex) & (+0.50 dex) \\
\hline
 Si\,{\sc i} & +0.01  & +0.00  & +0.01  & -0.02  & ~ & ~ & +0.04   & +0.04  & +0.01 & -0.01 \\ 
 Ca\,{\sc i} & +0.02  & -0.01  & +0.00  & -0.03  & ~ & ~ &  ..     & ..     & ..    & ..    \\
 Ti\,{\sc i} & +0.05  & +0.00  & +0.00  & -0.01  & ~ & ~ &  ..     & ..     & ..    & ..    \\ 
 Ti\,{\sc ii}& +0.01  & +0.10  & +0.01  & -0.01  & ~ & ~ & +0.00   & +0.14  & +0.03 & -0.01 \\ 
 Fe\,{\sc i} & +0.03  & +0.00  & +0.00  & -0.02  & ~ & ~ &  +0.07  & +0.00  & +0.01 & -0.03 \\ 
 Fe\,{\sc ii}*& -0.03  & +0.07  & +0.00  & -0.01  & ~ & ~ &  -0.08  & +0.04  & -0.02 & -0.02 \\
 \hline \hline
 \end{tabular}
\label{tab3a}
}
\end{table*}

\begin{table*}
\small
\caption{Sensitivity of the derived abundances to uncertainties in the model atmospheric parameters in optical+NIR wavelength region.}
\centering
\resizebox{0.95\textwidth}{!}{
\begin{tabular}{cccccc|ccccc}
\hline \hline \\
\multicolumn{11}{c}{$\Delta log \epsilon (\textbf{Optical+NIR})$} \\ 
\hline
\multicolumn{6}{c}{Sun} & \multicolumn{5}{c}{HD\,115617}\\
\hline
 & 5790 & 4.40 & 0.00 & 0.90 &  &  &5675 & 4.65 & 0.12 & 0.43 \\
Species & $\Delta\,T_{eff}$ & $\Delta\,log\,g$ & $\Delta[Fe/H]$ & $\Delta \xi$ &  & &$\Delta\,T_{eff}$ & $\Delta\,log\,g$ & $\Delta[Fe/H]$ & $\Delta \xi$ \\
  & (+118 K) & (+0.20 cgs) & (+0.08 dex) & (+0.50 dex) &  &  &(+132 K) & (+0.16 cgs) & (+0.12 dex) & (+0.50 dex) \\
\hline
Na\,{\sc i}   & +0.05  & -0.04  & +0.01  & -0.04  & ~ & ~ & +0.07   & -0.01  & +0.01 & -0.01 \\
Si\,{\sc i}   & +0.01  & +0.00  & +0.00  & -0.03  & ~ & ~ &  +0.01  & +0.02  & +0.02 & -0.02 \\
Ca\,{\sc i}   & +0.05  & -0.02  & +0.00  & -0.07  & ~ & ~ &  +0.08  & -0.07  & +0.01 & -0.07 \\
Sc\,{\sc ii}  & +0.01  & +0.05  & +0.03  & -0.10  & ~ & ~ &  +0.00  & +0.10  & +0.03 & -0.05 \\
Ti\,{\sc i}   & +0.07  & -0.02  & -0.01  & -0.11  & ~ & ~ &  +0.13  & -0.03  & +0.01 & -0.09 \\
 Ti\,{\sc ii} & +0.00  & +0.05  & +0.01  & -0.10  & ~ & ~ &  +0.00  & +0.08  & +0.03 & -0.07 \\
 Cr\,{\sc i}  & +0.07  & -0.02  & +0.00  & -0.17  & ~ & ~ &  +0.11  & -0.06  & +0.01 & -0.11 \\
 Cr\,{\sc ii} & -0.01  & +0.05  & +0.02  & -0.12  & ~ & ~ & -0.04   & +0.08  & +0.03 & -0.08 \\
 Fe\,{\sc i}  & +0.05  & -0.01  & +0.00  & -0.07  & ~ & ~ &  +0.08  & -0.02  & +0.02 & -0.07 \\ 
 Fe\,{\sc ii} & -0.03  & +0.05  & +0.01  & -0.09  & ~ & ~ &  -0.05  & +0.07  & +0.03 & -0.07 \\
 Ni\,{\sc i}  & +0.05  & +0.00  & +0.01  & -0.08  & ~ & ~ &  +0.06  & -0.02  & +0.02 & -0.07 \\  
 \hline \hline
\end{tabular}
\label{tab3b}
}
\end{table*}

Surface gravity ($\log g$) was determined by analyzing the iron abundances calculated using MOOG. This analysis enforced ionization equilibrium, requiring consistency between the abundances derived from the Fe I and Fe II lines. Subsequently, the metallicity ([Fe/H]) was iteratively updated until the derived iron abundance matched the initial value assumed for the model atmosphere.

The uncertainty in the effective temperature was calculated from the error in the slope of the relationship between the abundances derived from the Fe I lines and their lower-level excitation potentials (LEP). For the ESPRESSO spectrum of HD\,115617, the variation in this slope corresponds to a temperature uncertainty of $\pm$140 K (Figure \ref{fig1}). Furthermore, a 1$\sigma$ difference between the abundances obtained from neutral and ionized iron lines indicates a surface gravity uncertainty of approximately $\pm$0.16 dex for the ESPRESSO spectrum. The same method was applied to the IGRINS spectrum of HD\,115617, yielding uncertainties consistent with the values reported in Table \ref{tab1}, which summarizes all model parameters and their error ranges derived from the line analyses of both the ESPRESSO and IGRINS spectra.

The model atmospheric parameters obtained for HD\,115617 in this study are listed in Table \ref{tab1}. Figure \ref{fig3} visually compares the parameters derived from the optical (ESPRESSO), NIR (IGRINS), and combined optical+NIR analyses. While the parameters from the optical spectrum alone align with the literature values, a key finding emerges from the combined analysis: the optical+NIR parameters show a noticeable deviation. This contrasts with the findings of \cite{Senturk2024} and suggests that line selection and systematic effects may play a more substantial role in the combined analysis of HD 115617.

Uncertainties in the effective temperatures obtained from the spectroscopic excitation technique may originate from systematic errors in the oscillator strengths (log gf) as a function of the excitation potential. Similarly, errors in the equivalent width (EW) measurements can introduce systematic shifts in the parameters. To mitigate such effects,   differential abundance analysis was performed using a star with well-determined parameters, the Sun (\citealt{Fulbright2006, Koch2008}). This approach enhanced the reliability of the parameters derived for HD\,115617 and aimed to minimize discrepancies in the literature.

A significant challenge in determining model atmospheric parameters is the covariance and degeneracy between parameters, particularly between the effective temperature and surface gravity \citep{McWilliam1995}. To independently verify our spectroscopic temperatures and investigate potential parameter degeneracies, we applied the line-depth ratio (LDR) method, a technique less sensitive to the absolute continuum placement, to both the solar and HD 115617 spectra. The line pairs reported by \cite{Fukue2015} were tested on both the KPNO-NIR solar spectrum and the IGRINS spectrum of HD\,115617.  The IR-based temperature derived for the Sun from line pairs 2, 3, 6, and 9 was $T_{\rm eff}$ = 5685$\pm$99 K, which is consistent with the spectroscopically derived $T_{\rm eff}$ = 5780$\pm$55 K. When applied to HD\,115617, the same line pairs yielded $T_{\rm eff}$ = 5636$\pm$15 K using the LDR method. This value lies between the classical spectroscopic results from the optical (5500$\pm$140 K) and NIR (5750$\pm$140 K) regions.

\begin{figure}
    \centering
    \includegraphics[width=0.6\linewidth]{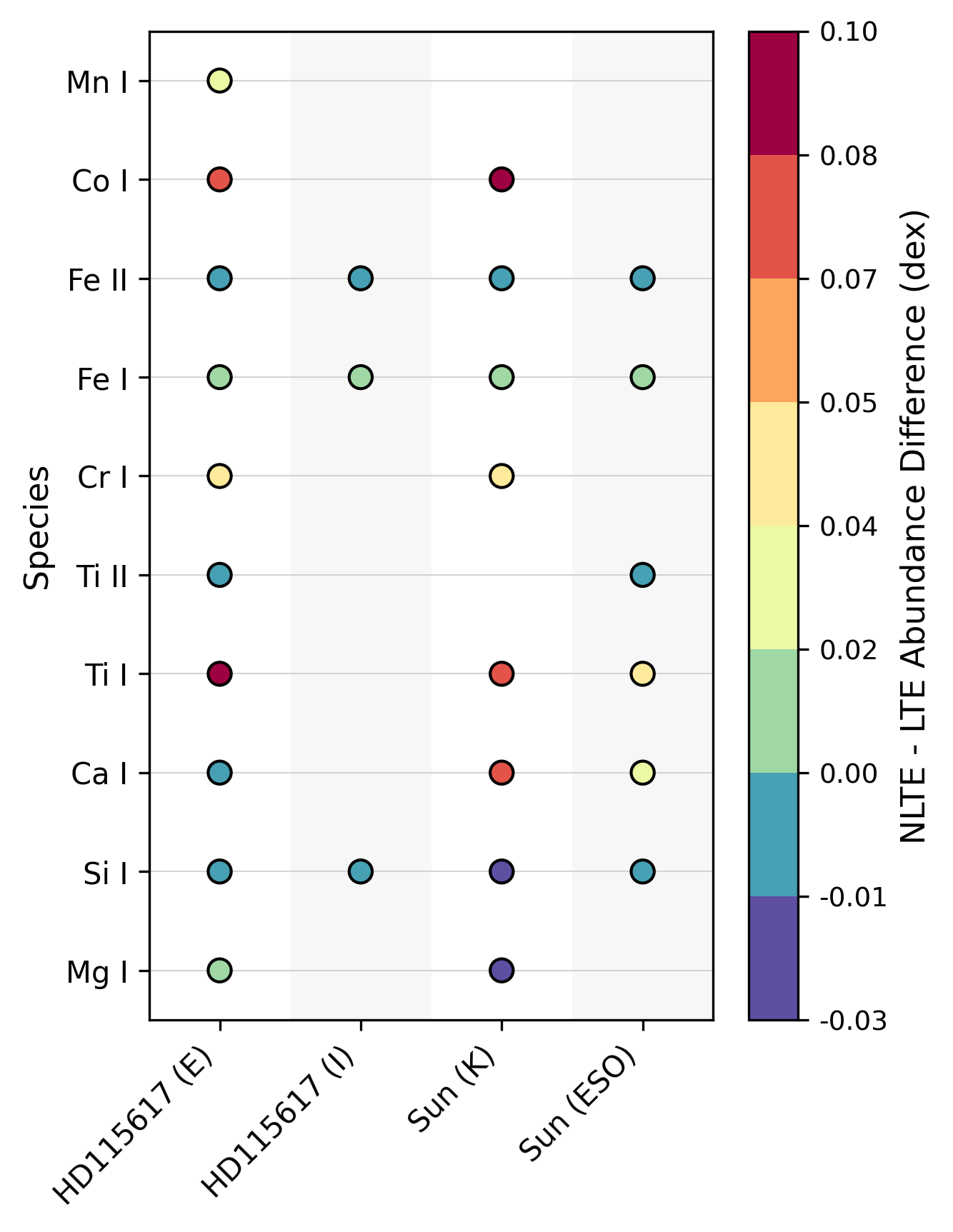}
    \caption{NLTE abundance corrections (in dex) for various atomic species in the Sun and HD\,115617. The vertical axis lists the elements and their ionization states, and the horizontal axis shows the stellar identifiers. The color indicates the magnitude of the NLTE correction, which is defined as the abundance difference between the NLTE and LTE calculations for each element at each point. Corrections were computed using previously published data. Positive values indicate an increase in the inferred abundance when non-LTE effects are considered.}
    \label{fig4}
\end{figure}

Convection and non-local thermodynamic equilibrium (NLTE) effects in stellar atmospheres directly impact the reliability of the abundance results obtained from spectral analyses. To quantify the potential influence of convection, we computed two different mixing-length parameters ($\alpha$) using the prescriptions of \cite{Ludwig1999} (based on a 2D hydrodynamic model) and \cite{Magic2015} (based on a 3D hydrodynamic model), which yielded values of 1.60 and 1.99, respectively. The corresponding ATLAS9 models \citep{Castelli2004} were constructed for each $\alpha$ value. A comparison of the resulting synthetic spectra with the observed spectrum of HD\,115617 revealed no significant differences; however, the spectrum generated using the \cite{Magic2015} $\alpha$ value showed a marginally better agreement. To finalize the elemental abundances using this preferred convective model, the atmospheric parameters required slight refinement: the microturbulence velocity was adjusted by +0.05 km s$^{-1}$ to ensure abundance independence from the equivalent width, while the effective temperature and surface gravity were fine-tuned by +10 K and +0.03 dex, respectively, to re-establish excitation and ionization balance. The final model parameters (5505 K, 4.38 cgs, 0.62 km s$^{-1}$) induced a negligible mean abundance change of 0.01 $\pm$ 0.01 dex across 26 species. This minimal impact is consistent with expectations for F-G-K dwarfs, where the mixing-length parameter has been shown to affect metallicity by less than 0.02 dex \citep{Song2020}. Specifically, for the $\alpha = 1.99$ model, the abundance of ionized Fe decreased by only –0.01 dex, with the largest variations observed for Ti\,{\sc ii} and Sc\,{\sc ii} at +0.02 dex. Here, convection was tested using the mixing-length parameter in the ATLAS9 models, and NLTE corrections from the INSPECT and MPIA databases were applied to the upper atmospheric layers, where the LTE assumptions break down. This methodology aligns with the approaches of \cite{Marismak2024}, \cite{Sahin2024}, and \cite{Senturk2024}, which aimed to reduce systematic errors from convection and NLTE effects.

Figure \ref{fig4} shows the differences between the LTE and NLTE analyses for various atomic species. NLTE corrections were more pronounced for neutral species (e.g., Fe\,{\sc i}, Ti\,{\sc i}, and Mn\,{\sc i}), consistent with the literature, which shows that lines from neutral species are more sensitive to NLTE departures in the upper atmospheric layers \citep{Asplund2005, Bergemann2014}.

\begin{figure*}
    \centering
    \includegraphics[width=0.88\linewidth]{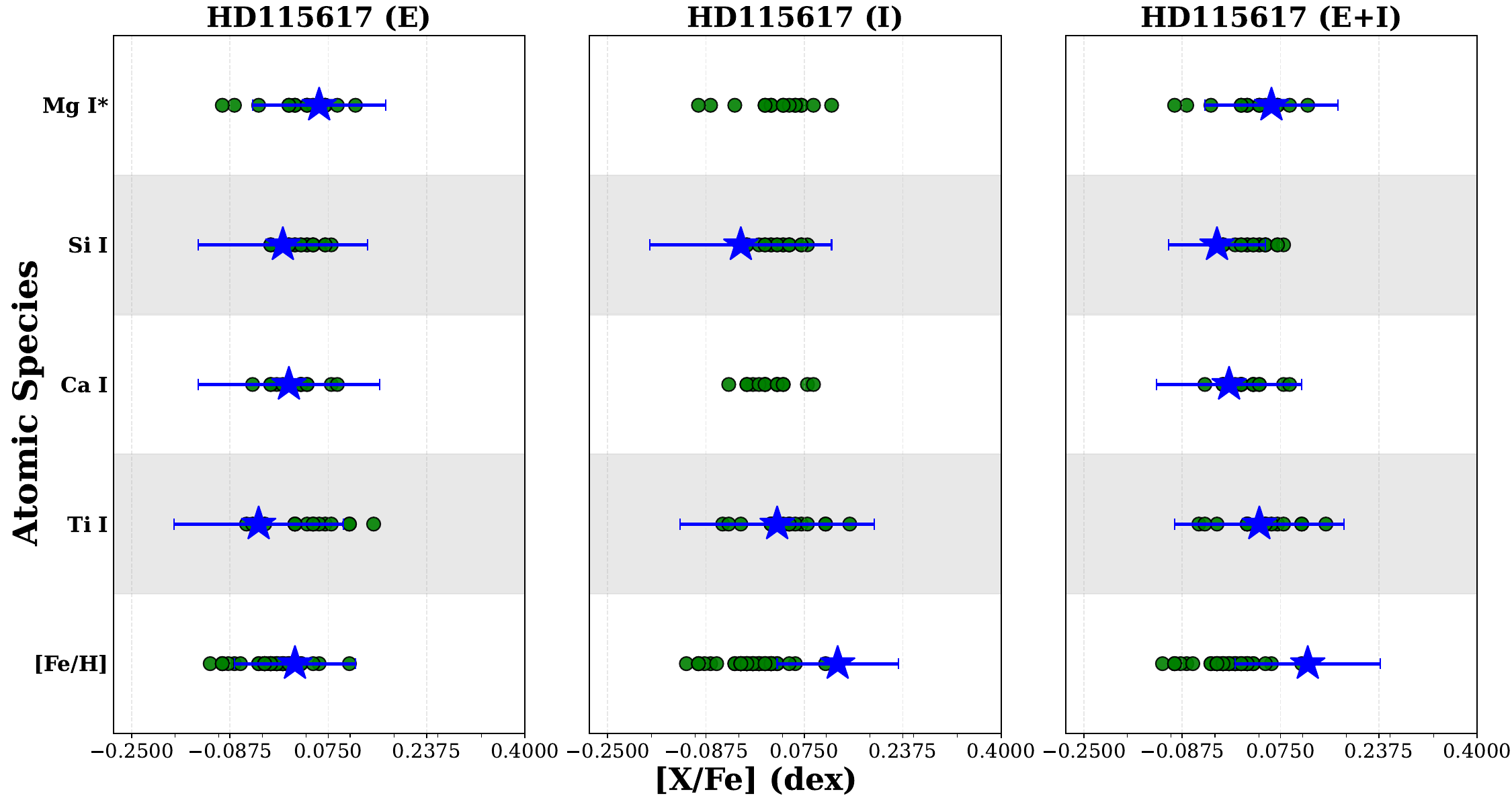}
    \caption{Comparison of logarithmic abundances derived from ESPRESSO (E), IGRINS (I) and ESPRESSO+IGRINS (E+I) analyses for the star HD\,115617 with metallicity and $\alpha$-element abundance values from the literature. Asterisks on the species label denote abundances derived through spectrum synthesis. \textbf{Missing data points for Mg\,{\sc i} and Ca\,{\sc i} in the IGRINS-based panel indicate that reliable measurements could not be obtained; the lines were too weak and suffered from low signal-to-noise.}}
    \label{fig5}
\end{figure*}

The elemental abundances for HD\,115617 are expressed relative to solar values, using the standard spectroscopic notation and presented in Table \ref{tab2} where log $\epsilon$(X)\footnote{The logarithmic absolute abundance of an element X is calculated as the number of atoms of X per 10$^{12}$ hydrogen atoms, log$\epsilon$(X) = log$_{10}$(N$_X$/N$_H$) + 12.0} is the logarithmic abundance of X. The errors reported in log $\epsilon$(X) abundance are represented by the 1$\sigma$ line-to-line scatter of the abundance. The number of lines used in this analysis is also mentioned. For an element X, its abundance relative to hydrogen is given by [X/H] = log$\epsilon$(X)$_{star}$ - log$\epsilon$(X)$_{\odot}$. The key ratio of element X to iron is then [X/Fe] = [X/H]$-$[Fe/H].
The $\sigma$[X/Fe] is the error in [X/Fe], and the square root of the sum of the quadratures of the errors in [X/H] ($\sigma$[X/H]) and [Fe/H]. The $\sigma$[X/H] presents the error in [X/H] and is the square root of the sum of the quadrature of the errors in the stellar and solar logarithmic abundances. Table \ref{tab2} also includes a comparison with solar logarithmic abundances from the literature \citep{Asplund2021, Sahin2024}. The estimated formal errors for the abundances arising from the uncertainties in the atmospheric parameters $T_{\rm eff}$, $\log g$, and $\xi$ are listed in Tables \ref{tab3a} and \ref{tab3b}. 

The abundances from the optical (ESPRESSO) and NIR (IGRINS) regions agreed within  0.13–0.16 dex (e.g., $\Delta$log$\epsilon$ (Si\,{\sc i}) = 0.13 dex; $\Delta$log$\epsilon$ (Ti\,{\sc ii}) = 0.16 dex; $\Delta$log$\epsilon$ (Fe\,{\sc ii}) = 0.13 dex), with infrared results showing higher logarithmic abundances. This difference appears to stem from the higher effective temperature derived from the NIR region ($T_{\rm eff}$ = 5750 K) compared to the optical value  ($T_{\rm eff}$ = 5500 K), which influences the abundance calculations. Figure \ref{fig5} shows that the $\alpha$-element abundances (Mg, Si, Ca, Ti) derived in this work are consistent with the overall [$\alpha$/Fe] trend reported in the literature for this star.

To evaluate the sensitivity of abundance calculations to model atmospheric parameters, deviations of $\pm 1 \sigma$ from the nominal values were applied to each fundamental parameter  ($T_{\rm eff}$, $\log g$, $\xi$, and [Fe/H]), and their effects on the abundances were examined. Each parameter was varied individually, while the others were kept constant at their average values. For example, for the effective temperature, a variation within $T_{\rm eff}$ = 5500 $\pm$140 K was applied, increasing the value to 5640 K, and the abundances were recalculated. This was repeated for $\log g$, $\xi$, and [Fe/H].

Error propagation analyses were applied separately to the optical spectrum (ESPRESSO), NIR 
region (IGRINS), and combined analyses. The resulting logarithmic abundances were compared with the nominal values in Table \ref{tab2}, with deviations listed in Tables \ref{tab3a} (optical and NIR) and \ref{tab3b} (optical+NIR). This approach quantified the sensitivity of each spectral region to systematic uncertainties and was incorporated into the total abundance uncertainty estimates.

\section{SED Analysis via Synthetic Stellar Atmospheric Models}
\label{sec:sed}

We used spectrAl eneRgy dIstribution bAyesian moDel averagiNg fittEr-{\texttt{ARIADNE}} \citep{2022MNRAS.513.2719V} code to analyse the spectral energy distribution (SED) of HD\,115617. We selected photometric observations of the star in the broad wavelength range between {$\sim$}3500 and {$\sim$}52\,000 \AA\, that were used with synthetic stellar atmospheric models {\citep{1993KurCD..13.....K, 2003IAUS..210P.A20C, 2013AA...553A...6H, 2012RSPTA.370.2765A}} to fit the shape of stellar SED. The star  was observed in broad-band photometry bandpasses (\textit{2MASS JHKs; Johnson UBV; Strömgren ubvy; \textit{Gaia} DR3 G, BP, RP; GALEX FUV; TESS)}. The photometric data of the star were obtained from the VizieR{\footnote{\url{https://vizier.cds.unistra.fr/}}}, MAST, and \textit{Gaia} DR3{\footnote{\url{https://www.cosmos.esa.int/web/gaia/data-release-3}}} databases using {\texttt{astroquery}}{\footnote{\url{https://astroquery.readthedocs.io}}} and are listed in Table~\ref{sedtable}.

\begin{figure}
\includegraphics[width=0.9\columnwidth]{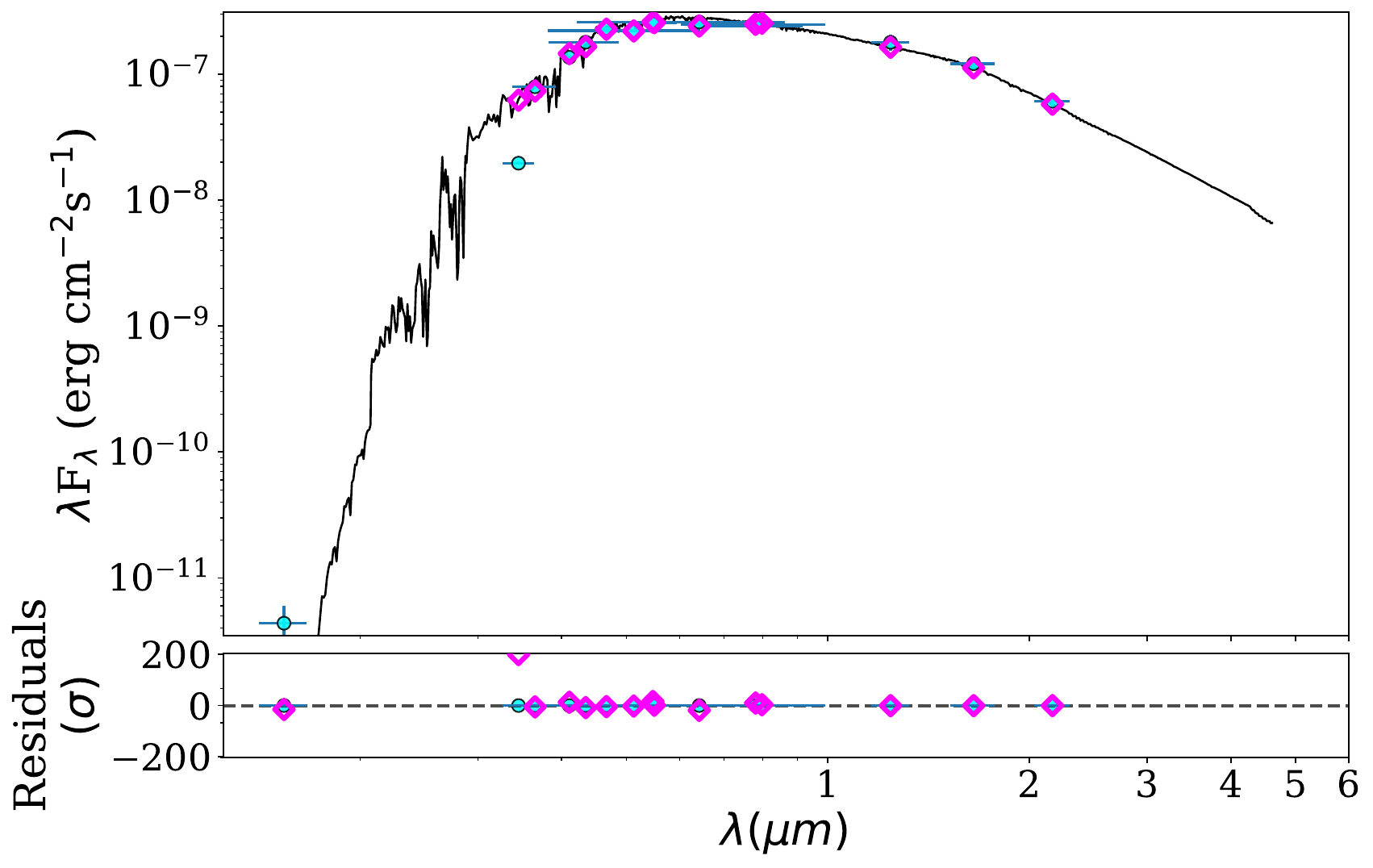}
    \caption{ \textit{Top:} \texttt{ARIADNE} reddened model of SED and
    photometric observations of HD\,115617. The SED includes (\textit{2MASS JHKs; Johnson UBV; Strömgren ubvy; \textit{Gaia} DR3 G, BP, RP; GALEX FUV and TESS}
    photometric observations, and the error bars of the star (blue circles).
    The diamonds denote synthetic photometry. \textit{Bottom:} Residuals of the fit
    that are scaled to normalize the photometry errors.}
    \label{fig:sed}
\end{figure}

To determine the stellar physical and atmospheric parameters
(bolometric luminosity --L{\textsubscript{bol}}, T{\textsubscript{eff}},
logg, and [M/H]), and their probability for HD\,115617 for each value using model,
We used Bayesian analysis with the nested sampling algorithm from \texttt{ARIADNE}. The spectral parameters of the star for SED fitting as the initial parameters
for T{\textsubscript{eff}}, $\log g$ and $[M/H]$.

Interstellar extinction ($A_{\rm V}$) and
reddening (E(B-V)) values were estimated using \texttt{extinction}
package{\footnote{\url{https://github.com/kbarbary/extinction}}}
using the extinction law of \citet{1999PASP..111...63F} in SED analysis.

For synthetic stellar atmospheric models, we chose
Castelli \& Kurucz
\citep[hereafter C\&K2003]{2003IAUS..210P.A20C}, Kurucz {\citep{1993KurCD..13.....K}},
PHOENIX v2 {\citep{2013A&A...553A...6H}} and
BT-Settl {\citep{2012RSPTA.370.2765A}} stellar atmosphere models for \texttt{ARIADNE}.

\begin{figure*}
\includegraphics[width=0.85\columnwidth]{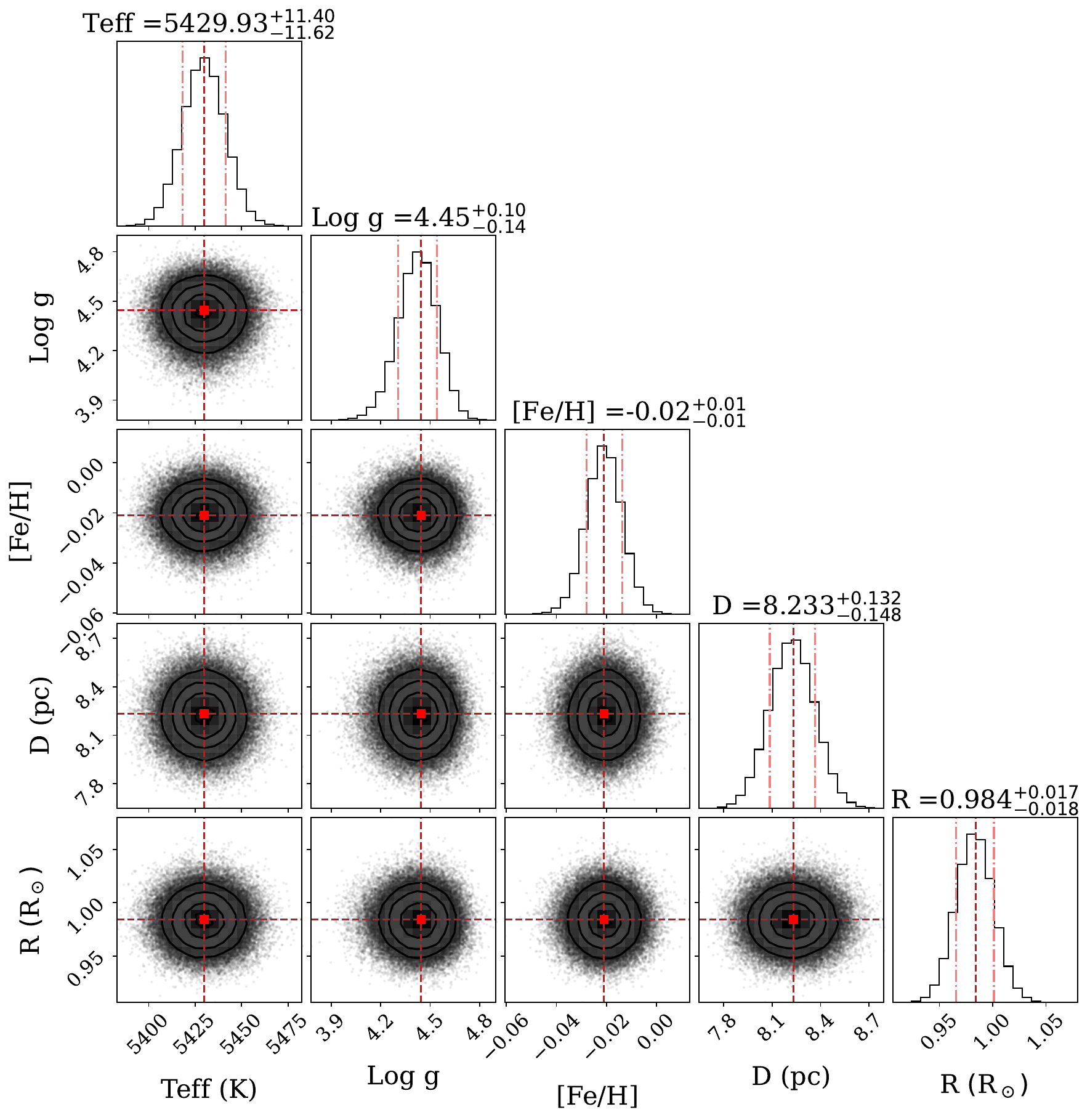}
    \caption{ \texttt{ARIADNE} SED fitting results of HD\,115617. The parameters and 
    accompanying the uncertainties in the SED model of the observed photometric data. 
   Effective temperature (Teff) in the K unit, logarithmic surface gravity(Logg) in the dex unit, metallicity ([Fe/H]) in the dex unit, distance (D) in the pc unit,
    radius (R) in the solar unit of the SED model results of HD\,115617 are plotted.}
    \label{fig:sed_corner}
\end{figure*}

As the star does not have sufficient observed
data on broadband photometry bandpass data in the infrared region; only 2MASS observations were included in the SED analysis.
The SEDs model  is shown in
Figure~\ref{fig:sed} along with the observed flux. Scattering in the observed fluxes of GALEX and Strömgren u is clearly visible in this figure.  
We use \texttt{ARIADNE} with the Bayesian Model Averaging method to analyse
SED of the star.  The fit parameters and their uncertainties were estimated using dynamic nested sampling with 100\,000 iterations and 64 threads for each sample. The parameters of HD\,115617 obtained from the SED fitting model are listed in Table~\ref{tab:sedtable}. The average distributions of the obtained parameters are illustrated in Figure~\ref{fig:sed_corner} where the lack of pronounced degeneracies in the parameter space reinforces the reliability of the derived stellar properties via SED analysis.

\begin{table}
\caption{Photometric observation data of HD\,115617 used in SED fitting.}
\centering
\resizebox{0.45\columnwidth}{!}{
\begin{tabular}{cc|c|c}
\hline
\hline
Filters & & Magnitude & Error \\
 ~   &  &    (mag)       &   (mag) \\
\hline
\hline
 & H  & 2.9740 & 0.1760 \\   
2MASS & J  & 3.3340 &  0.2000 \\   
 & K$_{s}$ & 2.9560 &  0.2360 \\   
\hline
& U	& 5.7100 & 0.0153 \\   
Johnson & V & 4.7390 & 0.0080 \\   
 & B	& 5.4480 & 0.0106 \\   
\hline
 & u & 6.8730 & 0.0118 \\   
Strömgren & v & 5.8580 & 0.0060 \\  
 & b & 5.1700 & 0.0028 \\    
 & y & 4.7360 & 0.0020 \\  
\hline
 & G & 4.5325 & 0.0040 \\   
\textit{Gaia} DR3 & R$_{p}$ & 4.0062 & 0.0037 \\   
& B$_{p}$ & 4.9047 & 0.0028 \\   
\hline
GALEX & FUV & 18.0220 & 0.0630 \\   
\hline
TESS & T$_{mag}$ & 4.0850 & 0.0082 \\   
\hline
\end{tabular}
\label{sedtable}
}
\end{table}

\begin{table}
\centering
\caption{SED results for HD\,115617. The T$_{\rm eff}$, logg, [Fe/H], D, R are denoted by effective temperature, surface gravity, metallicity, distance and radius of the star, respectively.}
\begin{tabular}{ccccc}
\hline
\hline
T$_{\rm eff}$ & logg & [Fe/H] & D & R \\
 (K)   & (dex)  & (dex) & (pc) & (R$_\odot$) \\
\hline
5430 $\pm$ 12 & 4.45 $\pm$ 0.14 & -0.02 $\pm$ 0.01 & 8.23 $\pm$ 0.15 & 0.98 $\pm$ 0.02 \\   
\hline
\end{tabular}
\label{tab:sedtable}
\end{table}

\section{Asteroseismic Analysis}
\label{sec:astero}

The mean stellar density derived from asteroseismology \citep{1986ApJ...306L..37U} is more accurate than the mean 
density derived from any method for solar-like oscillating stars because the large separation between oscillation frequencies ($\Delta\nu$) is related to the sound travel time throughout the stellar radius ($R$). 
The frequency of the maximum amplitude 
($\nu_{\rm max}$) that is related with acoustic cutoff frequency ($\nu_{\rm ac}$) 
is obtained by applying a gaussian fit to the power spectrum obtained from the oscillation frequencies
\citep{1991ApJ...368..599B}
and is directly related to
($T_{\rm eff}$)
and ($logg$)
, which are the atmospheric parameters of the star
($\nu_{\rm max} \simeq g/{T_{\rm eff}}^{-0.5}$).

Using these asteroseismic quantities
({$\Delta$}${\nu}$ and $\nu_{\rm max}$), 
if $T_{\rm eff}$ is also obtained with high precision, 
the asteroseismic stellar mass ($M_{seis}$) 
and radius ($R_{seis}$) 
relative to Sun
are obtained accurately using asteroseismic scaling relations \citep{1995AA...293...87K}: 

\begin{equation}
    \frac{M_{seis}}{M_\odot}=\left(\frac{\nu_{\rm max}}{\nu_{\rm max_\odot}}\right)^3\left(\frac{{\Delta}
    {\nu}}{{\Delta}{\nu_\odot}} \right)^{-4}\left(\frac{T{\textsubscript{eff}}}{T{\textsubscript{eff}_\odot}}\right)^{1.5},
	\label{eq:sca_mass}
\end{equation}

\begin{equation}
    \frac{R_{seis}}{R_\odot}=\left(\frac{\nu_{\rm max}}{\nu_{\rm max_\odot}}\right)\left(\frac{{\Delta}
    {\nu}}{{\Delta}{\nu_\odot}}\right)^{-2}\left(\frac{T{\textsubscript{eff}}}{T{\textsubscript{eff}_\odot}}\right)^{0.5}.
	\label{eq:sca_radius}
\end{equation}

We use the observed asteroseismic solar values as 
$\nu_{\rm max_\odot}= 3050$ ${\mu}$Hz and
${\Delta}{\nu_\odot}=136$ ${\mu}$Hz \citep{2019MNRAS.489.1753Y} 
 and also $T{\textsubscript{eff}_\odot}=5772$ K \citep{2016AJ....152...41P} 
in Eq.~\ref{eq:sca_mass} and \ref{eq:sca_radius}.

\begin{figure*}[h!]
\centering
\includegraphics[width=0.9\linewidth]{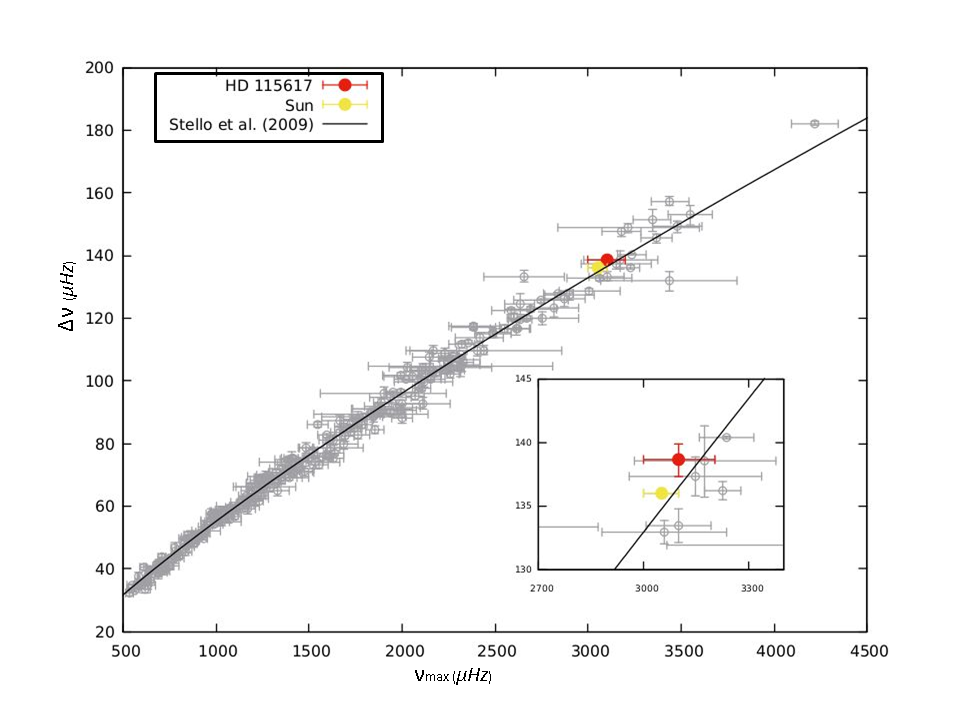}
    \caption{${\Delta}{\nu}$ (${\mu}$Hz) vs. $\nu_{\rm max}$ (${\mu}$Hz) for Sun and HD\,115617. The line represents the Main-Sequence from \citet{Stelloetal2019}.}
    \label{fig:dnu_numax}
\end{figure*}

We computed the asteroseismic stellar mass ($M_{seis}$) and radius ($R_{seis}$) of the star
using the $\nu_{\rm max}$ and 
{$\Delta$}${\nu}$ that 
were given as 
3099.8 $\pm$ 101 ${\mu}$Hz 
and 138.6 $\pm$ 1.3 ${\mu}$Hz in
\citep{2025AA...701A.285L}
and 
as spectroscopic T{\textsubscript{eff}} 
(T{\textsubscript{effs}}, see details in Sect.~\ref{sec:spec})
$M_{seis}$ and $R_{seis}$ for the star were obtained as 
0.98 $\pm$ 0.19 ${M_\odot}$ and 0.98 $\pm$ 0.09 ${R_\odot}$, respectively. This astroseismic radius is in accordance with 
those reported in the literature by \citet[1.0 $R_{\rm \odot}$]{Mennesson2014} and \citet[][R=0.9867 $\pm$ 0.0048]{vonBraun2014}. The radius value reported by \cite{Mennesson2014} is based on interferometric observation 
of the star. Accordingly, the asteroseismic 
gravity ($\log g_{\rm seis}$) was 4.45 $\pm$ 0.04 
cm s${^{-2}}$.
The uncertainties of $M_{\rm seis}$, $R_{\rm seis}$ and $\log g_{\rm seis}$ were computed using the method described by \citet{2019AA...622A.130B}.

The position of HD\,115617 on the $\Delta\nu~$vs.$~\nu_{max}$ diagram (Figure \ref{fig:dnu_numax}), when compared to a population of main-sequence and RGB stars, confirms its main-sequence evolutionary status. Its location, slightly offset from the solar values, is consistent with its classification as a solar analog rather than a perfect solar twin. The observed asteroseismic parameters ($\nu_{max}$ = 3099.8 $\mu$ Hz, $\Delta\nu$ = 138.6 $\mu$ Hz) were used to derive a stellar radius of 0.98 $\pm$ 0.09 $R_\odot$, which is in excellent agreement with both the SED fitting (0.98 $\pm$ 0.02 $R_{\odot}$) and interferometry (0.9867 $\pm$ 0.0048 $R_{\odot}$) results. This robust radius determination provides a strong foundation for evolutionary modeling.

We discuss the asteroseismic parameters along with the SED and spectroscopic analysis results in Section~\ref{sec:dis}.

\section{Age}
\label{age}

The target of this study, HD\,115617 (61 Virginis), is a bright G7V dwarf star \citep{2006AJ....132..161G,2020MNRAS.491.2280S} and one of its closest solar analogs. The precise determination of its age is paramount for placing its physical characteristics and properties of its known planetary system in an evolutionary context. Previous estimates of the age of HD\,115617 have been derived using several methods, including gyrochronology, chemical diagnostics, and isochrone fitting.

Early kinematic analysis proposed HD\,115617 as a member of the Ursa Major Moving Group (UMa MG) \citep{Vican2012}, which implies an age of approximately $500 \pm 100$ Myr \citep{King2003}. However, this hypothesis is strongly contradicted by the results of direct physical diagnostics. Gyrochronology, based on a measured rotation period of $8.91 \pm 0.36$ days, indicates an age of several billion years \citep{Metcalfe2010}. 

Similarly, the measured atmospheric lithium abundance provides an independent, physics-based constraint on the star's evolutionary stage. For a dwarf star such as HD 115617, lithium is gradually destroyed via mixing processes in the stellar interior throughout its main-sequence lifetime. Empirical calibration from open clusters establishes a well-defined relationship: lithium abundance decreases monotonically with age for stars of a given mass and temperature \citep[e.g.,][]{2005A&A...442..615S}. For HD\,115617, high-resolution spectroscopic studies report depleted lithium contents of $\log\epsilon(\mathrm{Li}) = 1.67$ dex \citep{Mallik1999} and 1.58 dex \citep{Luck2018}. This low value is consistent with typical inter-study systematic uncertainties and is a clear signature of main sequence depletion over several billion years. Using the lithium-age calibration empirically derived for solar-type dwarfs by \cite{2005A&A...442..615S} and applying it to this abundance and the star's effective temperature yields an age estimate consistent with the several-Gyr timescale indicated by gyrochronology and optical isochrone fitting. Therefore, the definitively low lithium measurement robustly invalidates the implausibly young age derived from the combined optical+NIR parameter set.

\begin{figure*}
    \centering
    \includegraphics[width=0.8\linewidth]{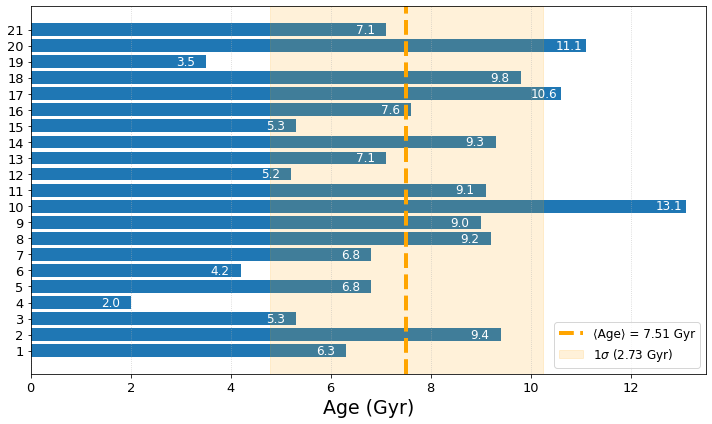}
    \caption{Horizontal bar chart showing the 21 literature age values for HD\,115617 retrieved from the Vizier database. The orange dashed vertical line indicates the mean age of the dataset ($(Age)$ = 7.51 Gyr), whereas the shaded orange region represents the $\pm 1\sigma$ standard deviation interval (2.73 Gyr). The numerical labels on the y-axis correspond to the following sources: [1] \cite{Valenti2005}, [2] \cite{Takeda2007}, [3] \cite{Holmberg2009}, [4] \cite{Ghezzi2010}, [5] \cite{Gonzalez2010}, [6] \cite{Ramirez2012}, [7] \cite{DaSilva2012}, [8] \cite{Ramirez2013}, [9] \cite{Tsantaki2013}, [10] \cite{Bensby2014}, [11] \cite{DaSilva2015}, [12] \cite{Bonfanti2016}, [13] \cite{Brewer2016}, [14] \cite{Stassun2017}, [15] \cite{Yee2017}, [16] \cite{Rich2017}, [17] \cite{Luck2018}, [18] \cite{Soto2018} (\citeyear{Soto2018}, F), [19] \cite{Chavero2019}, [20] \cite{Chen2020}, [21] \cite{Baum2022}.}
    \label{age_distribution_hist}
\end{figure*}

A compilation of age determinations from the literature, primarily based on spectroscopic analyses and isochrone fitting, is presented in Table~\ref{tab:A1}. The values listed in the t-column show a significant spread, ranging from $\sim$2 Gyr to over 13 Gyr (Figure \ref{age_distribution_hist}). However, the majority of these estimates clustered in the range of $5$–$7$ Gyr, with a median value of $7.51 \pm 2.73$ Gyr. This is consistent with the most robust previous constraints, such as those of \citet{Valenti2005} ($6.52 \pm 2.27$ Gyr) and \citet{Sousa2008} ($5.8 \pm 1.5$ Gyr), which were later refined by the \textit{Gaia} DR3 asteroseismic analysis to $5.5 \pm 1.1$ Gyr \citep{Creevey2023}. The consensus from these independent techniques robustly indicates that HD\,115617 is a middle-aged main-sequence star, with an age comparable to or slightly older than that of the Sun.

In this study, we determined the age of HD\,115617 using two distinct yet complementary modern methodologies to derive the most consistent and reliable age estimates by comparing the stellar evolution models with observational data. The isochrone fitting technique compares a star's position on the Hertzsprung-Russell diagram with theoretical evolutionary tracks. The MESA stellar evolution models provide sophisticated analyses based on physical parameters. In the following sections, each of these methodologies is discussed in detail, and the results obtained from each method are presented.

\subsection{Isochrone Fitting Technique}
\label{MCMC}

In this work, we revisit the age of HD\,115617 using a Markov Chain Monte Carlo (MCMC) method based on isochrone fitting. The PAdova and tRieste Stellar Evolutionary Code (PARSEC) isochrones \citep{Bressan2012} were employed, and the parameter space was sampled using high-resolution grids along the age ($6 \leq \log(\tau)$ (yr) $\leq 10.13$) and initial metallicity ($0 \leq Z \leq 0.03$) axes. To enhance statistical reliability, the maximum likelihood function was processed using 22 independent walkers for 5000 iterations, and the median age value along with $\pm 1\sigma$ confidence intervals were derived from the resulting posterior distributions (see \citealt{Cinar2025, Yolalan2025} for details).

\begin{figure*}
    \centering
    \includegraphics[width=0.49\linewidth]{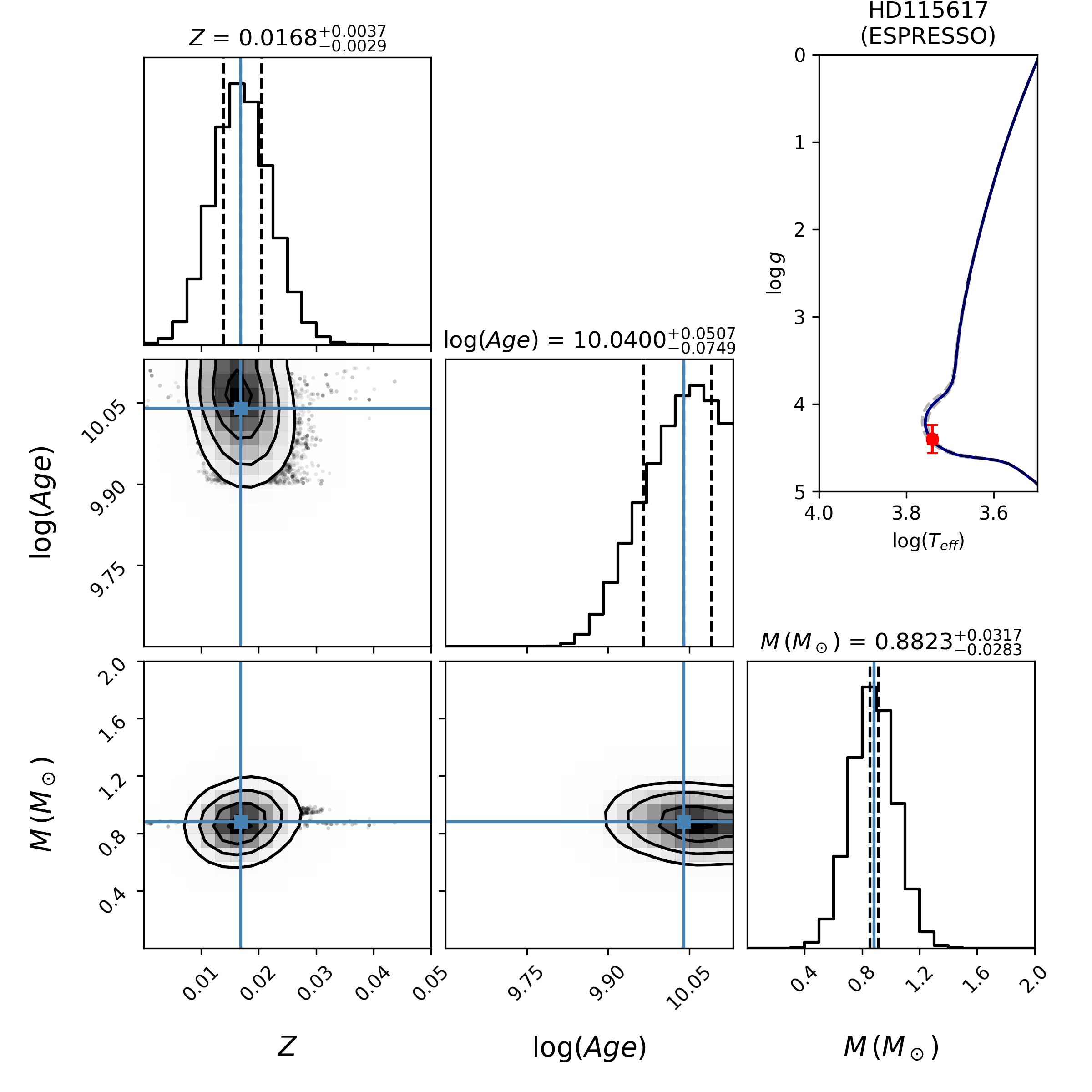}
    \includegraphics[width=0.49\linewidth]{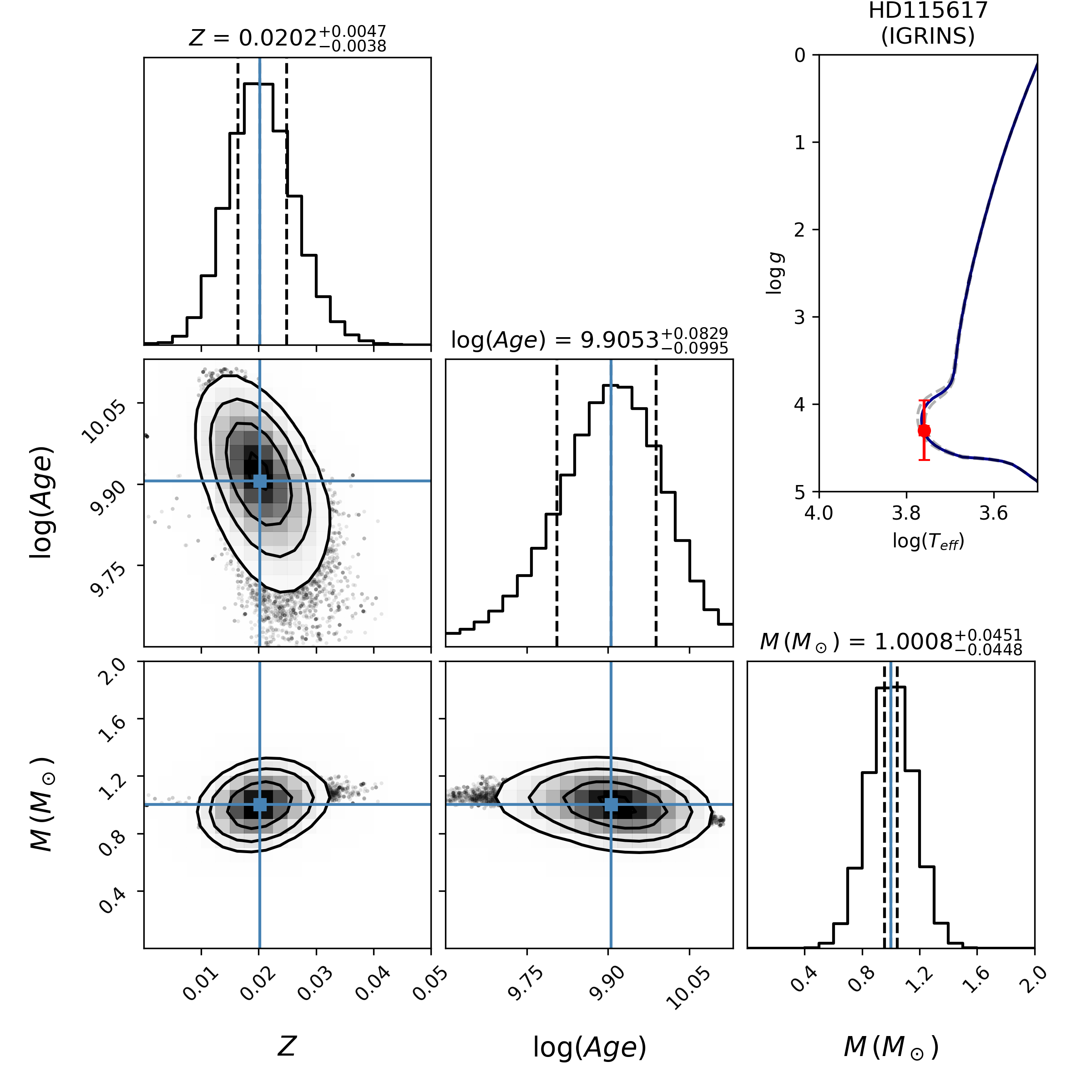}
    \caption{Corner plot displaying the posterior probability distributions for HD\,115617 (E, I), with confidence levels marked at 68$\%$, 90$\%$, and 95$\%$. The 1D marginal distributions indicate the median values and 16th and 84th percentiles. The right side of each panel shows the corresponding positions of the stars on the Kiel diagram.
}
    \label{fig6}
\end{figure*}

The analyses were applied to two distinct parameter sets: optical only (ESPRESSO) and NIR only (IGRINS) ( Table \ref{tab1}). The resulting age values were found to be $10.97_{-1.74}^{+1.35}$ Gyr (ESPRESSO) and $8.04_{-1.64}^{+1.69}$ Gyr (IGRINS) (see Figure~\ref{fig6}). The old age derived from the ESPRESSO data is consistent with the upper extreme of the literature distribution (e.g., $\sim$13 Gyr in Table~\ref{tab:A1}), whereas the NIR (IGRINS) results are broadly consistent with the literature consensus of $\sim$5.5–7 Gyr.

\begin{table*}
\centering
\caption{MESA model results of HD\,115617 and HD\,76151.}
\resizebox{1.00\textwidth}{!}{
\begin{tabular}{ccrccccccc}
\hline \hline
M & R & Age  & T$_{\rm eff}$ & logg & [Fe/H] & Method & T$_{\rm eff}$ & logg & [Fe/H] \\
\cline{1-6}
\cline{8-10}

(M$_\odot$) & (R$_\odot$) & (Myr) & (K) & (dex) & (dex) & & (K) & (dex) & (dex) \\
\hline \hline
\multicolumn{10}{c}{\textbf{HD\,115617}}\\
\cline{1-6}
\cline{8-10}
0.96 $\pm$ 0.19 & 1.01 $\pm$ 0.10 & 21.7 $\pm$ 8.68 & 4785 $\pm$ 115 & 4.40 $\pm$ 0.04 & 0.00 $\pm$ 0.20 & ESPRESSO & 5500 $\pm$ 100 & 4.40 $\pm$ 0.16 & 0.00 $\pm$ 0.16 \\
0.98 $\pm$ 0.20 &  1.02 $\pm$ 0.10 & 22.5 $\pm$ 9.00 & 4977 $\pm$ 119 & 4.41 $\pm$ 0.04 & 0.00 $\pm$ 0.20 & ESPRESSO & 5500 $\pm$ 100 & 4.40 $\pm$ 0.16 & 0.00 $\pm$ 0.16 \\
1.00 $\pm$ 0.21 &   1.02 $\pm$ 0.10 & 27.5 $\pm$ 11.0 & 5522 $\pm$ 132 & 4.42 $\pm$ 0.04 & 0.00 $\pm$ 0.20 & ESPRESSO & 5500 $\pm$ 100 & 4.40 $\pm$ 0.16 & 0.00 $\pm$ 0.16 \\
\cline{1-6}
\cline{8-10}
0.96 $\pm$ 0.19 &   1.13 $\pm$ 0.11 & 11.5 $\pm$ 4.60 & 4148 $\pm$ 99 & 4.31 $\pm$ 0.04 & 0.13 $\pm$ 0.07 & IGRINS & 5750 $\pm$ 140 & 4.30 $\pm$ 0.30 & 0.13 $\pm$ 0.08 \\
0.98 $\pm$ 0.20 &   1.16 $\pm$ 0.12 & 10.8 $\pm$ 4.32 & 4178 $\pm$ 100 & 4.30 $\pm$ 0.04 & 0.13 $\pm$ 0.07 & IGRINS & 5750 $\pm$ 140 & 4.30 $\pm$ 0.30 & 0.13 $\pm$ 0.08 \\
1.00 $\pm$ 0.21 &   1.17 $\pm$ 0.12 & 11.1 $\pm$ 4.44 & 4221 $\pm$ 101 & 4.30 $\pm$ 0.04 & 0.13 $\pm$ 0.07 & IGRINS & 5750 $\pm$ 140 & 4.30 $\pm$ 0.30 & 0.13 $\pm$ 0.08 \\
\cline{1-6}
\cline{8-10}
0.91 $\pm$ 0.18 &   0.96 $\pm$ 0.10 & -- & 5500 $\pm$ 132 & 4.43 $\pm$ 0.04 & -- & astero+E & 5500 $\pm$ 100 & 4.40 $\pm$ 0.16 & 0.00 $\pm$ 0.16 \\
0.98 $\pm$ 0.20 &   0.98 $\pm$ 0.10 & -- & 5780 $\pm$ 139 & 4.45 $\pm$ 0.05 & -- & astero+I & 5780 $\pm$ 140 & 4.15 $\pm$ 0.34 & 0.18 $\pm$ 0.10 \\
\hline \hline
M & R & Age  & T$_{\rm eff}$ & logg & [Fe/H] & Method & T$_{\rm eff}$ & logg & [Fe/H] \\
\cline{1-6}
\cline{8-10}
(M$_\odot$) & (R$_\odot$) & (Gyr) & (K) & (dex) & (dex) & & (K) & (dex) & (dex) \\
\hline \hline
\multicolumn{10}{c}{\textbf{HD\,76151}}\\
\cline{1-6}
\cline{8-10}
1.06 $\pm$ 0.21 &   1.14 $\pm$ 0.11 & 6.5 $\pm$ 2.6 & 5779 $\pm$ 139 & 4.35 $\pm$ 0.04 & 0.14 $\pm$ 0.06 & HARPS & 5780 $\pm$ 88 & 4.35 $\pm$ 0.16 & 0.14 $\pm$ 0.08 \\
1.09 $\pm$ 0.22 &   1.21 $\pm$ 0.12 & 6.9 $\pm$ 2.8 & 5768 $\pm$ 138 & 4.31 $\pm$ 0.03 & 0.19 $\pm$ 0.12 & IGRINS & 5780 $\pm$ 178 & 4.31 $\pm$ 0.25 & 0.19 $\pm$ 0.17 \\
\hline
\hline
\end{tabular}
\label{tab:modeltable}
}
\end{table*}

\subsection{Age via MESA Stellar Evolution Models}
\label{MESA}

We construct interior models of HD\,76151 and HD\,115617 with MESA evolution code 
\citep[version r-24.08.1]{2011ApJS..192....3P, 2013ApJS..208....4P, 2015ApJS..220...15P, 2019ApJS..243...10P}. The prevailing standard of mixing length theory 
\citep[$\alpha$]{1958ZA.....46..108B} was used for convection treatment. OPAL opacity 
tables are taken from \citet{1993ApJ...412..752I,  1996ApJ...464..943I}. 
In nuclear reaction rates, \citet{1999NuPhA.656....3A} with updated by 
\citet{2002ApJ...567..643K} and \citet{2010ApJS..189..240C} were used in the models.  
For the purposes of simplicity, the 
\texttt{$simple\_atmospheres$} option in MESA is 
selected for the stars (see \citet{2011ApJS..192....3P, 2013ApJS..208....4P, 2015ApJS..220...15P, 2019ApJS..243...10P} for details). Element diffusion is 
included with MESA default option and is taken into account for the stars in the 
models. For the solar values, the initial hydrogen abundance $X = 0.70358$, 
metallicity $Z = 0.0172$, age $t = 4.57$ \textit{Gyr}, 
and the mixing length parameter 
($\alpha = 2.175$) was used in MESA. These values are obtained according to the 
model that fits helioseismic parameters.

The ADIPLS package \citep{2008ApSS.316..113C} is utilised within the MESA module to 
facilitated the calculation of adiabatic oscillation frequencies for interior models. 
We compute model $\Delta\nu$, and $\nu_{\rm max}$ that is from  
\citep{1991ApJ...368..599B} with the solar values that are given in 
Section~\ref{sec:astero} using the model oscillation frequencies to fit the 
observed values. In the near-surface region of stars, the lower sound speed makes it 
It is difficult to simulate this with stellar evolution codes; therefore, surface corrections must be implemented in the future. In this study, we applied surface correction using the ADIPLS package \citep{2008ApJ...683L.175K}.

Our modelling strategy employed for HD\,76151 and HD\,115617 involves the utilisation of 
input parameters of the MESA evolution code, namely, $M$, $Y$, $Z$ and $\alpha$. 
$Z$ is computed from observed metallicity and the mixing length parameter is 
taken as the solar value ($\alpha=2.00$). 
During the calibration procedure, $M$ and $t$ is  
appropriately altered in order to align models with asteroseismic and non-asteroseismic 
constraints. 
In the case of stars, the spectral $T_{\rm{eff}}$ and $logg$ are derived from the 
asteroseismic scaling relation as a function of spectral $T_{\rm{eff}}$ and $\nu_{\rm max}$ 
(see details in Section ~\ref{sec:astero}).

In order to calibrate the interior models of the stars according to observed spectral and 
observational asteroseismic constraints, we try to fit model of the star to the observed 
spectral parameters with errors ($T_{\rm eff}$ and $logg$) in Kiel diagram and also satisfy 
asteroseismic constraints ($\nu_{max}$ and $\Delta\nu$) simultaneously.
The best-fitting model is determined by applying the classical 
$\chi^{2}$ method. In order to 
minimise the $\chi^{2}$, it may be necessary to make slight adjustments to the model $log$ 
and $T_{\rm eff}$ respectively. For asteroseismic constraints, density is the pivotal parameter that facilitates the fitting of the observed $\Delta\nu$. However, it is important to note that different combinations 
of $M$ and $R$ may utilise the same mean density, but may employ entirely different 
interior models for each of $M$ and $R$ combinations. Therefore, we paid special attention 
to the use of both asteroseismic constraints in our model.

To independently determine the age of 
HD\,115617, we first validated the reliability 
of our MESA-based approach by applying it to 
the star HD\,76151, a benchmark target 
analyzed by \citet{Senturk2024}.
They estimated that
the age of the star was $5.5^{+2.5}_{-2.1}$ Gyr using a Bayesian framework. In this 
study, we derived the stellar age for HD\,76151 as $5.6 \pm 1.4$ Gyr using MESA
stellar evolution code shows excellent 
agreement with this previous Bayesian 
estimate, thereby cross-validating 
both techniques. 

After confirming the robustness of the method,
we applied it to our primary target, 
HD\,115617. The resulting age, along with its 
uncertainty, is presented in the summary 
table (Table~\ref{tab8}) and
an important independent constraint 
in our multifaceted age-related analysis.

\begin{table*}
\caption{Fundamental Astrophysical Parameters and Kinematic and Dynamic Orbital Parameters of HD\,115617 Calculated from Photometric, Astrometric, and Spectroscopic Data.}
\centering
\resizebox{0.63\textwidth}{!}{
\small
\begin{tabular}{c|c|c|c}
\hline
\hline
\multicolumn{2}{c}{Parameter} & \multicolumn{2}{c}{Values} \\ 
\hline
\multicolumn{2}{c}{($\alpha$, $\delta$)$_{J2000}$ (sexagesimal)} & \multicolumn{2}{c}{13:18:24.31, -18:18:40.30} \\  
\multicolumn{2}{c}{(l, b)$_{J2000}$ (decimal)} & \multicolumn{2}{c}{311.851528, +44.088629}  \\  
\multicolumn{2}{c}{( $\mu_\alpha cos\,\delta$, $\mu_\beta$) (mas yr$^{-1}$)}  & \multicolumn{2}{c}{-1070.202$\pm$0.153, -1063.849$\pm$0.123} \\  
\multicolumn{2}{c}{$\varpi$ (mas)} & \multicolumn{2}{c}{117.1726$\pm$0.1456} \\  
\multicolumn{2}{c}{d (pc)} & \multicolumn{2}{c}{8.53$\pm$0.01} \\  
\multicolumn{2}{c}{Spectral type} & \multicolumn{2}{c}{G7V} \\  
\multicolumn{2}{c}{V (mag)} & \multicolumn{2}{c}{4.74}  \\  
\multicolumn{2}{c}{U - B (mag)} & \multicolumn{2}{c}{0.27} \\  
\multicolumn{2}{c}{B - V (mag)} & \multicolumn{2}{c}{0.7} \\  
\multicolumn{2}{c}{E(B - V) (mag)} & \multicolumn{2}{c}{0} \\  
\multicolumn{2}{c}{G (mag)} & \multicolumn{2}{c}{4.5325 $\pm$0.001} \\  
\multicolumn{2}{c}{G$_{BP}$ - G$_{RP}$ (mag)} & \multicolumn{2}{c}{0.8984 $\pm$0.004} \\  
\multicolumn{2}{c}{M$_{G}$ (mag)} & \multicolumn{2}{c}{4.877 $\pm$0.029} \\  
\hline
\hline
\multicolumn{2}{c}{}   & ESPRESSO & IGRINS \\
\hline
\multicolumn{2}{c}{$T_{\rm eff}$ (K)} & 5500$\pm$140 & 5750$\pm$140  \\  
\multicolumn{2}{c}{$\log g$ (cgs)} & 4.40$\pm$0.16 & 4.30$\pm$0.34 \\  
\multicolumn{2}{c}{$[Fe/H]$ (dex)} & 0.02$\pm$0.10 & 0.13$\pm$0.10  \\  
\multicolumn{2}{c}{$\xi$ (km s$^{-1}$)} & 0.52$\pm$0.50 & 2.30$\pm$0.50 \\  
\multicolumn{2}{c}{Z} & 0.01814 & 0.02259  \\  
\multicolumn{2}{c}{V$_R$ (km s$^{-1}$)} & -7.81$\pm$0.06 & -8.13 \\  
\multicolumn{2}{c}{M (M$_\odot$)$_{iso}$ (section \ref{MCMC})} & 0.900$_{-0.129}^{+0.028}$ & 1.061$_{-0.049}^{+0.055}$  \\  
\multicolumn{2}{c}{M (M$_\odot$) (section \ref{evo})} & 0.854$_{-0.012}^{+0.613}$ & 0.564$_{-0.004}^{+0.412}$  \\  
\multicolumn{2}{c}{R (R$_\odot$)$_{iso}$ (section \ref{MCMC})} & 1.064$_{-0.077}^{+0.037}$ & 1.407$_{-0.195}^{+0.366}$  \\  
\multicolumn{2}{c}{R (R$_\odot$) (section \ref{evo})} & 0.966$_{-0.050}^{+0.300}$ & 0.880$_{-0.003}^{+0.278}$  \\  
\multicolumn{2}{c}{log L (L$_\odot$)$_{iso}$ (section \ref{MCMC})} & 0.001$_{-0.088}^{+0.030}$ & 0.301$_{-0.131}^{+0.083}$ \\  
\multicolumn{2}{c}{log L (L$_\odot$) (section \ref{evo})} & 0.769$_{-0.012}^{+0.552}$  & 0.764$_{-0.007}^{+0.556}$ \\  
\multicolumn{2}{c}{t (Gyr)$_{MCMC}$ (section \ref{MCMC})} & 10.97$_{-1.74}^{+1.35}$ & 8.04$_{-1.64}^{+1.69}$  \\ 
\multicolumn{2}{c}{t (Gyr)$_{MESA}$ (section \ref{MESA})} & 0.0275$_{-0.0110}^{+0.0110}$ & 0.0111$_{-0.0044}^{+0.0044}$  \\
\multicolumn{2}{c}{R$_{Birth}$ (kpc)} & 7.97$\pm$0.14 & 7.43$\pm$0.09 \\
\hline
\hline
\multicolumn{2}{c}{(U, V, W)$_{LSR}$ (km s$^{-1}$)} & \multicolumn{2}{c}{-14.62$\pm$0.24, -33.17$\pm$0.35, -24.91$\pm$0.21} \\ 
\multicolumn{2}{c}{S$_{LSR}$ (km s$^{-1}$)} & \multicolumn{2}{c}{43.99$\pm$0.47} \\ 
\multicolumn{2}{c}{R$_a$, R$_p$ (pc)} & \multicolumn{2}{c}{8041$\pm$5, 5634$\pm$5} \\ 
\multicolumn{2}{c}{Z$_{max}$ (pc)} & \multicolumn{2}{c}{375$\pm$1} \\ 
\multicolumn{2}{c}{e} & \multicolumn{2}{c}{0.176$\pm$0.001} \\
\multicolumn{2}{c}{P (Myr)} & \multicolumn{2}{c}{190$\pm$1} \\ 
\hline
\hline
\end{tabular}
\label{tab8}
}
\end{table*}

\section{Kinematic and Orbital Dynamic Analysis}
\label{sec:kinematic}

The space velocity components of HD\,115617 were calculated using the method described by \cite{JohnsonSoderblom}. Because the star is so close to the Sun, a differential rotation correction \citep{MihalasBinney} was deemed unnecessary for this study. The key input data included the star’s parallax and proper motions from \textit{Gaia} DR3 \citep{DR3}, along with spectroscopically measured radial velocity.



\begin{figure*}
    \centering
    \includegraphics[width=0.75\linewidth]{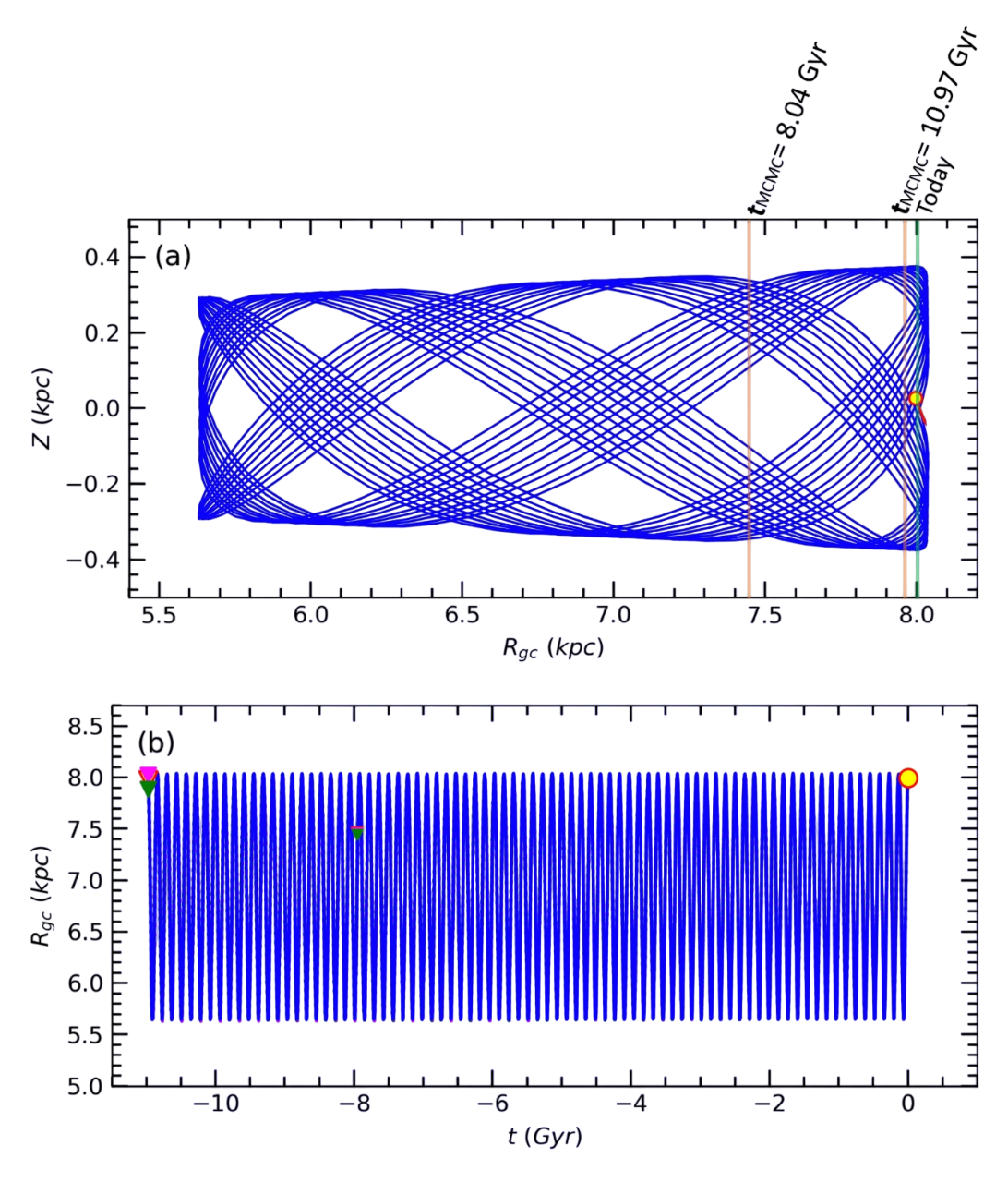}
    \caption{Galactic orbit and birth radii of HD\,115617 from ESPRESSO model parameters in the Z × R$_{gc}$ (a) and R$_{gc}$ × t (b) diagrams. The filled yellow circles and triangles represent present and birth positions, respectively. The red arrow indicates the motion vector of HD\,115617. The green and pink dotted lines show the orbit when parameter errors of the input values are considered, while the green and pink filled triangles represent the birthplaces of the HD\,115617 based on the lower and upper error estimates.}
    \label{fig:Rbirth}
\end{figure*}

The obtained (U, V, W) velocities, measured relative to the Sun, were transformed into the local standard of rest (LSR) frame using the solar space velocity values from \cite{Coskunoglu2011}. The resulting LSR velocity components were determined to be (U, V, W)$_{LSR}$ = (-14.62$\pm$0.24, -33.17$\pm$0.35, -24.91$\pm$0.21) kms$^{-1}$, with a total space velocity of S$_{LSR}$ = 43.99$\pm$0.47 kms$^{-1}$. This total velocity aligns with the kinematic properties of the thin-disc population described by \cite{Leggett1992}. The relevant inputs and results are listed in Table \ref{tab8}.

The orbital parameters of the stars were computed using the MWPotential2014 model of the galpy \citep{galpy} library. The calculations were based on standard Galactic assumptions: R$_0$ = 8.0 kpc for the Galactocentric distance and V$_0$ = 220 km s $^{-1}$ for the rotational velocity \citep{Majewski1993}. Furthermore, the Sun's distance from the Galactic plane was taken as 27$\pm$4 pc, as reported by \cite{Chen2000}. The orbit was simulated backward for 13 Gyr with a time step of 5 Myr to reflect the measurement uncertainties. The derived orbital parameters, including the perigalactic distance (R$_{peri}$), apogalactic distance (R$_{apo}$), eccentricity (e), and maximum vertical distance from the Galactic plane (Z$_{max}$), indicate a low eccentricity orbit that remains close to the Galactic plane. These characteristics are typical of stars belonging to the thin disk (see Figure \ref{fig:Rbirth}), further supporting the initial kinematic classification.

To determine the probable birth location of the star, the two age solutions obtained in Section \ref{age} were used separately, and for each age, the R$_{\rm gc}$(t) curve was traced backward. Accordingly, the Galactic birth radius of HD\,115617 from the Galactic center was calculated as R$_{\rm birth}$ = 7.97$\pm$0.14 kpc (ESPRESSO) and 7.43$\pm$0.09 kpc (IGRINS) (see Figure \ref{fig:Rbirth}). These values indicate that the star originated from a birth location within the Galactic thin disk, which is consistent with the typical R$_{gc}$ range in the vicinity of the Sun.

\section{The Evolutionary Status and Fundamental Parameters of HD\,115617}
\label{evo}

Determining the precise evolutionary stage and fundamental parameters of HD\,115617 is essential for understanding its chemical composition and spectral discrepancies identified in this study. This section synthesizes evidence from spectral energy distribution fitting, asteroseismology, spectroscopy, and theoretical stellar evolution models to determine the exact position of a star on the Hertzsprung-Russell diagram. We then independently determined the luminosity, radius, and mass using precise \textit{Gaia} DR3 photometry and astrometric data.

The primary goal of this study was to determine the precise evolutionary stage of HD\,115617, which was achieved by integrating the results of photometric, spectroscopic, asteroseismic, and theoretical analyses. The convergence of these independent methods provides a robust framework for unambiguously classifying the star as a main-sequence solar analog.

Foundational observational constraints place HD\,115617 firmly on the main sequence. SED fitting using broadband photometry from the ultraviolet to the infrared yielded an effective temperature of 5430 $\pm$ 12 K, a surface gravity of 4.45 $\pm$ 0.14 cgs, and a radius of 0.98 $\pm$ 0.02 R$_\odot$ (Table \ref{tab:sedtable}). This radius is corroborated by independent asteroseismic scaling relations applied to the TESS data. Using the observed large frequency separation, $\Delta \nu$ = 138.6 $\pm$ 1.3 ${\mu}$Hz, and the frequency of maximum power, $\nu_{\rm max} =$ 3099.8 $\pm$ 101 ${\mu}$Hz, the asteroseismic radius was calculated to be 0.98 $\pm$ 0.09 R$_\odot$ (Section \ref{sec:astero}). The remarkable agreement between the stellar energy distribution (SED) and asteroseismic radii (0.98 $\pm$ 0.02 vs. 0.98 $\pm$ 0.09 R$_\odot$) provides a strong and consistent anchor for the star's physical size. Furthermore, the star's position on the $\Delta \nu$ vs. $\nu_{\rm max}$ diagram (Figure \ref{fig:dnu_numax}) unequivocally places it within the main-sequence star locus and distinct from the region occupied by red giant branch stars.

Spectroscopic analyses offer complementary, albeit complex, insights. The atmospheric parameters derived from the optical ESPRESSO spectrum were fully consistent with those of a solar-composition main-sequence star: $T_{\rm eff} =$  5500 $\pm$ 140 K, $\log g =$ 4.40 $\pm$ 0.16 cgs, and $[Fe/H] =$ 0.02 $\pm$ 0.10 (Tables \ref{tab1} and \ref{tab8}). While the near-infrared IGRINS analysis suggests a higher effective temperature of 5750 K, its derived surface gravity of 4.30 cm s$^{-2}$ remains indicative of a dwarf star. Importantly, the asteroseismic surface gravity of 4.45 $\pm$ 0.04 cgs, which probes the star's interior structure, agrees well with the optical spectroscopic value and the SED result, which reinforces the main-sequence classification.

Theoretical stellar evolution models calculated using MESA provide the final critical evidence. Those constrained by asteroseismic parameters ($\Delta \nu$, $\nu_{\rm max}$) and spectroscopically derived $T_{\rm eff}$ and $[Fe/H]$ successfully reproduce the star's observed properties, yielding a mass close to that of the Sun. For example, the best-fit MESA model using the optical (ESPRESSO) parameters finds a mass of 0.98 $\pm$ 0.20 M$_{\rm \odot}$ and a radius of 1.02 $\pm$ 0.10 R$_{\rm \odot}$ (Table \ref{tab:modeltable}). Most decisively, the evolutionary tracks for these models place HD\,115617 within the main-sequence phase of its evolution.

Having established the main-sequence evolutionary status of HD\,115617 from the confluence of SED, asteroseismic, spectroscopic, and theoretical evidence, we now proceed to independently determine its fundamental parameters, luminosity, radius and mass. This was achieved through a detailed analysis of its precise \textit{Gaia} DR3 parallax and multiband photometry (Johnson, \textit{Gaia}, and TESS). The values derived here serve as a crucial quantitative verification of the stellar properties inferred in the previous section and provide the final astrophysical parameters for this solar analog star.

\begin{equation}
    M_\xi = \xi + 5\,log\,\varpi + 5 - A_\nu
\label{evo_1}    
\end{equation}

Using \textit{Gaia} DR3, Johnson and, TESS photometric data (see in Table~\ref{sedtable})
and the precise parallax ($\varpi$ = 117.1726$\pm$0.1456 mas), its absolute magnitudes were obtained from a distance modulus of 0.3441 mag (Eq. \ref{evo_1}) were calculated as M$_G$ = 4.877$\pm$0.029 mag, M$_{GBP}$ = 5.249$\pm$0.035 mag, M$_{GRP}$ = 4.350$\pm$0.020 mag, M$_B$ = 5.784$\pm$0.041 mag, M$_V$ = 5.084$\pm$0.031 mag, and M$_{TESS}$ = 4.429$\pm$0.021 mag, respectively, with an assumed interstellar extinction of zero. In Eq. \ref{evo_1},  $\varpi$ symbolizes the parallax in parsec units, and $\xi$ and A$_\xi$ (0.000 mag) stand for the apparent magnitude and interstellar extinction. The multiband absolute magnitude errors were computed as follows:
\begin{equation}
    \Delta M_\xi = \sqrt{(\Delta m_\xi)^2 + (5\,log\,e \frac{\sigma_\varpi}{\varpi})^2 + (\Delta A_\xi)^2}
 \label{evo_2}   
\end{equation}
Based on the information provided by \citep{Eker2023}, the errors of A$_\xi$ were assumed to be 0.029, 0.035, 0.020, 0.041, 0.031, and 0.021 for the \textit{Gaia} G, G$_{BP}$, G$_{RP}$, B, V, and TESS bands, respectively. These absolute magnitudes were then converted to bolometric magnitudes (M$_{Bol}$; Eq. \ref{evo_3}) using a bolometric correction term (BC$_\xi$; Eq. \ref{evo_4}) that is a function of the star's effective temperature ($T_{\rm eff}$ = 5500 K), obtained from the spectroscopic analysis of the ESPRESSO spectrum. 
\begin{equation}
    M_{Bol (\xi)} = M + BC_\xi
 \label{evo_3}
\end{equation}
\begin{equation}
   BC_\xi = a + b.X + c.X^2 + d.X^3 + e.X^4
 \label{evo_4}
\end{equation}
For coefficients a, b, c, d, and e in Eq. \ref{evo_4}, we adopted the values provided for the bands by \citet[Table 2]{Eker2023}. The RMS errors listed in the table correspond to the uncertainties in the bolometric correction terms. Using these, we computed BC$_G$ = 0.102 mag, BC$_{GBP}$ = –0.185 mag, BC$_{GRP}$ = 0.611 mag, BC$_{B}$ = -0.696 mag, BC$_{V}$ = 0.036 mag, and BC$_{TESS}$ = 0.564 mag, yielding bolometric magnitudes of M$_{G,Bol}$ = 4.979 $\pm$ 0.114 mag, M$_{GBP,Bol}$ = 5.065 $\pm$ 0.131 mag, M$_{GRP,Bol}$ = 4.961 $\pm$ 0.111 mag, M$_{B,Bol}$ = 5.088 $\pm$ 0.142 mag, M$_{V,Bol}$ = 5.120 $\pm$ 0.124 mag, and M$_{TESS,Bol}$ = 4.993 $\pm$ 0.113 mag.

Although HD\,115617 was also included in \cite{Eker2023}, significant methodological differences exist, particularly in reporting uncertainties. The reported uncertainty for the mean bolometric magnitude is the standard error. Furthermore, while they adopted an effective temperature of $\text{T}_{\text{eff}} = 5577$ K from the literature \citep{Ecuvillon2006}, our study used a spectroscopically determined $\text{T}_{\text{eff}}$ for this star, providing more reliable atmospheric parameters for analysis.

In this study, the uncertainties in the multiband absolute magnitudes were calculated as follows: the first term under the square root corresponds to the uncertainty in the apparent magnitude, whereas the second and third terms account for the contributions from parallax and interstellar extinction, respectively. 

These three values were combined into a weighted mean (1/$\sigma^2$) bolometric magnitude of M$_{Bol}$ = 5.026$\pm$0.302 mag. Using this value, the solar bolometric magnitude (M$_{Bol \odot}$ = 4.74 mag), and L$_{\odot}$ = 3.828 × 10$^{26}$ W in Eq. \ref{evo_5}, the luminosity of HD\,115617 was calculated to be L/L$_{\odot}$ = 0.769$_{-0.187}^{+0.246}$.
\begin{equation}
    M_{Bol} = M_{Bol \odot} - 2.5\,log \frac{L}{L_\odot}
 \label{evo_5}    
\end{equation}
Subsequently, the stellar radius was determined from this luminosity and the effective temperature using the Stefan–Boltzmann relation, yielding R/R$_{\odot}$ = 0.966\(_{-0.125}^{+0.144}\) (Eq. \ref{evo_6}). A comparison of our radius with the values in the literature is presented in Table \ref{tab:A1}. Finally, the stellar mass was estimated by combining this radius with the spectroscopically determined surface gravity ($\log g$ = 4.40  cgs) and adopting T$_{\odot}$ = 5772 K in the mass-radius relation (g $\approx$ GM/R$^{2}$). This calculation resulted in a mass of M/M$_{\odot}$ =0.854\(_{-0.165}^{+0.217}\). It is noteworthy that \cite{Eker2023} did not report a mass value for this star. A comparative analysis of the fundamental parameters derived in this study with those from \cite{Eker2023} reveals the following differences (TS - \citealt{Eker2023}): $\Delta M_{bol}(avg)$ = -0.003 mag, $\Delta L=$ +0.003 $L_{\rm \odot}$, and $\Delta R =$ +0.028 $R_{\rm \odot}$.
\begin{equation}
    \frac{R}{R_\odot} = \sqrt{\frac{(\frac{L}{L_\odot})}{(\frac{T_{eff}}{T_\odot})^4}}
  \label{evo_6}   
\end{equation}
For validation, the process was repeated using only the G$_{RP}$-band magnitude, which is less affected by extinction (A$_{GRP}$=0.020 mag), and the results were consistent with those derived from the mean bolometric magnitude. In conclusion, the derived parameters of luminosity, radius, and mass confirm that HD\,115617 is a main-sequence star with properties very similar to those of the Sun, although it is slightly less massive and less luminous than the Sun.

\section{Discussion and Conclusion}
\label{sec:dis}

In this study, the high-resolution optical (ESPRESSO) and NIR (IGRINS) spectra of HD\,115617 were analyzed together for the first time, allowing for a comprehensive examination of the star's atmospheric parameters, chemical abundances, evolutionary properties, and kinematic structure. This integrated analysis provides detailed insights into the fundamental parameters, chemical composition, age, and galactic origin of solar-type stars, contributing to our understanding of the environments in which planetary systems are formed.

The optical spectrum suggests an old star (10.97 Gyr), whereas the NIR spectrum suggests an intermediate-age star (8.04 Gyr). \cite{Bensby2014} reported an age of 13.1$^{+0.4}_{-5.0}$ Gyr for HD\,115617, which, as shown in Figure \ref{age_distribution_hist} and Table \ref{tab:A1}, is the highest literature value for this star. \cite{Bensby2014} determined the age using Yonsei-Yale (Y2) isochrones \citep{Demarque2004} via the isochrone fitting method. In this procedure, they constructed an age probability density function (APD) for each star, considering the errors in the effective temperature, surface gravity, and metallicity. The most probable age was derived from APD. However, it has been reported that age determinations based on probabilistic methods may contain systematic uncertainties \citep{Nordstrom2004}. To mitigate these effects, Bayesian-based methodologies are often preferred (e.g., \citealt{Jorgensen2005}). \cite{Bensby2014} did not adopt a Bayesian approach and noted that the APD-based ages could be refined \citep{Bensby2011}. In our study of HD\,115617, we employed an MCMC-based approach for isochrone fitting. Using atmospheric parameters derived from the ESPRESSO optical spectrum, this method yielded an age of 10.97$\pm$1.55 Gyr, which is consistent within errors with the value from \cite{Bensby2014}. Conversely, the MCMC-based age derived from the IGRINS spectrum (8.04$\pm$1.67 Gyr) agrees with the median literature age of 7.51 $\pm$ 2.73 Gyr for this star. Regarding mass, the value obtained for HD\,115617 through our MCMC-based isochrone fitting, i.e., 0.90$^{+0.03}_{-0.13}$ M$_{\rm \odot}$, is consistent with the APD-based value of 0.89$^{+0.04}_{-0.03}$ M$_{\rm \odot}$ reported by \cite{Bensby2014}.

Spectroscopic analysis of the ESPRESSO spectrum yielded precise astrophysical parameters ($T_{\rm eff}$ = 5500$\pm$140 K, $\log g$ = 4.40$\pm$0.16 cgs, and [Fe/H] = 0.02$\pm$0.10), confirming that HD\,115617 is a G-type main-sequence star with solar metallicity. This finding is supported by the fundamental parameters derived from \textit{Gaia} DR3, Johnson, and TESS photometry (L/L$_\odot$ = 0.769\(_{-0.012}^{+0.552}\), R/R$_\odot$ = 0.966$(_{-0.050}^{+0.300})$, M/M$_\odot$ = 0.854$^{+0.613}_{-0.012}$), indicating a star slightly less massive than the Sun (see Section \ref{evo}).

The parallax of the star was obtained from third-release data (\textit{Gaia} DR3) as  117.17 $\pm$ 0.15 \textit{mas} \citep{2023AA...674A...1G}. Using synthetic stellar atmospheric models with the \textit{Gaia} parallax, the best fit of the stellar radius of the SED model was determined to be 0.98 $\pm$ 0.02 {$R_\odot$} ( Fig. ~\ref{fig:sed_corner}). Comparing the SED (0.98$\pm$0.02 {$R_\odot$}, Table~\ref{tab:sedtable}) and asteroseismic results of the stellar radius(0.98 $\pm$0.09 {$R_\odot$}), the agreement appears to be reasonable within the error range. The effective temperature (5500 K), surface gravity ($\log g$ = 4.40 cgs), and metallicity (solar) derived from the optical spectrum of HD\,115617 are similar to those of the Sun.

\begin{figure*}
    \centering
    \includegraphics[width=0.4\linewidth]{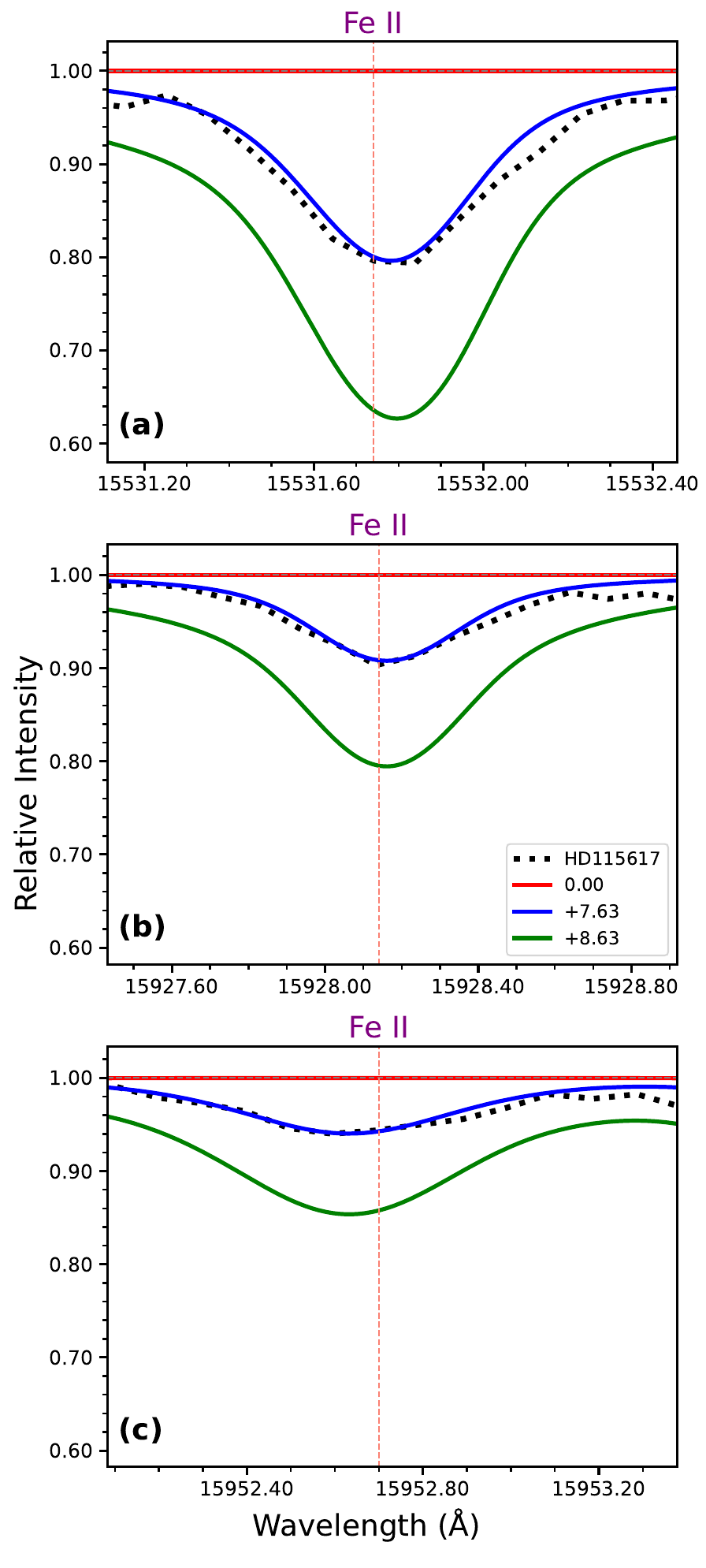}
    \caption{Spectrum synthesis of key H-band Fe\,{\sc ii} lines for HD\,115617. The panels show three distinct Fe\,{\sc ii} lines: (a) Fe\,{\sc ii} at 15531 \AA, (b) Fe\,{\sc ii} at 15928 \AA, and (c) Fe\,{\sc ii} at 15952 \AA. The synthetic spectra (colored lines) were computed for atmospheric parameters derived from the IGRINS analysis. The red curve in each plot models the spectrum with the respective Fe\,{\sc ii} line omitted, demonstrating the specific contribution of this feature. Our classification of these blended features (listed as both Fe\,{\sc i} and Fe\,{\sc ii} in APOGEE DR17) as purely Fe\,{\sc ii} lines was a critical choice for determining the ionization balance, which significantly influenced the higher effective temperature derived from the near-infrared spectrum compared to the optical analysis.}
    \label{fig:fe2_sentez}
\end{figure*}

The central finding of this study is the significant 250 K discrepancy in the effective temperature derived from classical spectroscopy in the optical versus the NIR. Although methodological systematics remain a consideration, the independent LDR method provides a critical test of its nature.

An examination of the common multiplets in the optical and NIR regions revealed a consistent line-to-line abundance pattern for Fe and $\alpha$-elements. Specifically, a comparison of the line lists from \citet{Sahin2024, Senturk2024} showed that Multiplet No. 266 was the only transition group common to both regions. The optical line at 7540.44 \AA\ yielded a solar iron abundance of $\log \epsilon$(Fe) = 7.51 dex, whereas the NIR line at 11882.86 \AA\, yielded $\log \epsilon$(Fe) = 7.54 dex. The agreement between these abundances indicates no significant wavelength-dependent systematic effects for the multiplet (266), weakening the possibility that the optical--NIR temperature discrepancy stems solely from atomic data issues for this multiplet. However, no multiplet information is available in the NIR for elements other than Fe\,{\sc i} \citep{Nave1994}, preventing similar tests for the remaining Fe\,{\sc ii} NIR lines.

Our finding of a \textit{250 K} increase in the effective temperature derived from the NIR spectrum is a central result that warrants an astrophysical explanation beyond pure methodological systematics. We consider several physical mechanisms that could explain this wavelength-dependent signature. First, the NIR continuum forms in deeper, hotter atmospheric layers; a genuine vertical temperature gradient could therefore yield a higher spectroscopically derived $T_{\rm eff}$ in the infrared. Second, magnetic activity can differentially alter line formation across wavelengths. Finally, the unique environment of HD\,115617—its known debris disk and multi-planet system—may induce subtle effects, such as a marginal NIR excess or localized atmospheric perturbations. The following discussion evaluates these possibilities against our spectroscopic and photometric constraints.

The independent verification of this higher temperature via the LDR method in the NIR region strengthens the case for its physical origin. It can be proposed that the known debris disk around HD\,115617 \citep{Wyatt2012,2025MNRAS.tmp.2118M} may plausibly contribute to this effect. Debris disks may produce a measurable excess flux in the NIR and mid-infrared regions. If not fully accounted for in the SED modeling used for our initial parameter checks or in the continuum normalization of the IGRINS spectrum, this excess flux could bias the spectroscopic analysis in the NIR toward a higher effective temperature than the actual value of the star. The NIR continuum, which forms in deeper, hotter atmospheric layers, can also be differentially affected by the star's magnetic activity. Although our IGRINS observations were obtained years after a documented activity burst, the star's general activity level could still influence the formation of specific spectral lines, particularly in the NIR, where features from different atmospheric depths are probed. The treatment of the blended ionization states of iron in the NIR line list, which yielded consistent results for the less active star HD\,76151 but divergent ones for HD\,115617 (as discussed in Section \ref{sec:dis}), further hints at an activity- or environment-related sensitivity of this method.

Given that the optical (ESPRESSO) and NIR (IGRINS) spectra were obtained using different instruments and processed through independent pipelines, we explicitly evaluated the potential impact of data quality and reduction in inhomogeneities on the derived parameters. The ESPRESSO spectrum has a nominal resolving power $R \approx 140,000$ with a per-pixel S/N $> 200$ in the continuum around 6000~\AA. The IGRINS spectrum has $R \approx 45,000$ with reported S/N of 235 at 2.2~$\mu$m. To ensure a fair comparison, we degraded the ESPRESSO spectrum to match the IGRINS resolution using a Gaussian kernel and remeasured the equivalent widths of the selected Fe\,{\sc i} and Fe\,{\sc ii} lines. The resulting atmospheric parameters shifted by +50K in $T_{\mathrm{eff}}$ and +0.02 dex in $\log g$, indicating that resolution differences alone cannot account for the 250K offset.

\begin{figure*}
    \centering
    \includegraphics[width=0.57\linewidth]{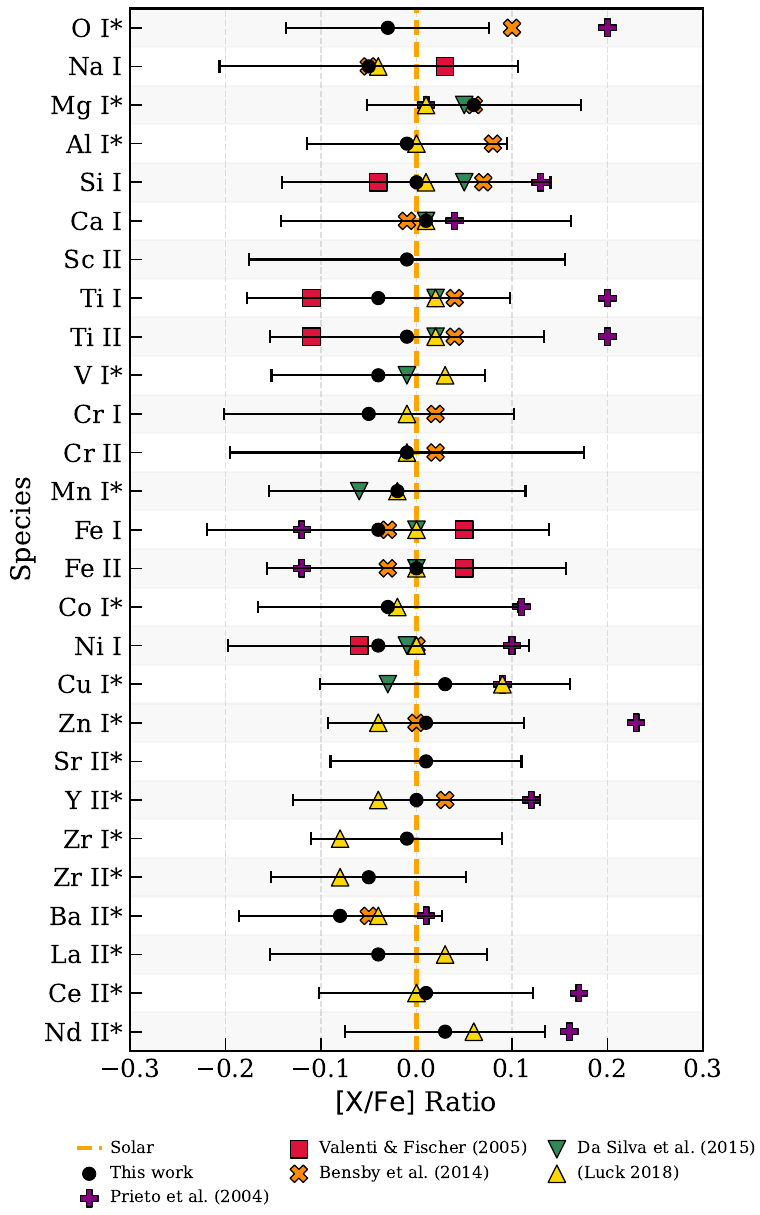}
    \caption{[X/Fe] pattern of HD\,115617 (black markers) compared with literature measurements from selected studies \citep{Prieto2004, Valenti2005, Bensby2014, DaSilva2015, Luck2018}. Asterisks on the species labels denote abundances derived through spectrum synthesis, whereas unmarked species were obtained from equivalent width analysis. Compared against solar-scaled composition ( [X/Fe]=0, orange dashed line), the [X/Fe] ratios derived here show a dispersion generally within $\pm$0.10 dex, supporting the interpretation that HD\,115617 has a chemical composition remarkably similar to that of the Sun.}
    \label{fig:lit_compare}
\end{figure*}

The presence of close-in massive planets \citep{Vogt2010} adds another layer of complexity to this problem. Although there is no direct evidence for planetary infall or surface pollution in our abundance analysis, subtle star-planet interactions could potentially influence the atmospheric structure or magnetic activity of the host star. While the multi-planet system around HD\,115617 is well established from radial velocity campaigns, our single-epoch, high-resolution spectra are not suited for detecting any planetary signatures in the spectra (e.g., the dynamical RV wobble). However, the phases of the observed spectra (0.81 - IGRINS and 0.34 - ESPRESSO) for 61 Vir b suggest a promising outlook for future research because of their significant differences. Furthermore, our chemical abundance analysis (Section \ref{sec:spec}, Figure \ref{fig:lit_compare}) revealed a composition remarkably similar to that of the Sun, with no significant anomalies in the refractory-to-volatile ratios that would suggest the large-scale accretion of planetary material. Therefore, our data show no direct chemical signatures of planetary ingestion. However, the intriguing 250 K discrepancy between the optical and NIR-derived effective temperatures warrants consideration as a potential indirect signature of the star's complex environment, such as a debris disk around the star \citep{Wyatt2012,2025MNRAS.tmp.2118M} 

Our analysis further validates the atmospheric parameters derived from the optical ESPRESSO spectrum by a detailed spectral synthesis of key diagnostic lines. As demonstrated in Appendix Figure \ref{fig:synthetic_spectra}, the synthetic spectra computed using the optical parameters (\(T_{\text{eff}} = 5500\) K, \(\log g = 4.40\), [Fe/H] = 0.02) show an excellent match with the observed ESPRESSO spectrum of HD\,115617. This is particularly evident for the highly temperature-sensitive Balmer lines H\(\alpha\) and H\(\beta\), where the synthetic profiles align precisely with the observed profiles (Figure \ref{fig:synthetic_spectra}). Furthermore, the synthesis of the surface-gravity-sensitive Mg\,{\sc i} triplet lines also revealed remarkable agreement with the observed spectrum. The synthetic spectra across these distinct diagnostic features, each probing different atmospheric layers and physical conditions, strongly reinforce the reliability and internal consistency of the model atmospheric parameters determined from optical spectral analysis.

 We applied the line depth ratio (LDR) method as an independent test to determine the effective temperatures. We have previously reported results from the NIR region, which suggested a higher temperature. The LDR method predicted a higher temperature, that is, $T_{\rm eff} =$ 5636$\pm$15 K, for HD\,115617 in the NIR region, consistent within errors with the value of $T_{\rm eff} =$ 5750$\pm$140 K from the classical spectroscopic method. To ensure a fair evaluation and ascertain whether the same method could reveal the difference between the two regions, we applied the LDR method to the optical domain of the ESPRESSO spectrum. For this purpose, we used 112 line pairs calibrated by \cite{Kovtyukh2003} for 181 dwarf stars. By testing the method on the high-resolution solar spectrum from \cite{Baker2020}, we identified 64 line pairs that fell within the reported effective temperature (5780 K) and error margin ($\pm$130 K). The mean temperature derived from these lines was 5764$\pm$72 K, which is in excellent agreement with the solar temperature reported by \cite{Senturk2024}. Applying the same line pairs to the ESPRESSO spectrum of HD\,115617 yielded a result of 5553$\pm$73 K, which is highly consistent with the spectroscopically determined temperature of 5500$\pm$140 K for this star. 
 
A pivotal aspect of this analysis is the atomic line list and its application in establishing the ionization equilibrium between Fe\,{\sc i} and Fe\,{\sc ii} in the NIR spectrum. During our previous investigation of the solar analog HD\,76151 \citep{Senturk2024}, we identified a set of lines in the APOGEE H-band library that were listed as both neutral and ionized Fe transitions. For HD\,76151, the choice to treat these specific features exclusively as Fe\,{\sc i} lines was made; notably, adopting them as Fe\,{\sc i} lines instead did not significantly alter the derived model atmosphere parameters, and the resulting optical and NIR parameters were in excellent agreement. As illustrated in Figure \ref{fig:fe2_sentez}, our spectrum synthesis of key H-band Fe\,{\sc ii} lines demonstrates the impact of line classification on ionization equilibrium. The features shown are included in the APOGEE DR17 line list with dual assignments as both Fe\,{\sc i} and Fe\,{\sc ii} transitions, reflecting their blended nature and uncertain identification in previous catalogs. By consistently treating these lines as pure Fe\,{\sc ii} features, a choice supported by the quality of the synthetic fits presented here, we obtained a significantly higher effective temperature from the near-infrared spectrum than that from the optical analysis. This sensitivity highlights the critical impact of ambiguous atomic data in the infrared, particularly for blended or multiply classified lines, on the derived stellar parameters. Such effects may be amplified in stars such as HD\,115617, where atmospheric inhomogeneities or activity could subtly alter the line formation conditions. In our earlier spectral analysis of the solar analog HD\,76151 \citep{Senturk2024}, the lines at 15\,531\AA, 15\,928 \AA, and 15\,952 \AA\, were examined as potential neutral and ionized Fe features. Spectrum synthesis for the first two transitions indicated that a neutral Fe classification was more appropriate, while the 15\,952 \AA\, line could be reasonably treated as either neutral or ionized Fe. In the present study, of HD\,115617, which, unlike HD\,76151, hosts a multi‑planet system, we note a striking similarity between the NIR (IGRINS) spectra of the two stars (Figure \ref{fig:spectrum_comparison}). To explore the impact of line classification, we repeated the same tests on the HD\,76151 spectrum. When all three lines were forced to be interpreted as ionized Fe\,{\sc ii}, the IGRINS analysis yielded an effective temperature roughly \textit{250 K} higher than that derived from the optical (ESPRESSO) spectrum. However, if only the 15\,952 \AA\, line was used as Fe\,{\sc ii} to establish ionization equilibrium, the resulting surface gravity remained within the 1$\sigma$ uncertainty of the optical value, and the temperature changed by only $\approx$30 K. For HD\,115617, the measured equivalent width (EW) of the 15\,952 \AA\, line is 35 m\AA. Using the model atmosphere parameters derived for the star, the theoretical EW computed with MOOG’s ewfind interface was 38 m\AA, which showed good agreement. Similarly, for the 15\,531 \AA\, line, the measured and theoretical EWs are 128 m\AA\, and 118 m\AA, respectively; for the 15\,928 \AA\, line, both values are 52 m\AA. These consistencies support the reliability of the line list and the EW measurements. These lines were rigorously tested on a high-resolution solar spectrum, yielding consistent abundances without introducing significant line-to-line scatter, which affirmed their reliability for abundance analysis. Intrigued by the systematic differences we began to uncover, we deliberately adopted a different approach for HD\,115617. In the NIR (IGRINS) spectrum of this star, we classified the same spectral features as Fe\,{\sc ii} lines to determine the ionization balance. The outcomes were remarkable. Unlike the case of HD\,76151, this classification was a critical factor in driving the NIR-derived atmospheric parameters, most notably a \textit{250 K} higher effective temperature, to values inconsistent with those derived from the optical (ESPRESSO) spectrum. This indicates that the sensitivity to the treatment of these blended ionization states is not uniform across all solar-type stars. The fact that a consistent methodology produces consistent results for HD\,76151 but divergent results for HD\,115617 strongly suggests that the root cause is not solely a systematic error in atomic data. We posit that the atmospheric structure of HD\,115617 may be affected by secondary processes, such as its known multi-planetary system or magnetic activity, which could subtly alter the formation depths and strengths of these specific lines in the NIR, thereby influencing the spectroscopic diagnostics in a way that was not observed in the less active single star HD\,76151.

The chemical abundance analysis revealed a surprisingly similar elemental distribution to solar values. Our derived abundance pattern, expressed as [X/Fe] for a range of species, is shown in Figure \ref{fig:lit_compare}. The ratios cluster within $\pm$0.10 dex of the solar-scaled composition ([X/Fe]=0), with no significant departures across elements measured via both equivalent width and spectrum synthesis methods. This tight agreement with solar abundances, consistent with previous studies \citep{Prieto2004, Valenti2005, Bensby2014, DaSilva2015, Luck2018}, confirms that HD\,115617 is a structural and chemical solar analog. Such homogeneity in the abundance pattern reinforces the interpretation that the star formed from material of solar-like composition, likely within the same region of the Galactic thin disk as the Sun, and has not undergone significant pollution or differential depletion processes. 

\begin{figure*}
    \centering
    \includegraphics[width=1.0\linewidth]{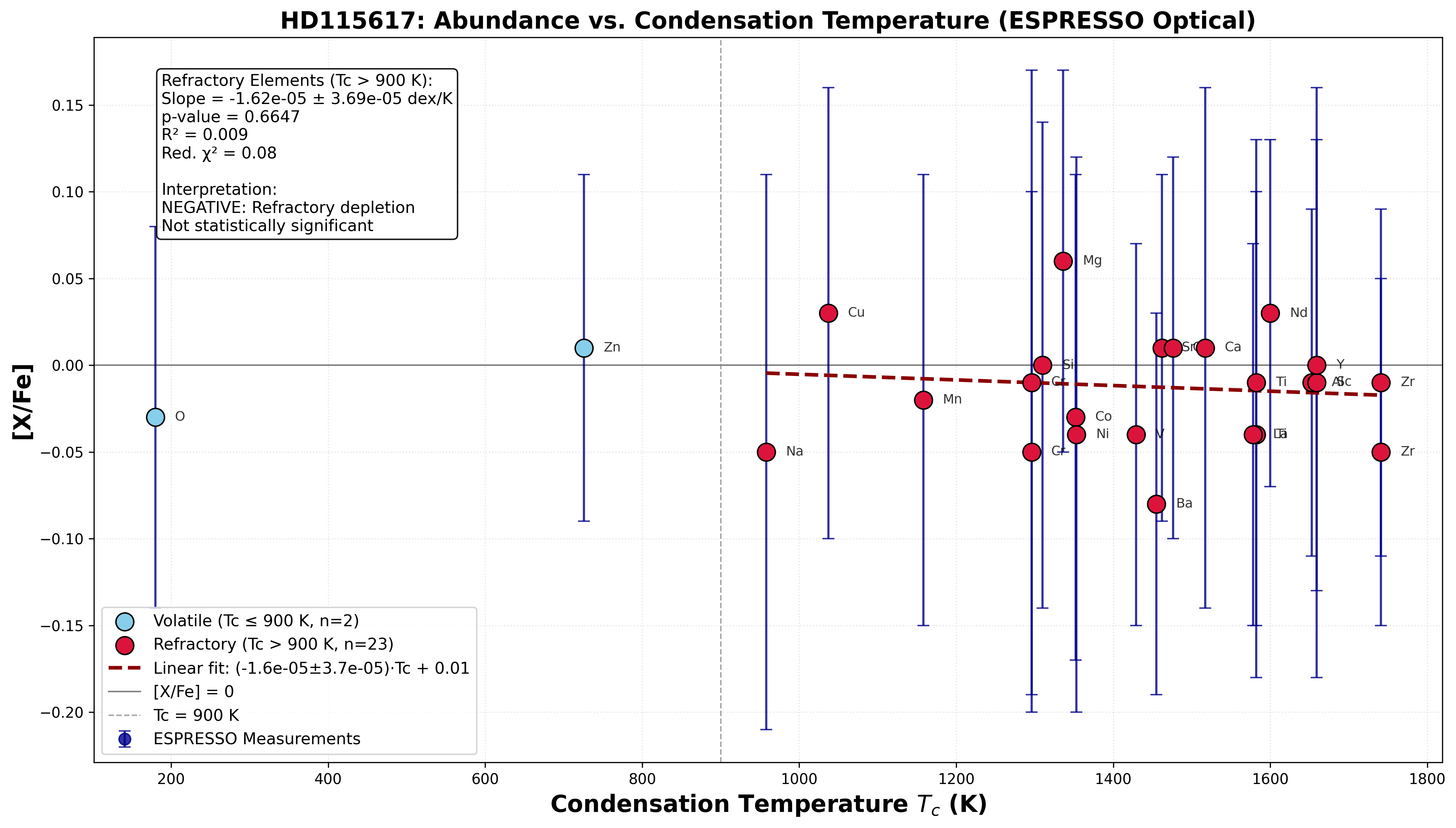}
    \caption{Elemental abundance ratios [X/Fe] as a function of condensation temperature ($T_{\mathrm{c}}$) for HD\,115617, derived from the high-resolution ESPRESSO optical spectrum. The red dashed line represents the linear regression fit to the data, which shows no statistically significant trend (slope $= -1.62 \times 10^{-5} \pm 3.69 \times 10^{-5}$\,dex\,K$^{-1}$, $p = 0.66$). The flat, solar-like pattern, with a mean [X/Fe] of $-0.014 \pm 0.031$\,dex, indicates the absence of a pronounced refractory-depletion or enrichment signature in this solar-analog host of multiple low-mass planets.}
    \label{fig:placeholder}
\end{figure*}

\subsection{Chemical Trends with Condensation Temperature}

Despite this overall solar-like chemical composition, which points to a similar formation environment, it remains valuable to search for more subtle temperature-dependent chemical signatures that could be linked to its known planetary system. Such signatures, often investigated via trends with condensation temperature, might persist even when bulk abundances appear to be solar-scaled. To further investigate the potential chemical signatures of planet formation in HD\,115617, we examined the trend of elemental abundances as a function of condensation temperature (\(T_c\)). In solar-type stars, deviations from a flat \([X/H]\) versus \(T_c\) relation—particularly a deficit of refractory elements (high \(T_c\)) relative to volatiles (low \(T_c\))—have been proposed as potential indicators of rocky planet formation, where refractories are sequestered into planetary material \citep{2009ApJ...704L..66M}.

Our analysis reveals no statistically significant trend between \([X/Fe]\) and \(T_c\) for the abundances derived from the ESPRESSO optical spectrum. A linear regression yielded a slope of \(-1.62 \times 10^{-5} \pm 3.69 \times 10^{-5}\) dex/K (\(p = 0.6647\), \(R^2 = 0.009\)). The mean \([X/Fe]\) across all measured elements is \(-0.014 \pm 0.031\) dex, with comparable values for the refractories (\(-0.015 \pm 0.031\) dex) and volatiles (\(-0.010 \pm 0.020\) dex). This flat, solar-like abundance–\(T_c\) pattern places HD\,115617 in an interesting context relative to the bimodal distribution of solar analogs identified by \cite{2021ApJ...907..116N}, where a majority “depleted” population shows refractory depletion (like the Sun) and a minority “non-depleted” population shows enrichment. HD\,115617 does not clearly belong to either extreme but occupies an intermediate position with a composition remarkably similar to that of the Sun.

The absence of a significant \(T_c\) trend in a star hosting three low-mass planets \citep{2010ApJ...708.1366V} challenges the simple predictions of planet–star chemical connections. This suggests that either the formation of its planetary system did not substantially alter the stellar composition or that any such signature has been homogenized within the convective envelope over time. This result aligns with the lack of a detected giant planet, as pressure traps induced by giant planet formation \citep{2020MNRAS.493.5079B} are often invoked to explain refractory depletion patterns. The solar-like composition, despite the presence of a multi-planet system and a debris disk, underscores the diversity of chemical patterns among solar analogs. This diversity may stem from variations in protoplanetary disk conditions, planet formation efficiency, and stellar mixing histories. Future high-precision, multi-element studies with larger samples are needed to disentangle these effects and clarify the relationship between stellar chemistry and planetary system architecture.

The absence of a significant condensation-temperature trend, coupled with its solar-like [X/Fe] pattern and kinematic signature, firmly places HD 115617 within the chemical and dynamical context of the Galactic thin disk. Returning to the core spectroscopic anomaly, we argue that the pronounced sensitivity of the NIR ionization balance in HD 115617, compared to the calm analog HD 76151, may itself be a signature of its complex astrophysical environment. Calculations of its birth radius (R$_{\rm birth}$) through backward orbital integration produced values between 5.7 and 8.0 kpc. These results strongly indicate that the star originated within the Galactic thin disk near the current orbital position of the Sun and has not undergone significant radial migration throughout its lifetime.

Having established that HD 115617 is a solar-analog star with a Galactic origin and evolution similar to that of the Sun, we return to the central spectroscopic anomaly of this study. Collectively, our analysis demonstrates that HD 115617, with well-determined basic properties, including radius, mass, bulk metallicity, and Galactic origin, is a solar-analog star. However, a detailed multiwavelength spectroscopic diagnosis revealed significant tension between the atmospheric parameters derived from different spectral regions. While part of this tension may stem from the persistent challenges of NIR spectrum analysis, the critical sensitivity of the NIR ionization balance to line classification choices is uniquely pronounced in HD 115617, compared to the calm solar analog HD 76151. Therefore, we argue that this sensitivity may be the spectroscopic signature of the complex astrophysical environment of HD 115617. To ascertain the precise nature of this signature, whether driven by magnetic activity, circumstellar disk interactions, or subtle effects from its planetary system, requires the systematic application of the methodology developed here to a large, homogeneous sample of solar-type stars. This study thus establishes a framework and a critical benchmark for stars like HD 115617 that have debris disks and planet companions via multi-wavelength spectroscopy in probing not just stellar composition, but also stellar environment and activity.

In conclusion, this comprehensive characterization of HD\,115617 underscores the critical importance of multi-wavelength analyses and reveals intriguing tensions that likely hold valuable astrophysical insights, paving the way for future studies of solar-type stars and their planetary systems.

\section{Concluding Remarks}
\label{sec:concluding}
This study presents the first comprehensive multiwavelength spectroscopic analysis of the solar-analog star HD\,115617, integrating high-resolution optical (ESPRESSO) and near-infrared (IGRINS) spectra with asteroseismic and photometric diagnostics. Our multiwavelength analysis revealed a significant (250 K) yet currently inconclusive discrepancy in the atmospheric parameters derived from the optical and NIR spectra. While the observed differences are intriguing and may point to wavelength-dependent systematics or underlying astrophysical processes, our analysis cannot definitively distinguish between these possibilities.

Complementing this, our detailed abundance analysis demonstrates that HD\,115617 has a fundamentally solar-like chemical composition. We found no statistically significant trend between [X/Fe] and condensation temperature (slope = -1.62 × 10$^{-5}$ $\pm$ 3.69 × 10$^{-5}$ dex/K, p = 0.6647), indicating the absence of systematic refractory element depletion or enrichment. With a mean [X/Fe] of -0.014 $\pm$ 0.031 dex across all measured elements, the star's chemical pattern closely resembles that of the Sun. This positions HD\,115617 as an intermediate case within solar analog populations, contrasting with the clear bimodality reported in larger surveys and challenging simplified models of planet-star chemical connections, given its three known low-mass planets.

The primary limitations of this work include the modest S/N of the NIR spectrum, remaining uncertainties in the NIR atomic line parameters, and the lack of time-resolved data to account for stellar activity. Therefore, we emphasize that our results are preliminary and highlight the necessity of homogeneous, high-quality multiwavelength datasets. Future systematic surveys of solar-type stars, encompassing both planet hosts and non-hosts, will be essential to determine whether such discrepancies correlate with stellar and planetary properties. This work underscores the potential of high-precision multi-wavelength spectroscopy to act as an indirect probe of stellar environments, including planetary systems, which is the focus of our ongoing research. Such efforts may eventually establish novel methodologies for exploring and characterizing exoplanetary environments, particularly when combined with high-precision abundance analyses that can detect the subtle chemical signatures of planet formation.

%
%

\ack
This study was supported by the Scientific and Technological Research Council of T\"{u}rkiye (T\"{U}B\.{I}TAK) under the project numbers MFAG-121F265.




\clearpage
\bibliographystyle{jphysicsB}
\bibliography{bibtex}

\appendix
\renewcommand{\thetable}{A\arabic{table}}
\section{Literature}

\renewcommand{\arraystretch}{1.2}
\begin{table*}
\caption{The model atmosphere parameters, [X/Fe] ratios from optical spectroscopic studies, radius, mass, and age parameters compiled from the literature for HD 115617. TS (E) and TS (I) denote the results of the optical and NIR spectroscopic analyses, respectively. Stellar radius, mass, and age were reported via MCMC and \textit{MESA} (in bold type face).}
\centering
\tiny
\begin{tabular}{c|c|c|c|c|c|c|c|c|c|c}
\hline
\hline
Teff & logg & [Fe/H] & [Mg/Fe] & [Si/Fe] & [Ca/Fe] & [Ti/Fe] & R & M & t & Ref.\\
\hline
(K) & (cgs) & (dex) & (dex) & (dex) & (dex) & (dex) & ($R_\odot$) & ($M_\odot$) & (Gyr) & \\
\hline
        5500 & 4.40 & 0.02 & 0.06 & 0.00 & 0.01 & -0.03 & 1.06$\pm$0.07/\textbf{1.02$\pm$0.10} & 0.90$\pm$0.08/\textbf{1.00$\pm$0.21} & 10.97$\pm$1.55/\textbf{0.03$\pm$0.01} & TS(E) \\  
        5750 & 4.30 & 0.13 & – & -0.03 & – & 0.03 & 1.41$\pm$0.28/\textbf{1.17$\pm$0.12} & 1.06$\pm$0.05/\textbf{1.00$\pm$0.21} & 8.04$\pm$1.67/\textbf{0.01$\pm$0.01} & TS(I) \\
        – & – & – & – & – & – & – & 1.0 & 0.9 & – & 1 \\  
        5585 & 4.50 & -0.02 & – & 0.03 & 0.03 & 0.11 & – & – & – & 2 \\  
        5590 & 4.23 & -0.03 & -0.06 & -0.06 & -0.06 & -0.06 & – & – & – & 3 \\  
        5552 & 4.33 & -0.03 & – & – & – & – & – & – & – & 4 \\  
        5600 & 4.50 & 0.02 & 0.05 & 0.02 & – & 0.03 & – & – & – & 5 \\  
        5623 & 4.52 & – & – & – & – & – & 1.0 & 0.9 & – & 6 \\  
        5600 & 4.35 & 0.03 & – & – & – & – & – & – & – & 7 \\  
        5483 & 4.52 & -0.12 & 0.01 & 0.13 & 0.04 & 0.20 & – & – & – & 8 \\  
        5585 & 4.35 & 0.02 & 0.07 & 0.00 & 0.01 & 0.03 & – & – & – & 9 \\  
        5577 & 4.34 & 0.01 & – & – & – & – & 0.9 & – & – & 10 \\  
        5537 & 4.38 & -0.07 & 0.04 & 0.07 & 0.04 & 0.15 & – & – & – & 11 \\  
        5571 & 4.47 & 0.05 & – & -0.04 & – & -0.11 & 1.0 & 1.0 & 6.3 & 12 \\  
        5515 & 4.53 & -0.03 & 0.02 & 0.05 & -0.02 & -0.06 & – & – & – & 13 \\  
        5720 & 4.67 & 0.11 & – & – & – & – & 1.0 & – & 9.4 & 14 \\  
        5487 & 4.49 & 0.01 & – & – & – & – & – & – & – & 15 \\  
        5558 & 4.36 & -0.02 & – & – & – & – & 0.9 & – & – & 16 \\  
        5572 & – & 0.00 & – & – & – & – & – & – & 5.3 & 17 \\  
        5558 & 4.36 & -0.02 & 0.02 & 0.02 & – & – & – & – & – & 18 \\  
        5618 & 4.52 & 0.06 & – & – & – & – & 1.1 & 0.9 & 2.0 & 19 \\  
        5533 & 4.29 & 0.01 & – & – & – & – & 0.9 & – & 6.8 & 20 \\  
        5539 & 4.35 & 0.02 & – & – & – & – & – & – & – & 21 \\  
        5558 & 4.36 & -0.02 & 0.07 & 0.03 & 0.08 & 0.07 & – & – & – & 22 \\  
        5566 & 4.41 & -0.01 & – & – & – & – & 0.9 & – & 4.2 & 23 \\  
        5400 & 4.57 & 0.00 & – & – & – & – & – & – & – & 24 \\  
        5587 & 4.41 & 0.00 & 0.06 & 0.00 & 0.00 & 0.02 & 0.9 & – & 6.8 & 25 \\  
        5571 & 4.42 & -0.02 & – & – & – & – & 0.9 & – & 9.2 & 26 \\  
        5537 & 4.38 & -0.03 & – & – & – & – & 0.9 & – & 9.0 & 27 \\  
        5469 & 4.36 & -0.03 & 0.06 & 0.07 & -0.01 & 0.04 & – & – & 13.1 & 28 \\  
        5534 & 4.41 & -0.01 & – & – & – & – & – & – & – & 29 \\  
        5469 & 4.40 & -0.03 & – & – & – & – & – & – & – & 30 \\  
        5544 & 4.32 & 0.00 & 0.05 & 0.05 & 0.01 & 0.02 & 0.9 & – & 9.1 & 31 \\  
        5490 & 4.40 & -0.10 & 0.10 & – & – & – & 0.9 & – & – & 32 \\  
        5651 & 4.46 & – & – & – & – & – & 1.0 & 0.9 & 5.2 & 33 \\  
        5490 & 4.40 & -0.10 & – & – & – & – & – & – & – & 34 \\  
        5562 & 4.44 & -0.04 & 0.03 & 0.03 & 0.03 & 0.03 & 0.9 & 1.0 & 7.1 & 35 \\ 
        5562 & 4.44 & -0.03 & – & – & – & – & 1.0 & 1.0 & – & 36 \\          
        5538 & 4.41 & 0.01 & – & – & – & – & 0.9 & 1.0 & 9.3 & 37 \\  
        5578 & 4.44 & -0.01 & 0.13 & 0.06 & 0.10 & 0.09 & - & 0.99 & 5.3 & 38 \\  
        5550 & 4.46 & -0.09 & – & – & – & – & – & – & – & 39 \\  
        5548 & 4.42 & 0.00 & 0.01 & 0.01 & 0.01 & 0.02 & 1.0 & – & 7.6 & 40 \\  
        5557 & 4.36 & 0.02 & 0.01 & 0.01 & 0.04 & -0.05 & 0.9 & 1.0 & 10.6 & 41 \\  
        5473 & 4.47 & -0.75 & -0.33 & 0.08 & 0.03 & 0.51 & 0.9 & 1.0 & 9.8 & 42 \\  
        5569 & 4.39 & 0.03 & – & – & – & – & 0.9 & – & – & 43 \\  
        5559 & 4.36 & -0.01 & – & – & – & – & 0.9 & – & – & 44 \\  
        5580 & 4.39 & 0.02 & – & – & – & – & 0.9 & – & – & 45 \\  
        5556 & 4.35 & -0.02 & – & – & – & – & 0.9 & – & – & 46 \\  
        5618 & 4.52 & 0.02 & – & – & – & – & 1.0 & – & 3.5 & 47 \\  
        5490 & 4.40 & -0.10 & – & – & – & – & 0.9 & – & 11.1 & 48 \\  
        5562 & 4.44 & -0.04 & – & – & – & – & – & – & 7.1 & 49 \\  
        5522 & 4.52 & -0.08 & – & – & – & – & – & – & – & 50 \\  
        5700 & 4.26 & -0.01 & – & – & – & – & 1.0 & – & – & 51 \\   
\hline
\hline
\end{tabular}
\begin{minipage}{14.7cm}
\scriptsize
\textbf{References:} \\
{[1] \cite{Zakhozhaj1979}, [2] \cite{Perrin1988}, [3] \cite{Edvardsson1993}, [4] \cite{Gratton1996}, [5] \cite{Thevenin1998}, [6] \cite{Prieto1999}, [7] \cite{Heiter2003}, [8] \cite{Prieto2004}, [9] \cite{Luck2005}, [10] \cite{Santos2005}, [11] \cite{Soubiran2005}, [12] \cite{Valenti2005}, [13] \cite{Reddy2006}, [14] \cite{Takeda2007}, [15] \cite{Ramirez2007}, [16] \cite{Sousa2008}, [17] \cite{Holmberg2009}, [18] \cite{Mena2010}, [19] \cite{Ghezzi2010}, [20] \cite{Gonzalez2010}, [21] \cite{Prugniel2011}, [22] \cite{Adibekyan2012}, [23] \cite{Ramirez2012}, [24] \cite{Maldonado2012}, [25] \cite{DaSilva2012}, [26] \cite{Ramirez2013}, [27] \cite{Tsantaki2013}, [28] \cite{Bensby2014}, [29] \cite{Nissen2014}, [30] \cite{Battistini2015}, [31] \cite{DaSilva2015}, [32] \cite{Sitnova2015}, [33] \cite{Bonfanti2016}, [34] \cite{Zhao2016}, [35] \cite{Brewer2016}, [36] \cite{Stassun2017}, [37] \cite{Yee2017}, [38] \cite{Luck2017}, [39] \cite{Rich2017}, [40] \cite{Luck2018}, [41] \citet[][F]{Soto2018}, [42] \citet[][H]{Soto2018}, [43] \citet[][F]{Sousa2018}, [44] \citeauthor{Sousa2018} (\citeyear{Sousa2018}, H), [45] \citet[][N]{Sousa2018}, [46] \citet[][U]{Sousa2018}, [47] \cite{Chavero2019}, [48] \cite{Chen2020}, [49] \cite{Baum2022}, [50] \cite{Contursi2024}, [51] \cite{Khalatyan2024} [TS] This Study}
\end{minipage}
\label{tab:A1}
\end{table*}

\section{Synthetic Spectrums}
\renewcommand{\thefigure}{B\arabic{figure}}
\begin{figure*}[ht]
    \centering
    \includegraphics[height=6.4cm, width=1\linewidth]{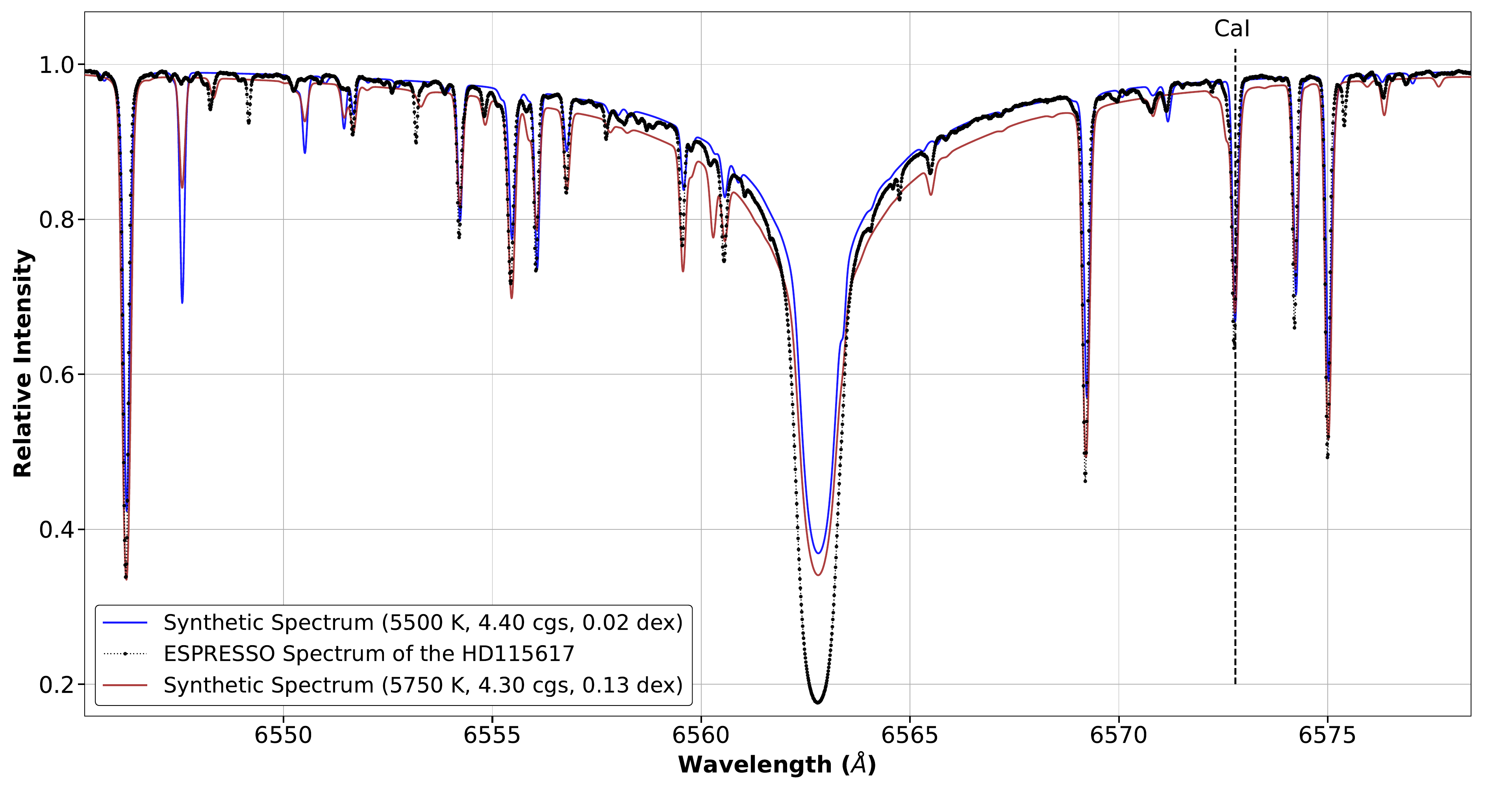}
    \includegraphics[height=6.4cm, width=1\linewidth]{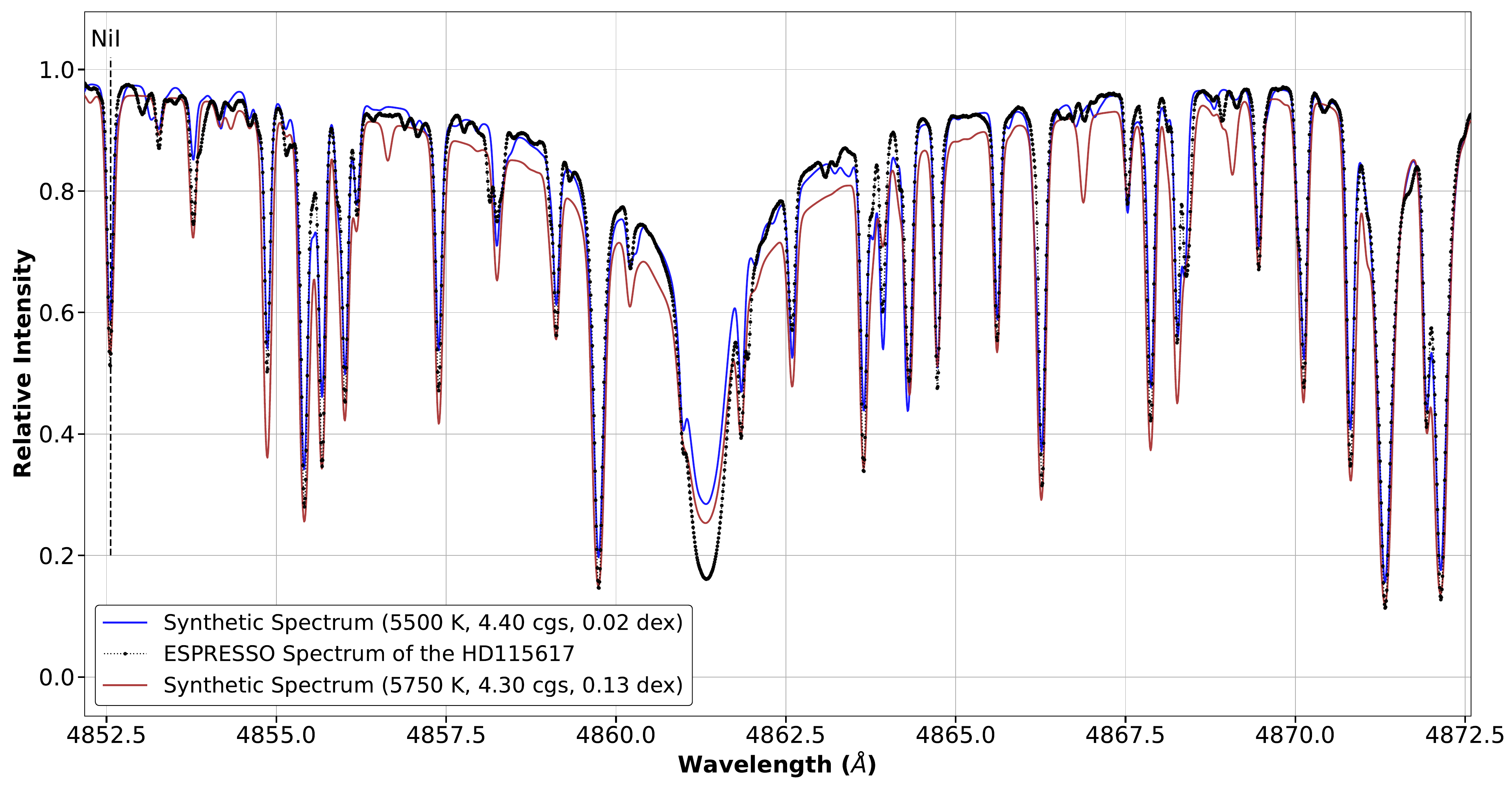}
    \includegraphics[height=6.4cm, width=1\linewidth]{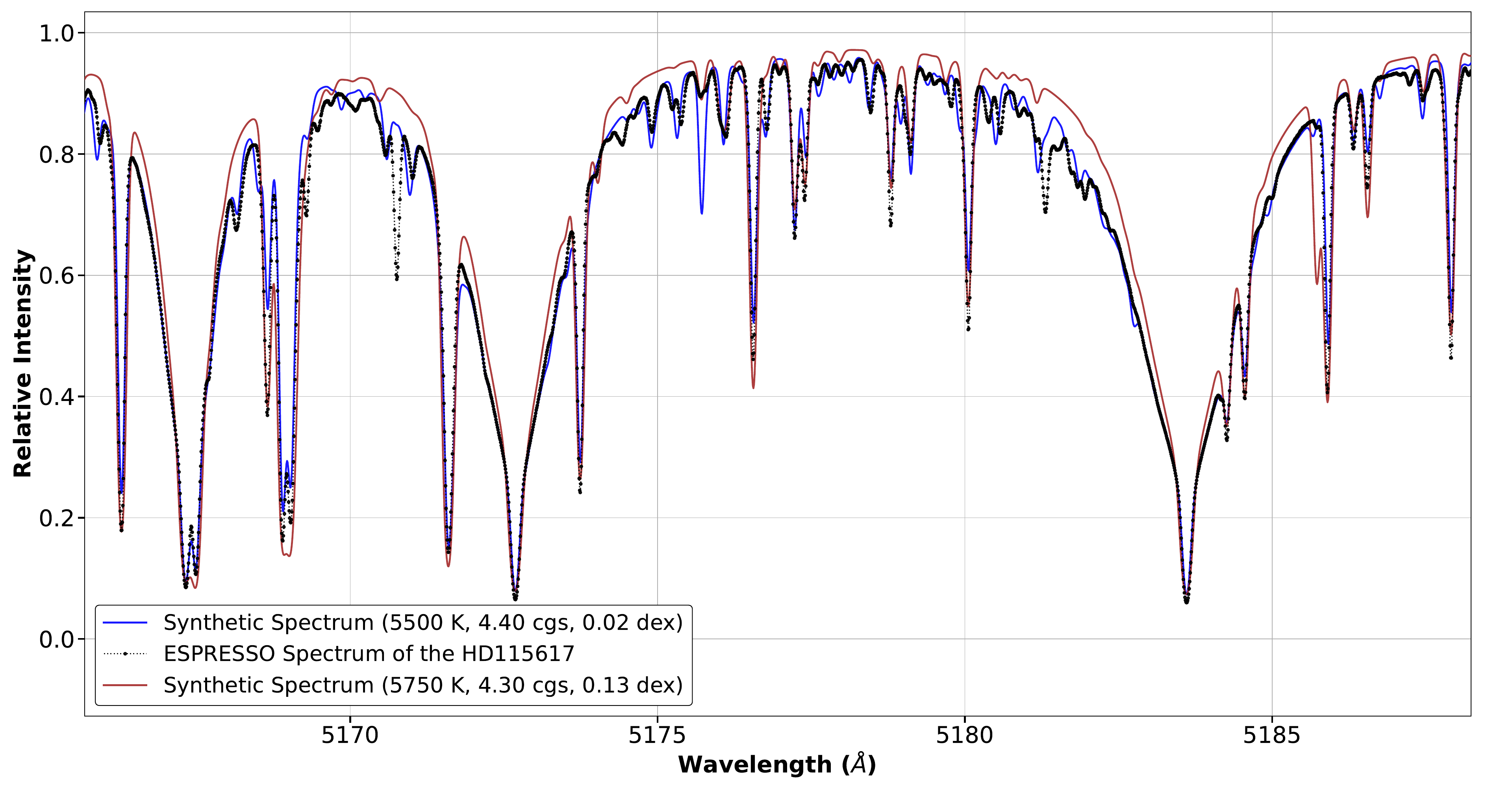}
    \caption{Comparison of the observed ESPRESSO spectrum (dotted black) of HD\,115617 with synthetic spectra calculated using atmospheric parameters obtained from optical (blue) and NIR (red) spectral analyses. The three panels show wavelength regions centered around known spectral regions: H$\alpha$ (top), H$\beta$ (middle), and Mg triplet (bottom). The H$\alpha$ and H$\beta$ lines are highly temperature-sensitive transitions, whereas the apparent sensitivity of the Mg triplet to surface gravity is well established. Therefore, these regions were chosen to demonstrate the extent to which the synthetic spectra can represent the observed line profiles and continuity behavior.}
    \label{fig:synthetic_spectra}
\end{figure*}

\begin{figure*}
    \centering
    \includegraphics[width=0.63\linewidth, angle=90]{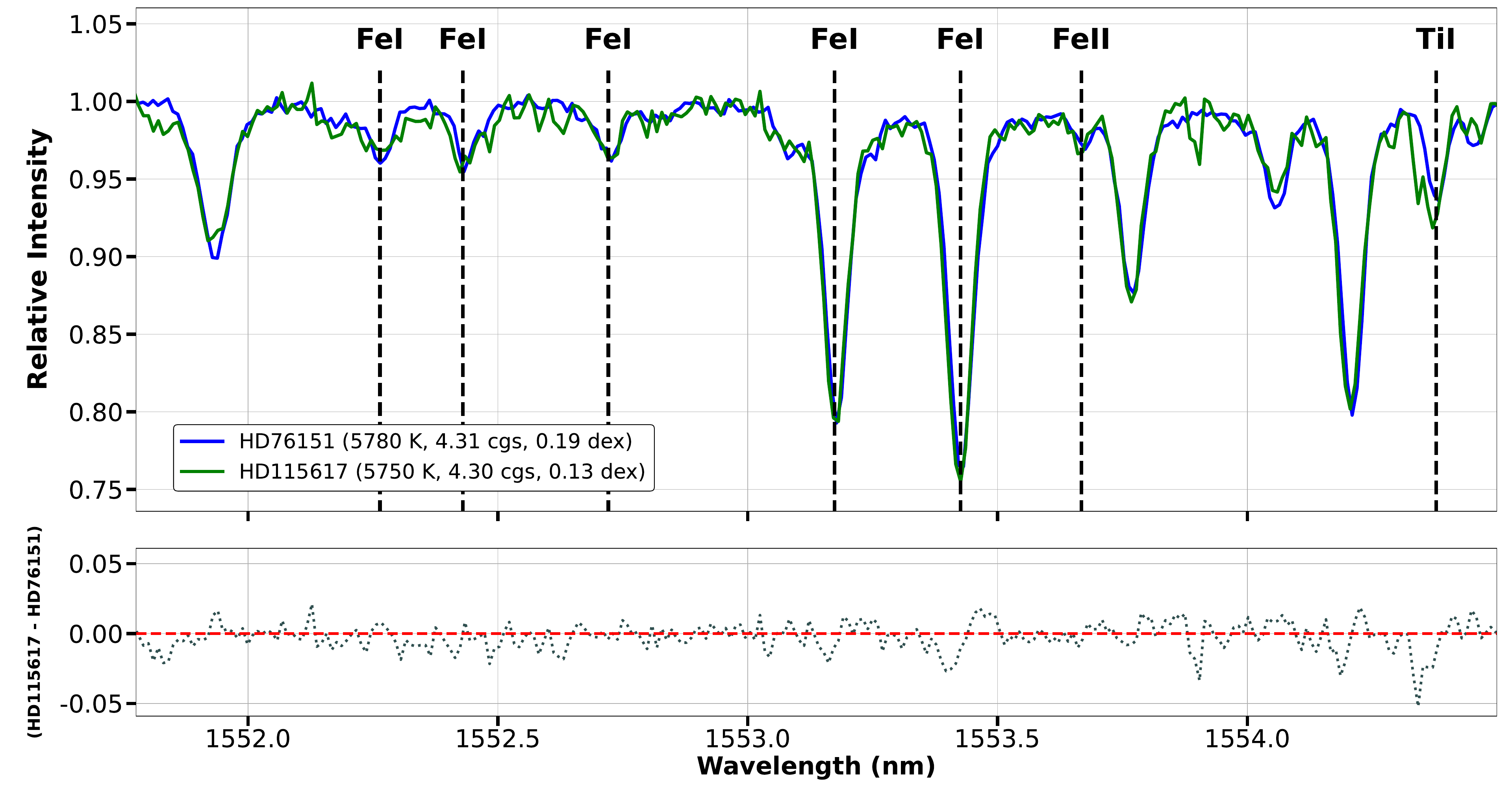}
    \includegraphics[width=0.63\linewidth, angle=90]{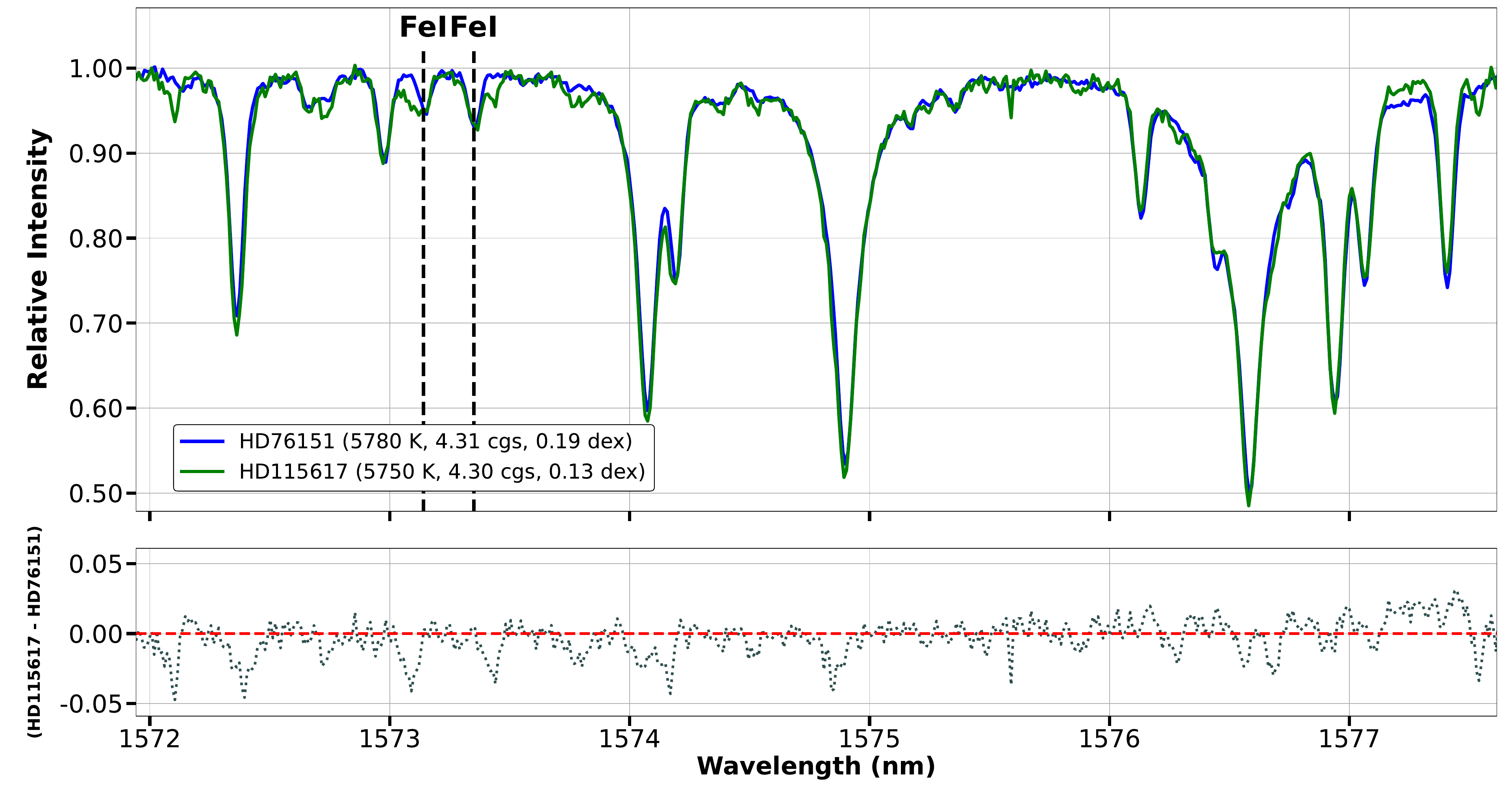}
    \includegraphics[width=0.63\linewidth, angle=90]{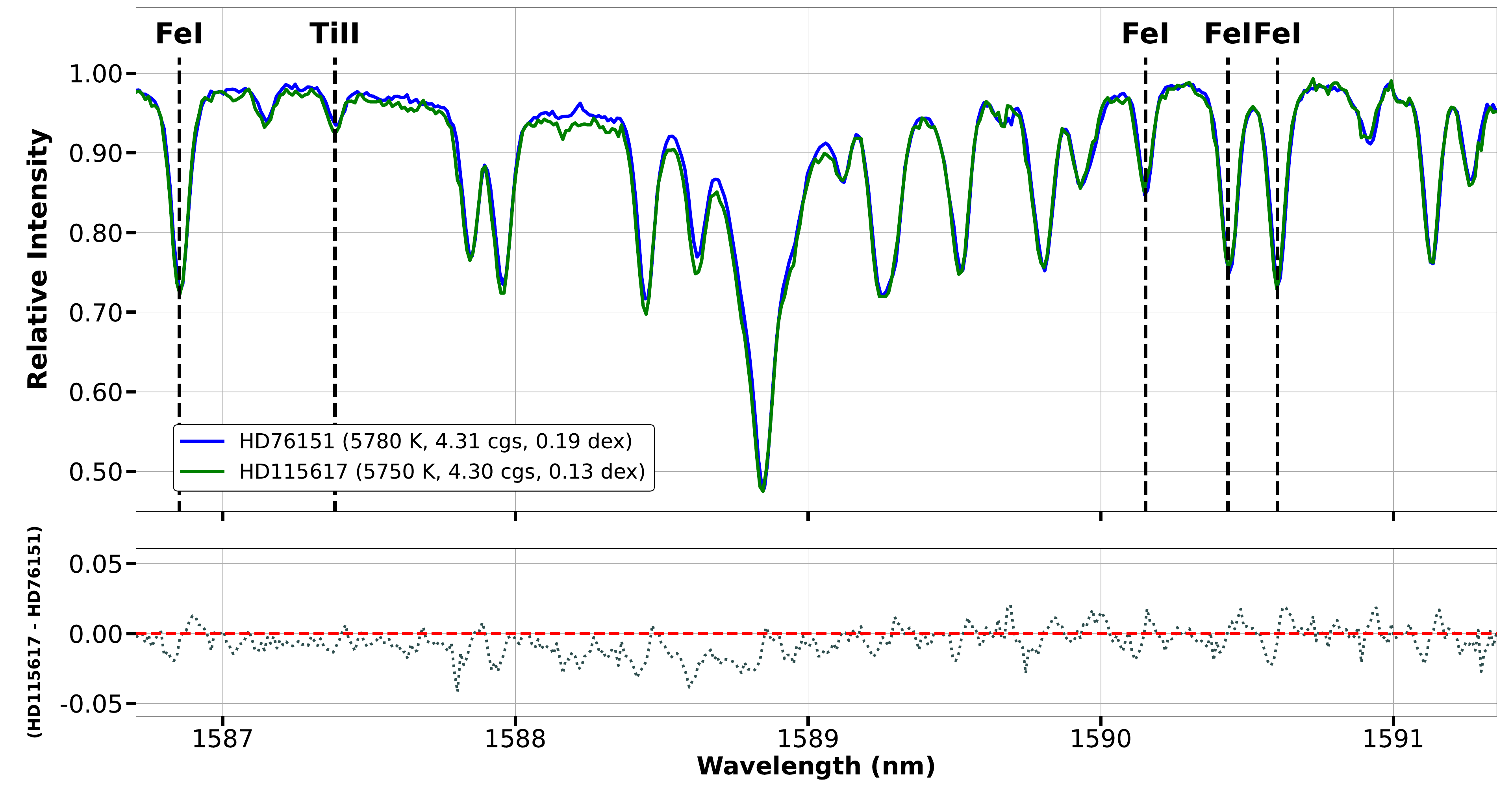}
    \includegraphics[width=0.63\linewidth, angle=90]{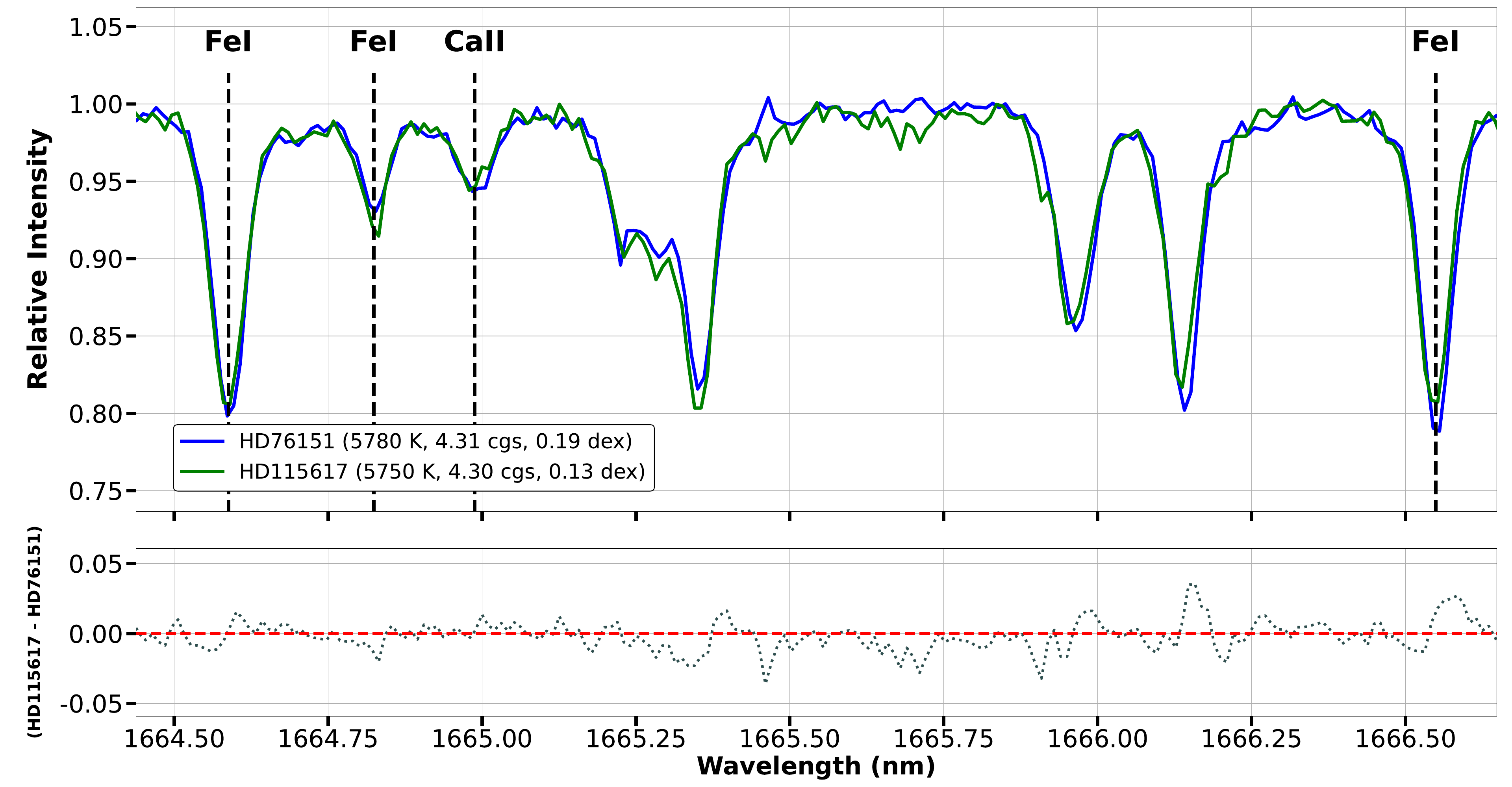}
    \includegraphics[width=0.63\linewidth, angle=90]{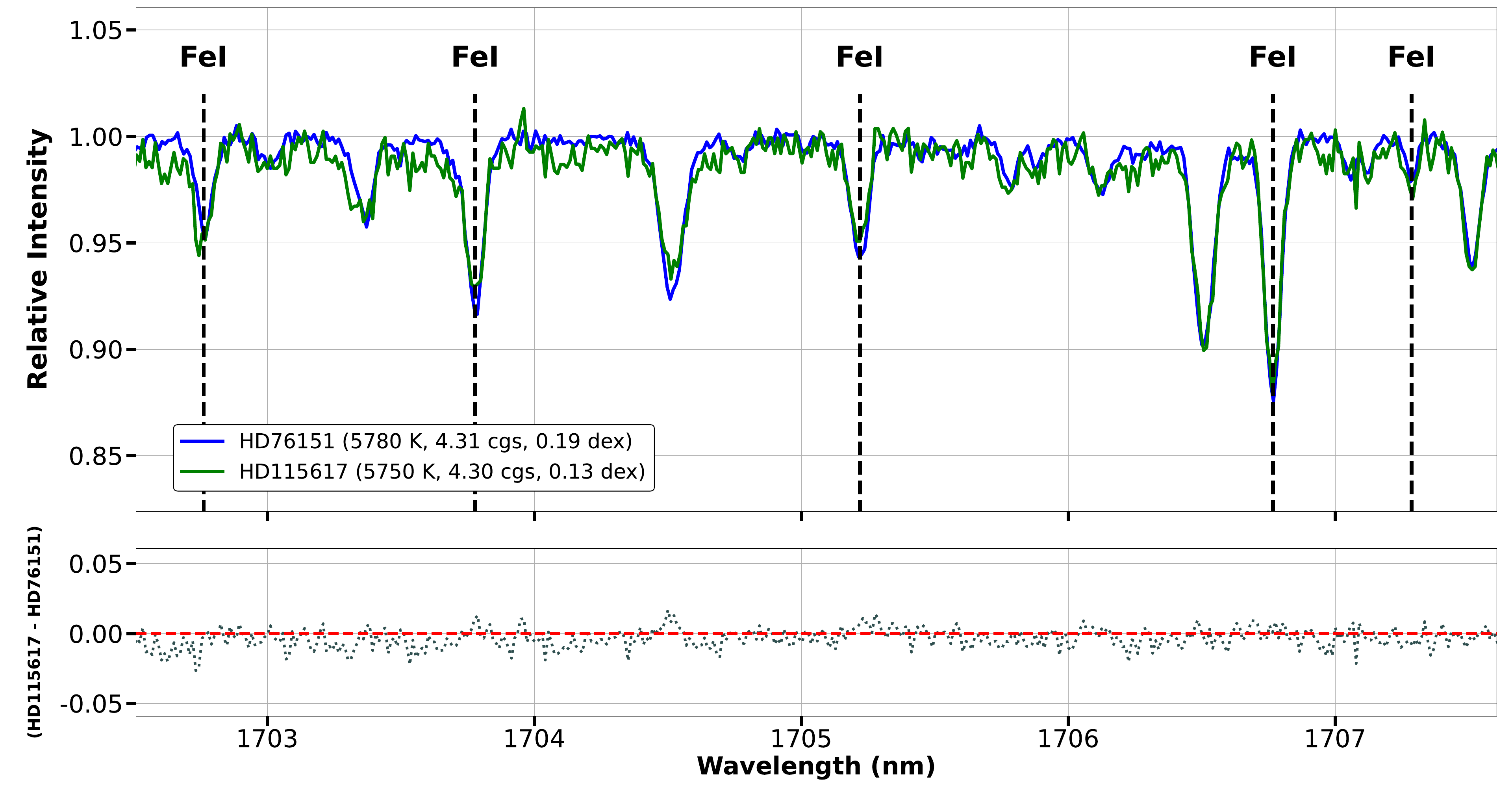}
    \includegraphics[width=0.63\linewidth, angle=90]{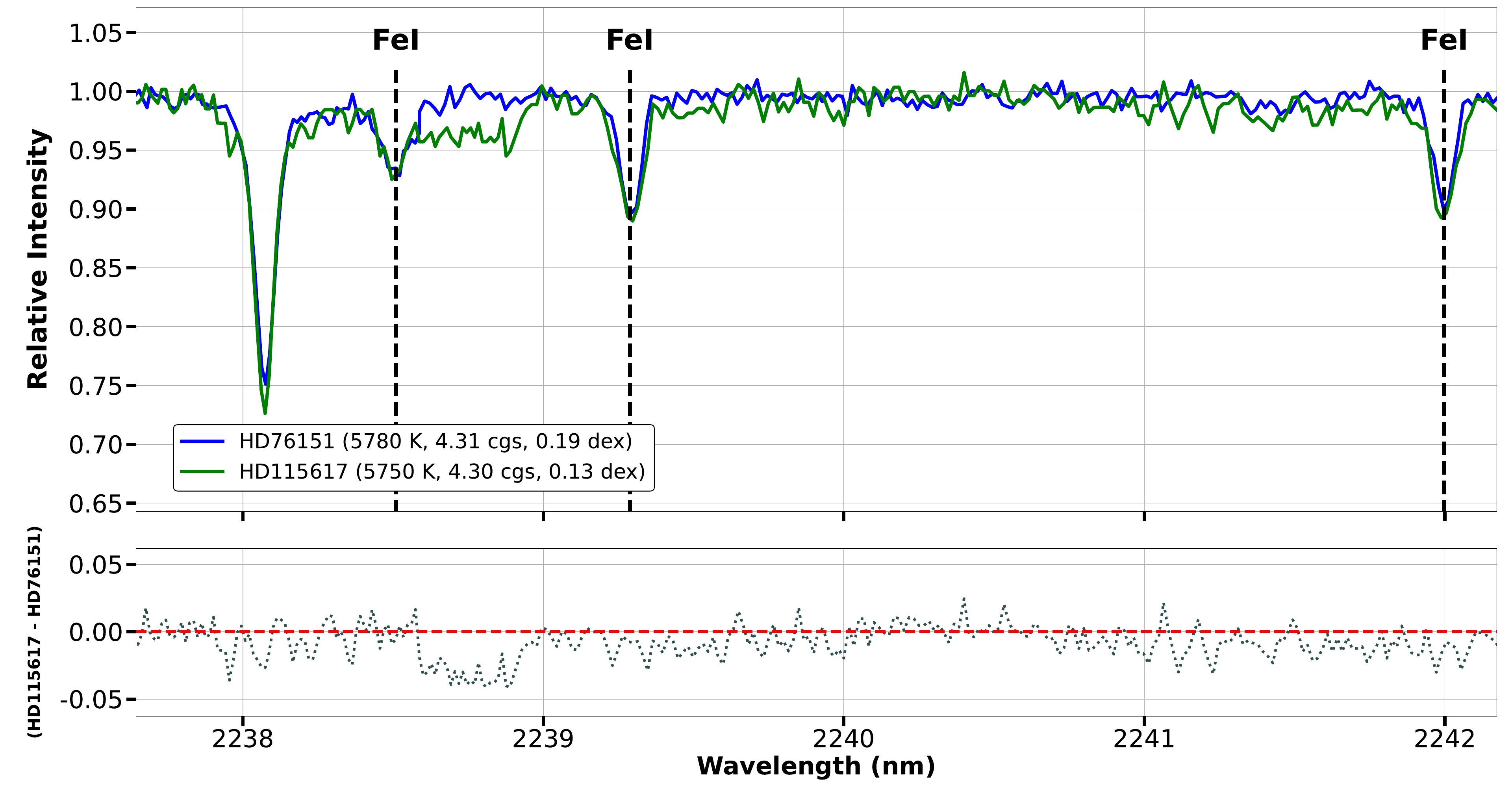}
    \caption{Comparison of the telluric- and RV-corrected H- and K-band spectra of HD\,76151 (blue) and HD\,115617 (green), taken from the IGRINS Spectral Library \citep{Park2018}. Selected regions around Fe\,{\sc i}, Fe\,{\sc ii}, Ca\,{\sc ii}, and Ti\,{\sc i} features are shown to illustrate the relative strengths and shapes of near-infrared lines used in this study.}
    \label{fig:spectrum_comparison}
\end{figure*}

\end{document}